\documentclass[12pt]{article}
\usepackage{amsmath}
\usepackage{amssymb}
\usepackage{psfig}
\date{}

\begin{document}

\title{A New Approach to Impulsive Rendezvous near Circular Orbit}

\author{ Thomas Carter \\
Professor Emeritus, Department of Mathematics,\\
Eastern Connecticut State University,\\
Willimantic, CT  06226 \thanks {e-mail: cartert@easternct.edu.  
} \\
and \\
Mayer Humi \\
Professor, Department of Mathematical Sciences, \\
Worcester Polytechnic Institute, \\
Worcester, MA 01609 \thanks {e-mail: mhumi@wpi.edu. }}

\maketitle

\thispagestyle{empty}

\newpage
\begin{abstract}

A new approach is presented for the problem of optimal impulsive rendezvous
of a spacecraft in an inertial frame near a circular orbit in a Newtonian
gravitational field. The total characteristic velocity to be minimized is 
replaced by a related characteristic-value function and this related
optimization problem can be solved in closed form. The solution of this 
problem is shown to approach the solution of the original problem in the 
limit as the boundary conditions approach those of a circular orbit. 
Using a form
of primer-vector theory the problem is formulated in a way that leads to
relatively easy calculation of the optimal velocity increments. A certain
vector that can easily be calculated from the boundary conditions
determines the number of impulses required for solution of the
optimization problem and also is useful in the computation of these
velocity increments.

Necessary and sufficient conditions for boundary
conditions to require exactly three nonsingular non-degenerate impulses
for solution of the related optimal rendezvous problem, and a means of
calculating these velocity increments are presented.  If necessary these
velocity increments could be calculated from a hand calculator containing
trigonometric functions. A simple example of a three-impulse rendezvous
problem is solved and the resulting trajectory is depicted.

Optimal non-degenerate nonsingular two-impulse rendezvous for the related 
problem is found to
consist of four categories of solutions depending on the four ways the
primer vector locus intersects the unit circle. Necessary and sufficient
conditions for each category of solutions are presented. The region of 
the boundary values that admit each category of solutions 
of the related problem are found,
and in each case a closed-form solution of the optimal velocity increments 
is presented. Some examples are simulated. Similar results are presented 
for the simpler optimal rendezvous that require only one-impulse.
For brevity degenerate and singular solutions are not discussed in detail,
but should be presented in a following study

Although this approach is thought to provide simpler computations
than existing methods, its main contribution may be in establishing a 
new approach to the more general problem.

\end{abstract}

\thispagestyle{empty}

\newpage

\section{Introduction}

It has been a convenient practice to model spacecraft orbital 
and trajectory problems having relatively high thrusts over short 
intervals of time by discontinuous jumps in velocity, but retaining 
continuity of the position vector at the time at which the discontinuity 
in the velocity appears. Problems that are modeled in this way are 
called {\bf impulsive orbital} or {\bf impulsive trajectory} problems.
The impulsive problems considered here are based on the restricted 
two-body problem, (i.e. a particle of mass in a Newtonian gravitational 
field emanating from a point source). These minimization problems consist 
of the determination of a finite set of velocity increments 
$\Delta {\bf v}_1,\,...\,\Delta {\bf v}_k$ and the values of the true 
anomaly  $\theta_1\, ... \, \theta_k$ at which they are applied, to minimize 
the total characteristic velocity $\Sigma |\Delta {\bf v}_i|$ subject to 
the two-body equations of motion, the initial position 
${\bf r}(\theta_0)={\bf r}_0$, initial velocity ${\bf v}(\theta_0)={\bf v}_0$, 
the terminal position  ${\bf r}(\theta_f)={\bf r}_f$
and the terminal  velocity ${\bf v}(\theta_f)={\bf v}_f$  where $\theta_0$  
and $\theta_f$ denote respectively the initial and terminal values of the 
true anomaly. If ${\bf r}_0,\,{\bf v}_0,\,{\bf r}_f,\, {\bf v}_f$ are 
specified points in $\mathbb{R}^3$ the minimization problem will be called 
an {\bf optimal rendezvous problem}.  If ${\bf r}_0,\,{\bf v}_0$ or 
${\bf r}_f,\,{\bf v}_f$ or both are arbitrary on a nontrivial arc of 
a fixed two-body orbit, the minimization problem will be called an 
{\bf optimal transfer problem}. Clearly, any solution of an optimal 
transfer problem found at the points ${\bf r}(\theta_0),{\bf v}(\theta_0),
{\bf r}(\theta_f),{\bf v}(\theta_f)$ also defines a solution of an 
optimal rendezvous problem having those end conditions.

Apparently the first significant publication of an impulsive orbital 
problem was the Hohmann Transfer [1], an elliptical arc, tangent to 
two circular orbits, that appeared in 1925. This was followed in 1929 
by the Oberth Transfer [2], a two-impulse transfer connecting a circular 
orbit to a hyperbolic orbit through an ellipse with apsides touching 
the circle and the center of attraction. The second impulse at (or near) 
the center of attraction sends the craft into hyperbolic speed.

Some of the earliest studies of orbital maneuvers were done 
by Contensou [3, 4] and Lawden [5]. Surveys of much early work were done 
by Edelbaum [6], Bell [7], Robinson [8], and Gobetz and Doll [9]. An 
excellent introduction to the subject is found in the book by Marec [10].

In the fifties it was discovered that the total characteristic velocity 
can be reduced over that of the Hohmann Transfer by the addition of a 
third impulse if one of the circular orbits is much larger than the other 
[11, 12, 13]. This might lead one to expect that some minimum total 
characteristic velocity problems for a single Newtonian gravitational 
source would require at least three impulses. In 1965 a paper appeared 
by Marchal [14] showing that a bounded version of the bi-elliptic transfer 
was indeed a three-impulse optimal transfer.

Recently this problem has been revisited by Pontani [15] who through 
calculus and a simple graphical technique separated the optimal two-impulse 
solutions (Hohmann) from the optimal three-impulse solutions (bi-elliptic).

Although three-impulse solutions to the optimal rendezvous problems and 
the optimal transfer problems exist, they are somewhat rare in the 
literature. Ting [16] in 1960 and Marchal [14] in 1965 discovered 
that at most three impulses are sufficient to solve the optimal rendezvous 
or transfer problem. There are solutions having more than three impulses, 
but these solutions are either degenerate or singular solutions, and also 
have a three-impulse solution containing the same total characteristic 
velocity as those having more than three-impulses.

Apparently the required number of impulses for optimality is dependent on 
the way the equations of motion are modeled. In relative-motion studies 
in which the equations of motion are linearized about a point in Keplerian 
orbit as many as four impulses can be required for optimality of the 
planar problem. It has been shown by Neustadt [17], and others [18-20] 
that the maximum number of impulses needed for linear equations is the same 
as the number of state variables, that is, four for planar problems. In 
1969 Prussing [21] displayed optimal four-impulse rendezvous maneuvers 
using equations linearized about a point in circular orbit. This problem 
was revisited much later by Carter and Alvarez [22]. Calculations of 
several optimal four-impulse rendezvous maneuvers and a degenerate 
five-impulse rendezvous maneuver were presented by Carter and Brient [20] 
in 1995 using equations linearized about elliptical orbits.

Optimal velocity increments in impulsive problems are generally calculated 
in one of three ways: The solution of a Lambert's problem, primer-vector 
analysis, or parameter optimization such as through a nonlinear programming 
algorithm. It is common practice to combine these methods in the solution 
of specific problems. In the present paper we employ a form of primer-vector 
analysis.

The primer vector was introduced by Lawden [5] as the part of the adjoint 
vector corresponding to the velocity that satisfied certain necessary 
conditions for optimality of a trajectory. Lion and Handelsman [23] 
summarized these necessary conditions and showed how to construct improvement 
in the velocity increments based on primer vector analysis. Prussing [19] 
showed that the necessary conditions of Lawden and Lion and Handelsman are 
also sufficient if the equations of motion are linear.

The present paper uses equations of motion that describe the restricted two-body problem, but the analysis is confined to near-circular orbits. This 
limitation is caused by an approximation which is highly accurate when 
the radial speed is small compared with the tangential speed. 
This simplification has been used successfully by the authors in studies that 
involve atmospheric drag [24, 25].  
We replace the total characteristic velocity which is the original 
cost function by a related cost function which is useful near a nominal
circular orbit, and present closed-form solutions to this related 
optimization problem. These results should be more accurate than those based
on the approximate Clohessy-Wiltshire equations [28] because the
linearization used herein is not approximate.  
We remark that the trajectories described 
between impulses are not approximations, and the solutions we present are 
optimum in terms of the model that we use. We may therefore use the 
terminology "optimal solutions". These velocity increments, however, are not 
accurate unless the trajectory is near a circular orbit.  

This paper uses a form of primer-vector analysis in a novel way. It has 
been well known for ages that in the restricted two-body problem, the 
equation of motion involving the magnitude $r$ of the position vector 
${\bf r}$ can be transformed to an equation linear in its reciprocal. 
This forms the basis of a set of linear equations that describe the 
motion of a particle of mass in the transformed variables. The impulsive 
minimization problem is then formulated in terms of the resulting linear 
equations, and is amenable to a specific theory of impulsive linear 
rendezvous developed by Carter and Brient [26, 20, 27]. The definition 
of the primer vector for linear equations is in a more general form than 
that of Lawden[5] or Lion and Handelsman[23], but the more general form is 
needed for use with the necessary and sufficient conditions for solution 
of the impulsive minimization problem studied in this paper.

In this new formulation the state variables $r$, the radial distance from 
the center of attraction, ${\dot r}$ its time rate of change, and $\theta$
, the true anomaly have been replaced by variables $y_1,y_2,y_3$
to be defined in the next section, called {\it transformed variables}. 
The transformed state vector ${\bf y}= (y_1,y_2,y_3)^{T}$ satisfies a 
linear differential equation. It is straightforward to define a 
transformed state transition matrix. The primer vectors are shown to 
be ellipses for this problem. It will be shown that for three-impulse 
solutions and two-impulse solutions the primer vectors are completely 
determined by the initial true anomaly $\theta_0$ and the terminal true 
anomaly $\theta_f$.

Given $\theta_0$ and $\theta_f$ the initial transformed state vector 
${\bf y}_0$ and the terminal transformed state vector ${\bf y}_f$ are used 
to define a generalized boundary point ${\bf z}_f(\bf{y}_0,\bf{y}_f)$. 
The geometric structure of the set of generalized boundary points 
associated with an optimal primer vector is a convex cone [20]. Given, 
$\theta_0$ and $\theta_f$ these convex cones partition the set of 
generalized boundary points into a simplex of convex conical sets. 
The three-dimensional cones contain 
the generalized boundary points that admit three-impulse solutions. 
Points on the boundary of such a cone admit degenerate three-impulse 
solutions, that is, one or more of the three velocity increments is zero. 
Although three-impulse solutions are somewhat rare in the literature for 
the restricted two-body problem, this analysis shows that there are plenty 
of them near circular orbits. It follows also that the two-dimensional 
cones define areas that admit two-impulse solutions, and the one-dimensional 
cones, that is, the straight lines admit the one-impulse solutions. 
Of course the vertex of any cone, that is, the origin, admits only 
the zero-impulse solution, a coasting trajectory.

For the original problem having boundary conditions in the vicinity of a 
nominal circular orbit, near optimal velocity increments can be obtained in
closed form. For more general boundary conditions the velocity increments 
calculated from the related problem could be useful as good initial guesses 
in a numerical optimization program, and should obviate the need to solve
Lambert's problem.

The paper begins with the general formulation of the problem and the related 
problem for boundary conditions near circular orbits, then presents some 
analysis and simulations of non-degenerate three-impulse solutions 
and various kinds of non-degenerate two-impuse solutions of 
the related problem and precisely determines the sets of boundary conditions
corresponding to each type,followed by some results on one-impulse solutions.
For brevity detailed discussion of degenerate and singular solutions is postponed to a later work.

\setcounter{equation}{0}
\section{Building the Model}

The equation of motion of a particle in a Newtonian gravitational field 
about a homogeneous spherical planet is
\begin{equation}
\label{2.1}
\ddot{r}= -\frac{\mu}{r^3}\bf{r}
\end{equation}
where ${\bf r}$ is the position vector with respect to the center of the 
planet,  
$\mu$ is the product of the universal gravitational constant and the mass 
of the spherical planet, the upper dot indicates differentiation with 
respect to time $t$ and $r=|{\bf r}|=({\bf r}\cdot{\bf r})^{1/2}$ where the inside 
dot denotes the usual inner product, and $|{\bf r}|$ represents the 
Euclidean norm of the vector ${\bf r}$.

\subsection{The Optimal Impulsive Rendezvous Problem} 

In polar coordinates (\ref{2.1}) becomes
\begin{equation}
\label{2.2}
r\ddot{r} +2\dot{r}\dot{\theta} =0
\end{equation}
\begin{equation}
\label{2.3}
\ddot{r} -r\dot{\theta}^2 = -\frac{\mu}{r^2}
\end{equation}
where $\theta$ represents the true anomaly. The radial and transverse 
velocities are denoted respectively by $v_r=\dot{r}$ and 
$v_{\theta}=r\dot{\theta}$. Throughout we shall use 
subscript notation to refer to any variable $x$ at time $t_i$ as
$x_i= x(t_i)$.

We shall consider the addition of velocity increments 
$\Delta {\bf v}_i = (\Delta v_{r_i},\Delta v_{\theta_i})^T$ at time $t_i$
for $i=1,...,k$ such that
\begin{equation}
\label{2.4}
\displaystyle \lim_{t \rightarrow t_i+} \dot{\bf r}(t) =
\displaystyle \lim_{t \rightarrow t_i-} \dot{\bf r}(t) +\Delta {\bf v}_i.
\end{equation}

In this formulation $r$ and $\theta$ are continuous everywhere and the 
differential 
equations (\ref{2.2}) and (\ref{2.3}) are satisfied everywhere on an 
interval $t_0 \le t \le t_f$ except at the instants where the velocity 
satisfies the jump discontinuities (\ref{2.4}). At the ends of the 
time interval, (\ref{2.4}) becomes
\begin{equation}
\label{2.5}
\displaystyle \lim_{t \rightarrow t_0+} \dot{\bf r}(t) = 
\dot{\bf r}(t_0) + \Delta {\bf v}_1,\,\,\, t_1=t_0
\end{equation}
\begin{equation}
\label{2.6}
\dot{\bf r}(t_f) =
\displaystyle \lim_{t \rightarrow t_f-} \dot{\bf r}(t) + 
\Delta {\bf v}_k,\,\,\, t_k=t_f. 
\end{equation}

The rendezvous problem requires that the initial conditions
\begin{equation}
\label{2.7}
{\bf r}(t_0) = {\bf r}_0,\,\,\, \theta(t_0)=\theta_0
\end{equation}
\begin{equation}
\label{2.8}
\dot{\bf r}(t_0) = {\bf v}_{r_0},\,\,\, {\bf v}_{\theta}(t_0)=
{\bf v}_{\theta_0}
\end{equation}
and the terminal conditions,
\begin{equation}
\label{2.9}
{\bf r}(t_f) = {\bf r}_f,\,\,\, \theta(t_f)=\theta_f
\end{equation}
\begin{equation}
\label{2.10}
\dot{\bf r}(t_f) = {\bf v}_{r_f},\,\,\, {\bf v}_{\theta}(t_f)=
{\bf v}_{\theta_f}
\end{equation}
be satisfied. The total characteristics velocity is defined by
\begin{equation}
\label{2.11}
c = \displaystyle\sum_{i=1}^{k} |\Delta {\bf v}_i|.
\end{equation}

The optimal impulsive rendezvous problem is stated as follows:

Find a positive integer k, a finite set $\{t_1,\,...\,t_k\}$ on the 
interval $\{t_0 \le t \le t_f\}$, and a set of velocity increments 
$\{\Delta {\bf v}_1,\,...\,\Delta {\bf v}_k\}$ such that the differential 
equations (\ref{2.2}) and (\ref{2.3}) are satisfied except on the finite set
$\{t_1,\,...\,t_k\}$ where (\ref{2.4})-(\ref{2.6}) are satisfied,
the boundary conditions (\ref{2.7})-(\ref{2.10}) are satisfied, and the 
total characteristic velocity (\ref{2.11}) is minimized.

Similarly one can state an optimal impulsive transfer problem by replacing 
the requirement that the initial conditions, as described by (\ref{2.7}), 
(\ref{2.8}) and the terminal conditions as described by (\ref{2.9}), 
(\ref{2.10}) be fixed with the requirement that either or both be 
contained on a Keplerian orbit. Mathematically, the optimal impulsive 
transfer problem is not significantly different from the
optimal impulsive rendezvous problem. In fact, an optimal solution 
satisfying the fixed end conditions (\ref{2.7})-(\ref{2.10}) also defines 
an optimal solution of the transfer problem from the Keplerian orbit 
containing the fixed point described by ({\ref{2.7}), (\ref{2.8}) to the 
Keplerian orbit that contains the fixed point described by (\ref{2.9}), 
(\ref{2.10}) if the time interval is sufficiently large.

\subsection{The Orbit Equation}

We sketch the derivation of the orbit equation which has been well known 
for many years.

Multiplying (\ref{2.2}) by $r$, we have the derivative of
\begin{equation}
\label{2.12}
r^2\dot{\theta} = h.
\end{equation}
The constant $h$ is the angular momentum. We may use (\ref{2.12}) and 
the chain rule to change the independent variable in (\ref{2.3}) to 
$\theta$. We use primes to indicate differentiation with respect to 
$\theta$. By this change of variable and noting from (\ref{2.12}) 
that $\theta$ is monotone in $t$, the fourth order system (\ref{2.2}), 
(\ref{2.3}) on the interval $t_0 \le t \le t_f$ is replaced by the 
following third order system on the interval 
$\theta_0 \le \theta \le \theta_f$:
\begin{equation}
\label{2.13}
r(\theta)r^{\prime\prime}(\theta)-2r^{\prime}(\theta)^2=r(\theta)^2-
\mu r(\theta)^3/h^2
\end{equation}
\begin{equation}
\label{2.14}
h^{\prime}(\theta) =0.
\end{equation}

\subsection{Transformation to Linear Equations}

Employing the well-known change-of-variable $y=1/r$ in the orbit equation 
(\ref{2.13}) and manipulating some symbols, we obtain the linear
differential equation:
\begin{equation}
\label{2.15}
y^{\prime\prime} +y =\mu/h^2.
\end{equation}

We now assign the state variables:
\begin{equation}
\label{2.16}
y_1=y
\end{equation}
\begin{equation}
\label{2.17}
y_2 =y^{\prime}=-\dot{r}/h
\end{equation}
\begin{equation}
\label{2.18}
y_3=\mu/h^2.
\end{equation}
In terms of these state variables, the equations (\ref{2.13}), (\ref{2.14}) 
are transformed into the following set of linear equations:
\begin{equation}
\label{2.19}
y_1^{\prime}=y_2
\end{equation}
\begin{equation}
\label{2.20}
y_2^{\prime}=-y_1+y_3
\end{equation}
\begin{equation}
\label{2.21}
y_3^{\prime}=0.
\end{equation}
The initial conditions are
\begin{equation}
\label{2.22}
y_1(\theta_0)=y_{10}=\frac{1}{r_0}
\end{equation}
\begin{equation}
\label{2.23}
y_2(\theta_0)=y_{20}=-\dot{r}(\theta_0)/h_0=-\frac{v_{r_0}}{r_0v_{\theta_0}}
\end{equation}
\begin{equation}
\label{2.24}
y_3(\theta_0)=y_{30}=\frac{\mu}{h_0^2}=\frac{\mu}{(r_0v_{\theta_0})^2}
\end{equation}
and the terminal conditions are
\begin{equation}
\label{2.25}
y_1(\theta_f)=y_{1f} =\frac{1}{r_f}.
\end{equation}
\begin{equation}
\label{2.26}
y_2(\theta_f)=y_{2f}=-\dot{r}(\theta_f)/h_f=-\frac{v_{r_f}}{r_fv_{\theta_f}}
\end{equation}
\begin{equation}
\label{2.27}
y_3(\theta_f)=y_{3f}=\frac{\mu}{h_f^2}=\frac{\mu}{(r_fv_{\theta_f})^2}
\end{equation}
where the subscript $0$ refers to the initial value of a variable and the
subscript $f$ refers to the terminal value.
Formulas for recovery of the original variables are
\begin{equation}
\label{2.28}
r=1/y_1
\end{equation}
\begin{equation}
\label{2.29}
\dot{\theta}=y_1^2\sqrt{\mu/y_3}
\end{equation}
\begin{equation}
\label{2.30}
\dot{r}=-y_2\sqrt{\mu/y_3}
\end{equation}
\begin{equation}
\label{2.31}
v_{\theta}=y_1\sqrt{\mu/y_3}.
\end{equation}

\subsection{Restatement of the Optimal Impulsive Rendezvous Problem}

At $\theta_i,\,\,i=1, ...,k$ we increment the velocity 
${\bf\dot{r}}(\theta_i)$ by 
$\Delta {\bf v_i}=(\Delta v_{r_i},\Delta v_{\theta_i})^T$ so that 
(\ref{2.4})-(\ref{2.6}) are satisfied. This causes
an instantaneous jump in the state variables $y_2$ and $y_3$ but $y_1$
is continuous. The increments in $y_2$ and $y_3$  caused by the increments 
in the velocity at $\theta_i$ are determined respectively from (\ref{2.17}) 
and (\ref{2.18}).
\begin{equation}
\label{2.32}
\Delta y_{2i}=-\frac{1}{h_i}(\Delta v_{r_i}-\frac{v_{r_i}}{v_{\theta_i}}
\Delta v_{\theta_i}),
\,\,\, i=1,...,k
\end{equation}
\begin{equation}
\label{2.33}
\Delta y_{3i}=-\frac{2\mu r_i}{h_i^3}\Delta v_{\theta_i},
\,\,\, i=1,...,k.
\end{equation}

We now consider the imposition of a velocity increment 
$\Delta{\bf v}_i=(\Delta v_{r_i},\Delta v_{\theta_i})^T$ at $\theta_i$ 
for $i=1,...,k$ on the interval $\theta_0 \le \theta \le \theta_f$ 
such that
\begin{equation}
\label{2.34}
\displaystyle \lim_{\theta \rightarrow \theta_i+} y_2(\theta) =
\displaystyle \lim_{\theta \rightarrow \theta_i-} y_2(\theta) +\Delta y_{2i},
\,\,\,
\displaystyle \lim_{\theta \rightarrow \theta_i+} y_3(\theta) =
\displaystyle \lim_{\theta \rightarrow \theta_i-} y_3(\theta) +\Delta y_{3i}.
\end{equation}

If $\theta_i$ is an end point (\ref{2.34}) becomes
\begin{equation}
\label{2.35}
\displaystyle \lim_{\theta \rightarrow \theta_0+} y_2(\theta) =
y_2(\theta_0) +\Delta y_{2i},
\,\,\,
\displaystyle \lim_{\theta \rightarrow \theta_0+} y_3(\theta) =
y_3(\theta_0) +\Delta y_{3i}
\end{equation}
\begin{equation}
\label{2.36}
y_2(\theta_f)=\displaystyle \lim_{\theta \rightarrow \theta_f-} y_2(\theta) 
+\Delta y_{2i},
\,\,\,
y_3(\theta_f)=\displaystyle \lim_{\theta \rightarrow \theta_f-} y_3(\theta)
= \Delta y_{3i}.
\end{equation}

The optimal impulsive rendezvous problem can now be restated as that
of finding a positive integer $k$, a finite set $K=\{\theta_1,...,\theta_k\}$
on the interval $\theta_0 \le \theta \le \theta_f$, and a set of velocity 
increments $\{\Delta {\bf v}_i \in \mathbb{R}^2, i=1,..,k\}$ to minimize the total 
characteristic velocity (\ref{2.11}) subject to the linear differential 
equations (\ref{2.19})-(\ref{2.21}) that are valid except on the
set $K$ where (\ref{2.32})-(\ref{2.36}) are satisfied, and that satisfy 
the boundary conditions (\ref{2.22})-(\ref{2.27}).

\subsection{Transformed Velocity Increments and the Related Problem}

We transform the velocity increments $\Delta v_{r_i}$ and 
$\Delta v_{\theta_i}$, $i=1,...,k$ as follows:
\begin{equation}
\label{2.37}
\Delta V_{1i} = \frac{1}{h_i}\left(\Delta v_{r_i} -\frac{v_{r_i}}{v_{\theta_i}}
\Delta v_{\theta_i}\right)
\end{equation}
\begin{equation}
\label{2.38}
\Delta V_{2i} = \frac{\mu r_i}{h_i^3} \Delta v_{\theta_i}.
\end{equation}
With this transformation, we define the increments
$\Delta {\bf y}_i \in \mathbb{R}^3,\,\,\, i=1,...,k$ from (\ref{2.32}) and 
(\ref{2.33}), caused by the velocity increments as
\begin{equation}
\label{2.39}
\Delta {\bf y}_i= B\Delta {\bf V}_i                 
\end{equation}
where $\Delta {\bf V}_i = (\Delta V_{1i}, \Delta V_{2i})^T, i=1,...,k$ and
\begin{eqnarray}
\label{2.40}
B = \left(\begin{array}{cc}
\;\;0 &0  \\
-1 &0 \\
\;\;0 &-2 \\
\end{array}
\right).
\end{eqnarray}
For $i=1,...,k$ the expressions (\ref{2.34})-(\ref{2.36}), in vector 
notation become
\begin{equation}
\label{2.41}
\displaystyle \lim_{\theta \rightarrow \theta_i+}{\bf y}(\theta) =
\displaystyle \lim_{\theta \rightarrow \theta_i-}{\bf y}(\theta) +
\Delta{\bf y}_i,\,\,\, \theta_0 < \theta_i < \theta_f,
\end{equation}
\begin{equation}
\label{2.42}
\displaystyle \lim_{\theta \rightarrow \theta_0+}{\bf y}(\theta) =
{\bf y}(\theta_0)+\Delta{\bf y}_i,\,\,\, \theta_i =\theta_0,
\end{equation}
\begin{equation}
\label{2.43}
{\bf y}(\theta_f)=\displaystyle \lim_{\theta \rightarrow \theta_0-}{\bf y}
(\theta) +\Delta{\bf y}_i,\,\,\, \theta_i =\theta_f.
\end{equation}

In vector form the boundary conditions (\ref{2.22})-(\ref{2.27}) are written
\begin{equation}
\label{2.44}
{\bf y}(\theta_0)={\bf y}_0
\end{equation}
\begin{equation}
\label{2.45}
{\bf y}(\theta_f)={\bf y}_f
\end{equation}
where  ${\bf y}_0=(y_{10},y_{20},y_{30})^T$ and 
${\bf y}_f=(y_{1f},y_{2f},y_{3f})^T$ are defined from the respective 
right-hand sides of (\ref{2.22})-(\ref{2.24}) and (\ref{2.25})-(\ref{2.27}). 
For $\theta \notin K$ the linear equations (\ref{2.19})-(\ref{2.21}) are
\begin{equation}
\label{2.46}
{\bf y}^{\prime}=A{\bf y}
\end{equation}
where
\begin{eqnarray}
\label{2.47}
A = \left(\begin{array}{ccc}
\;\;0 &1 &0 \\
-1 &0 &1\\
\;\;0 &0 &0 \\
\end{array}
\right).
\end{eqnarray}

We now replace the cost function (\ref{2.11}) by the related cost function
\begin{equation}
\label{2.48}
C=\displaystyle\sum_{i=1}^{k} |\Delta {\bf V}_i| .
\end{equation}
The related optimal impulsive rendezvous problem is that of finding a set 
$K=\{\theta_1,...,\theta_k\}$ on the interval $\theta_0 \le \theta \le\theta_f$
and velocity increments $\Delta{\bf V}_1,...,\Delta {\bf V}_k \in \mathbb{R}^2$
that minimize the related cost function $C$ subject to the linear 
differential equation (\ref{2.46}) that is satisfied everywhere on 
the interval except the set $K$ where the velocity increments
$\Delta{\bf V}_1,...,\Delta {\bf V}_k$ are applied subject to (\ref{2.39}) 
and (\ref{2.41})-(\ref{2.43}), and the initial condition (\ref{2.44}) and 
the terminal condition (\ref{2.45}) are satisfied.

It has been shown [20] that a solution of this problem exists under 
appropriate controllability conditions.

Since the differential equation (\ref{2.46}) is linear, it is known 
[17-20] that it is sufficient to set $k=3$. For this reason we set 
$k=3$ for the remainder of this paper. This agrees with the results of 
Ting [16] and Marchal [14] although the linearized model used by 
Prussing [21] allows $k=4$.

\subsection{Necessary and Sufficient Conditions}

A fundamental matrix solution $\Phi$ associated with a matrix A satisfies 
the matricial differential equation $\Phi^{\prime}=A\Phi$.
It is not difficult to see that
\begin{eqnarray}
\label{2.49}
\Phi(\theta) = \left(\begin{array}{ccc}
\;\;\cos\theta &\sin\theta &1 \\
-\sin\theta &\cos\theta &0\\
\;\;0 &0 &1 \\
\end{array}
\right)
\end{eqnarray}

is a fundamental matrix solution associated with (\ref{2.47}). The 
inverse of $\Phi$ is
\begin{eqnarray}
\label{2.50}
\Phi^{-1}(\theta) = \left(\begin{array}{ccc}
\;\;\cos\theta &-\sin\theta &-\cos\theta \\
\sin\theta &\cos\theta &-\sin\theta \\
\;\;0 &0 &1 \\
\end{array}
\right) .
\end{eqnarray}

Lawden's definition [5] of the primer vector is not adequate for the
work herein. We use the definition [20,26]:
\begin{equation}
\label{2.51}
{\bf p}(\theta) =R(\theta)^T{\bf \lambda}
\end{equation}
where ${\bf \lambda} \in \mathbb{R}^3$ and
\begin{equation}
\label{2.52}
R(\theta)=\Phi^{-1}(\theta) B.
\end{equation}

We briefly review some previous results [20]. 

If $\Delta {\bf V}_i$ is a velocity impulse at $\theta_i$ it follows 
from (\ref{2.39}), (\ref{2.46}), and (\ref{2.41})-(\ref{2.43}) that
\begin{equation}
\label{2.53}
{\bf y}(\theta)= \Phi(\theta)\Phi(\theta_i)^{-1}({\bf y}_i +B\Delta {\bf V}_i)
\end{equation}
on the interval $\theta_i < \theta < \theta_{i+1}$ if there is a 
velocity impulse at $\theta_{i+1}$, otherwise $\theta_i < \theta < \theta_f$. 
If there are a total of $k$ impulses it follows from (\ref{2.39}), 
(\ref{2.41})-(\ref{2.44}) and (\ref{2.53}) that
\begin{equation}
\label{2.54}
{\bf y}(\theta)= \Phi(\theta)\Phi(\theta_0)^{-1}{\bf y}_0 +
\Phi(\theta)\displaystyle\sum_{i=1}^{k}\Phi(\theta_i)^{-1} B\Delta {\bf V}_i
\end{equation}
where $\theta_k \le \theta \le \theta_{f}$. Applying the terminal condition
(\ref{2.45}) we obtain
\begin{equation}
\label{2.55}
{\bf y}_f= \Phi(\theta_f)\Phi(\theta_0)^{-1}{\bf y}_0 +
\Phi(\theta_f)\displaystyle\sum_{i=1}^{k}\Phi(\theta_i)^{-1} B
\Delta {\bf V}_i.
\end{equation}
Introducing the vector ${\bf z}_f$, we write this expression as
\begin{equation}
\label{2.56}
{\bf z}_f= 
\displaystyle\sum_{i=1}^{k}\Phi(\theta_i)^{-1} B\Delta {\bf V}_i
\end{equation}
where all of the information about the boundary conditions is contained 
in the definition:
\begin{equation}
\label{2.57}
{\bf z}_f=\Phi(\theta_f)^{-1}{\bf y}_f -\Phi(\theta_0)^{-1}{\bf y}_0.
\end{equation}

Setting $\alpha_i= |\Delta \bf{V}_i|$, $i=1 \ldots k$ we have all of the 
terminology needed to state the following result from previous work.[20] 

{\bf Theorem}: For a minimizing $k$-impulse solution of the related optimal 
impulsive 
rendezvous problem, it is necessary and sufficient that
\begin{equation}
\label{2.58}
\Delta {\bf V}_i=0,\,\,\, or \,\,\, \Delta {\bf V}_i=-p(\theta_i)\alpha_i,\,\,\, i=1,\ldots,k
\end{equation}
\begin{equation}
\label{2.59}
\alpha_i \ge 0,\,\,\, i=1,\ldots,k
\end{equation}
\begin{equation}
\label{2.60}
\displaystyle\sum_{i=1}^{k} R(\theta_i)p(\theta_i)\alpha_i = -z_f
\end{equation}
\begin{equation}
\label{2.61}
\Delta {\bf V}_i =0, \,\,\, or \,\,\, |p(\theta_i)| =1,\,\,\, i=1,\ldots,k
\end{equation}
\begin{equation}
\label{2.62}
\theta_i=\theta_0,\,\,\, or \,\,\, |p(\theta_i)|^{\prime}=0,\,\,\, or \,\,\, 
\theta_i=\theta_f,\,\,\, i=1,\ldots,k
\end{equation}
\begin{equation}
\label{2.63}
|p(\theta)| \le 1,\,\,\, \theta_0 \le \theta \le \theta_f.
\end{equation}

{\bf Proof}: See Sec $3.2$ and $3.3$ of Reference 20.

In this theorem $k$ is the specified number of impulses and can be any
non-negative integer sufficiently large that (\ref{2.58})-(\ref{2.63})
are satisfied. It has been shown [17-20] to be unnecessary for $k$
to be larger than the dimension of the state space (i.e. dim ${\bf z}_f$
which is $3$ for the present problem), although there can be degenerate 
solutions where $k > dim\, {\bf z}_f$. A {\bf degenerate solution} is 
a minimizing solution for which there is an equivalent minimizing solution 
where some $\alpha_i=0$ in (\ref{2.59}).  If a solution
is degenerate an equivalent minimizing solution exists having fewer
than k impulses so that, effectively, the number 
$k$ could be reduced. 

For many boundary conditions there can be minimizing solutions for 
$k < dim\, {\bf z}_f$. For this reason we shall set $k$ equal to the 
number of elements in the set
$$
K=\{\theta |\, |{\bf p}(\theta)| =0,\,\,\, \theta_0 \le \theta \le \theta_f\}
$$
if this set is finite. Since the function 
$f(\theta)={\bf p}^T(\theta){\bf p}(\theta) -1$
is analytic either $K$ is  finite (i.e. $K=\{\theta_1,\ldots,\theta_k\})$
or else $K$ is the entire interval $\theta_0 \le \theta \le \theta_f$.
In the latter case a minimizing solution is called ${\bf singular}$
and $k$ and $\theta_1,\ldots,\theta_k$ are arbitrary as long as $k$ is 
large enough that (\ref{2.58})-(\ref{2.63}) are satisfied, as they will be 
if $k = dim\, {\bf z}_f$.

In the present paper we shall isolate those boundary conditions where 
$k=3$ and investigate minimizing non-degenerate three-impulse solutions. 
In the following paper we shall investigate the non-degenerate two-impulse
and one-impulse solutions. In the final paper we shall present some results 
on singular and degenerate solutions.

We shall introduce some terminology based on (\ref{2.62}). We refer to 
$\theta_0$ as an {\it initial value} of $\theta$  and $\theta_f$ 
as a {\it terminal value} of $\theta$. If an optimal impulsive rendezvous 
has a velocity increment assigned at $\theta_0$ 
(i.e. $\Delta {\bf V}_i= \Delta {\bf V}_0)$ it is called an 
{\it initial velocity impulse}; if it has a velocity increment assigned 
at $\theta_f$ (i.e. $\Delta {\bf V}_i= \Delta {\bf V}_f$) it is called a 
{\it terminal velocity impulse}. A value $\theta_i$ is called a 
{\it stationary value} if $|p(\theta_i)|^{\prime}=0$ 
and $p(\theta_i)=1$. A velocity 
increment $\Delta {\bf V}_i$ assigned at a stationary value $\theta_i$ 
is called 
{\it a stationary velocity impulse}. If the one-sided derivative of
$|p(\theta)|$  is zero at $\theta_0$, then $\theta_0$ is both an 
{\it initial value} and a {\it stationary value}, and $\Delta {\bf V}_0$ is 
both an {\it initial} and a {\it stationary velocity impulse}. Similarly, 
if the one-sided derivative of $|p(\theta)|$ is zero at $\theta_f$, 
then $\theta_f$ is both a {\it terminal value} and a {\it stationary value}
and $\Delta {\bf V}_f$ is both a terminal and a stationary velocity increment.

\subsection{Optimal Impulsive Rendezvous near a Nominal Orbit of Low 
Eccentricity}

Given $\theta_0$, ${\bf y}_0$, $\theta_f$, ${\bf y}_f$ an analytical 
solution will be presented for the related optimal impulsive rendezvous 
problem based on the necessary and sufficient conditions recently stated. 
It is possible to display the velocity increments, their points of 
application, and the actual arcs of Keplerian orbits that minimize
the related cost function subject to (\ref{2.39})-(\ref{2.47}). A 
practical limitation of this result is that it is the total characteristic 
velocity (\ref{2.11}) that we seek to minimize, not the related cost 
function (\ref{2.48}).

We shall show that the total characteristic velocity is approximately 
proportional to the related cost function and approaches exactness as 
${\bf y}_0$ and ${\bf y}_f$ approach end conditions associated with a 
circular orbit.

Consider a state vector ${\bar{\bf y}}(\theta)$ associated with a 
nominal Keplerian orbit where ${\bar{\bf y}}(\theta_0)={\bar{\bf y}}_0$ 
and ${\bar{\bf y}}(\theta_f)={\bar{\bf y}}_f$. Since there are no impulses 
in the nominal orbit, (\ref{2.55}) shows that 
\begin{equation}
\label{2.64}
{\bar{\bf y}}_f=\Phi(\theta_f)\Phi(\theta_0)^{-1}{\bar{\bf y}}_0.
\end{equation}
Comparing an impulsive orbit with a nominal orbit,
\begin{equation}
\label{2.65}
{\bf y}(\theta)-{\bar{\bf y}}(\theta)=\Phi(\theta)\Phi(\theta_0)^{-1}(
{\bf y}_0-{\bar{\bf y}}_0)+ \Phi(\theta)
\displaystyle\sum_{i=1}^{k}\Phi(\theta_i)^{-1}B\Delta {\bf V}_i
\end{equation}
where $k=1, 2$, or $3$. From (\ref{2.56}) and (\ref{2.57}) we observe that
\begin{equation}
\label{2.66}
{\bf z}_f-\bar{{\bf z}}_f =\Phi(\theta_f)^{-1}({\bf y}_f-{\bar{\bf y}}_f)-
\Phi(\theta_0)^{-1}({\bf y}_0-{\bar{\bf y}}_0)
\end{equation}
where $\bar{{\bf z}}_f=0$ from (\ref{2.56}). This shows that ${\bf z}_f$ 
can be made arbitrarily small by making ${{\bf y}}_f$  sufficiently 
near ${\bar{\bf y}}_f$ and ${{\bf y}}_0$ sufficiently near 
${\bar{\bf y}}_0$. Upon multiplying both sides of (\ref{2.56}) by a 
very small positive number, we see that we can select $\epsilon > 0$
and $|\Delta {\bf V}_i| < \epsilon/3$, $i=1,2,3$ where $\epsilon$ is made arbitrarily 
small by selecting ${\bf z}_f$ sufficiently small. In this manner we can 
make the cost $C < \epsilon$ and the difference (\ref{2.65}) in the 
trajectories arbitrary small.

We shall choose the nominal trajectory ${\bar{\bf y}}(\theta)$ to be 
a circular orbit or a low-eccentricity elliptical orbit. Picking end 
conditions sufficiently near the end conditions of the nominal orbit 
we establish 
a sufficiently small bound on $\Delta {\bf V}_i,\,\,i=1,2,3$ (i.e. there is 
a sufficiently small positive number $M$ such that $|\Delta {\bf V}_i| < M$ 
for $i=1,2,3$). 
This establishes a bound $N$ on $|\Delta v_{\theta_i}|$, for $i=1,2,3$
(i.e. there is a number $N > 0$ such that $|\Delta v_{\theta_i}| < N$
for $i=1,2,3$). 
We have shown that the difference (\ref{2.65}) from the nominal circular 
or low-eccentricity orbit can be made as small as possible. This and 
({\ref{2.17}) show that $|v_{r_i}|$ is much less than 
$|v_{\theta_i}|$ near a low-eccentricity nominal orbit. The conclusion 
of these arguments is that the second term in the parenthesis on the 
right-hand side of (\ref{2.37}) can be made arbitrarily small by picking 
the end conditions sufficiently near those of the nominal circular orbit.

One way to select a radius defining a nominal circular orbit is by 
setting $\bar{r}=(r_0+r_f)/2$. It is not difficult to show that the 
angular momentum $\bar{h}$ of a circular orbit of radius $\bar{r}$ is 
\begin{equation}
\label{2.67}
\bar{h}=(\mu \bar{r})^{1/2}.
\end{equation}

Using $k=3$ we now show that the total characteristic velocity (\ref{2.11}) 
can be approximated to any desired accuracy using the related cost 
(\ref{2.48}) by picking the end conditions sufficiently near these of 
a nominal circular orbit.

It follows from (\ref{2.37}) that for $i=1,2,3$
\begin{equation}
\label{2.68}
|\Delta {\bf v}_i -h_i \Delta {\bf V}_i| = \left[ ( \frac{v_{r_i}}{v_{\theta_i}})^2 +(1-\frac{\mu r_i}{h_i^2})^2\right]^{1/2}|\Delta v_{\theta_i}|.
\end{equation}
Given any $\epsilon > 0$, , there are end points sufficiently near end points of a nominal circular orbit 
such that for $i=1,2,3$
\begin{equation}
\label{2.69}
\left|\frac{v_{r_i}}{v_{\theta_i}}\right| < \frac{\epsilon}{6N},\,\,\, 
\left|\frac{\mu \bar{r}}{\bar{h}^2}-\frac{\mu r_i}{h_i^2}\right| < \frac{\epsilon}{6N}.
\end{equation}
It follows from (\ref{2.67}) and the fact that $|\Delta v_{\theta_i}| < N$
for $i=1,2,3$ that
\begin{equation}
\label{2.70}
|\Delta {\bf v}_i -h_i \Delta {\bf V}_i| < \frac{\epsilon}{6}.
\end{equation}
To complete the arguments, we note that
\begin{equation}
\label{2.71}
|\displaystyle\sum_{i=1}^{3}|\Delta {\bf v}_i| -\bar{h}\displaystyle\sum_{i=1}^{3}
|\Delta {\bf V}_i||\le \displaystyle\sum_{i=1}^{3}|\Delta {\bf v}_i -\bar{h} \Delta {\bf V}_i|
\le \displaystyle\sum_{i=1}^{3}|\Delta {\bf v}_i -h_i \Delta {\bf V}_i| +
\displaystyle\sum_{i=1}^{3}|h_i -\bar{h}||\Delta {\bf V}_i|. 
\end{equation}

Picking initial and final conditions sufficiently near the nominal 
circular orbit, for each $i=1,2,3$ we have
\begin{equation}
\label{2.72}
|h_i -\bar{h}| < \frac{\epsilon}{6M}.
\end{equation}
This establishes the final result:
\begin{equation}
\label{2.73}
|\displaystyle\sum_{i=1}^{3}|\Delta {\bf v}_i| -\bar{h}\displaystyle\sum_{i=1}^{3}
|\Delta {\bf V}_i|| < \epsilon 
\end{equation}
For trajectories near a nominal orbit of low eccentricity we have the 
approximation
\begin{equation}
\label{2.74}
\displaystyle\sum_{i=1}^{3}|\Delta {\bf v}_i| \cong \bar{h}\displaystyle\sum_{i=1}^{3}|\Delta {\bf V}_i|. 
\end{equation}
Since $\bar{h}$ is a constant we may use the cost
\begin{equation}
\label{2.75}
C = \displaystyle\sum_{i=1}^{3}|\Delta {\bf V}_i| 
\end{equation}
in these situations. After solution of this related problem one can 
calculate the actual impulses $\Delta {\bf v}_i$ in terms of 
$\Delta {\bf V}_i$ 
through (\ref{2.37}) and (\ref{2.38}) where $v_{r_i}$ and $v_{\theta_i}$
are recovered from (\ref{2.28}) - (\ref{2.31}). If the end conditions are 
very close to a nominal circular orbit one can simplify and approximate 
(\ref{2.37}) and (\ref{2.38}) by
\begin{equation}
\label{2.76}
\Delta {\bf v}_i =\bar{h}\Delta {\bf V}_i ,\,\,\, i=1,2,3.
\end{equation}

\setcounter{equation}{0}
\section{Primer Vector Analysis}

The primer vector can be a very useful tool [5, 23, 19, 20] in the analysis 
of optimal impulsive problems. In this section we survey the various 
geometric arrangements of primer vector loci for the problem addressed, 
its degeneration in the case of singular solutions, and its use in the 
calculation of three-impulse trajectories.

\subsection{Geometry of Primer Vector Loci}

From (\ref{2.40}), (\ref{2.50}), (\ref{2.52}) and (\ref{2.51}) the primer 
vector is determined in terms of the vector ${\bf \lambda}$. 
Writing ${\bf \lambda}^T=(\lambda_1,\lambda_2,\lambda_3)$ and
${\bf p}(\theta)^T=(p_1(\theta),p_2(\theta))$  it follows that
\begin{equation}
\label{2.77}
p_1(\theta)=\lambda_1\sin\theta - \lambda_2\cos\theta
\end{equation}
\begin{equation}
\label{2.78}
p_2(\theta)=2\lambda_1\cos\theta+2\lambda_2\sin\theta -2\lambda_3
\end{equation}
We set $\lambda_1=\lambda \cos\phi$ and $\lambda_2=\lambda \sin\phi$
where $\lambda =(\lambda_1^2+\lambda_2^2)^{1/2}$. 
The primer vector is therefore described by
\begin{equation}
\label{2.79}
p_1(\theta)=\lambda\sin(\theta-\phi) 
\end{equation}
\begin{equation}
\label{2.80}
p_2(\theta)=2\lambda\cos(\theta-\phi) -2\lambda_3
\end{equation}

If $\lambda \ne 0$ then the primer vector loci are ellipses. The equations 
(\ref{2.79}) and (\ref{2.80}) can be combined and put in the standard form:
\begin{equation}
\label{2.81}
\frac{p_1^2}{\lambda^2}+\frac{(p_2+2\lambda_3)^2}{(2\lambda)^2} =1.
\end{equation}

The major axis is $2\lambda$, the minor axis is $\lambda$, and the center 
is at $(-2\lambda_3,0)$ where the $p_2$-axis is the ordinate and the 
$p_1$-axis is the abscissa. According to (\ref{2.61}) it is necessary that 
optimal velocity impulses are applied at the values of $\theta_i$ described 
by the intersections of the ellipse (\ref{2.81}) and the unit circle. 
Also (\ref{2.63}) shows that the primer vector arc must be contained 
inside the unit circle. This is depicted graphically in Figures $1$-$8$ for 
various types of three-impulse, two-impulse, and one-impulse solutions.
The arrows indicate the points of application of the velocity increments.

If $\theta$ is any value of the true anomaly we let $\bar{\theta}=
\theta-\phi$. Similarly for a subscript $i$ we let
$\bar{\theta}_i=\theta_i-\phi$. We therefore write (\ref{2.79}) and
(\ref{2.80}) as
\begin{equation}
\label{2.82a}
p_1(\theta)=\lambda \sin\bar{\theta}
\end{equation}
\begin{equation}
\label{2.83a}
p_2(\theta)=2\lambda( \cos\bar{\theta} -\kappa)
\end{equation}
where
\begin{equation}
\label{2.84a}
\kappa= \frac{\lambda_3}{\lambda}.
\end{equation}

If (\ref{2.82a}) and (\ref{2.83a}) are associated with an optimal solution 
for a generalized boundary point ${\bf z}_f$ and optimal velocity increments are
$\Delta \bf{V}_i$, $i=1,...,k$, we see that
\begin{equation}
\label{2.85a}
p_1(\theta)=-\lambda \sin\bar{\theta}
\end{equation}
\begin{equation}
\label{2.86a}
p_2(\theta)=-2\lambda( \cos\bar{\theta} -\kappa)
\end{equation}
are associated with the generalized boundary point $-{\bf z}_f$ and optimal
velocity increments are $-\Delta \bf{V}_i$, $i=1,...,k$. This latter 
solution for the generalized boundary point $-{\bf z}_f$ will be called the
${\bf antipodal}$ solution relative to the original solution associated
with ${\bf z}_f$. Obviously if the original solution has been determined,
the antipodal solution is available without calculation. Figure $2$
depicts the locus of a primer vector that is antipodal to that presented 
in Figure $1$. This concept is not found much in the literature although
Breakwell does refer to "mirror image" trajectories [29].

\subsection {Singular Solutions}

An arc of a trajectory is called {\it singular} if 
$|p(\theta)|=1$ identically on an interval. If $\lambda=0$, then 
(\ref{2.79}), (\ref{2.80}),and (\ref{2.61}) show that $\lambda_3=\pm 1/2$
and
\begin{equation}
\label{2.82}
p_1(\theta)=0
\end{equation}
\begin{equation}
\label{2.83}
p_2(\theta)= \pm 1
\end{equation}
everywhere on the interval $\theta_0 \le \theta \le \theta_f$. There 
are no other values of $\lambda_1,\lambda_2$, or $\lambda_3$ for which 
singular solutions occur. The two 
constant primer vectors described by (\ref{2.82}) and (\ref{2.83}) 
will define two optimal singular solutions on 
$\theta_0 \le \theta \le \theta_f$. From (\ref{2.58})-(\ref{2.60}) 
we find that
\begin{eqnarray}
\label{2.84}
\Delta {\bf V}_i = \pm \left(\begin{array}{c} 
0 \\
\alpha_i \\
\end{array}
\right),\,\,\, i=1,...,k
\end{eqnarray}
\begin{equation}
\label{2.85}
\displaystyle\sum_{i=1}^k \alpha_i\cos\theta_i=\pm z_{f_1}/2
\end{equation}
\begin{equation}
\label{2.86}
\displaystyle\sum_{i=1}^k \alpha_i\sin\theta_i=\mp z_{f_2}/2
\end{equation}
\begin{equation}
\label{2.87}
\displaystyle\sum_{i=1}^k \alpha_i=\pm z_{f_3}/2
\end{equation}
where $k$ is an arbitrary positive integer. We note that terms determined 
from the lower algebraic signs represent antipodal solutions.  
We also observe that 
although the optimal velocity increments must satisfy the form (\ref{2.84}), 
and the sums on the left-hand sides must match the specified end conditions 
on the right-hand sides, these velocity increments are otherwise arbitrary. 
In these singular cases there are non-unique optimal trajectories that 
satisfy the end conditions and minimize the related cost function 
(\ref{2.48}). We note that the left-hand side of (\ref{2.87}) is 
the minimum cost (\ref{2.48}). We note that for the singular solutions 
all velocity increments are stationary velocity impulses. In terms of the 
original variables $r$ and $\theta$, the singular solutions describe an 
outward (or inward) multi-impulse spiral from  $(\theta_0,r_0)$ to 
$(\theta_f,r_f)$.

\subsection{Three-Impulse Solutions}

For the remainder of this first paper we set $k=3$ in (\ref{2.58})-(\ref{2.63}).

According to (\ref{2.61}), for three-impulse solutions it is necessary 
that optimal velocity increments 
be applied where $|{\bf p}(\theta_i)|=1,\,\,i=1,2,3$. For primer vector 
equations (\ref{2.82a}) and (\ref{2.83a}) this becomes
\begin{equation}
\label{2.93}
\lambda^2[\sin^2\bar{\theta}_i+ 4(\cos\bar{\theta}_i -\kappa)^2] =1
\end{equation}
for $i=1,2,3$.

\subsubsection{\bf Stationary Solutions}

We first consider three-impulse solutions in which all the velocity increments
are stationary velocity increments. These occur at values $\theta_i$
where (\ref{2.62}) and (\ref{2.93}) are satisfied. Using (\ref{2.82a}) 
and  (\ref{2.83a}) with (\ref{2.62}) we find that if $\theta_i$ is a
stationary value then it is necessary that
\begin{equation}
\label{2.94}
\sin\bar{\theta}_i\cos\bar{\theta}_i - 4\sin\bar{\theta}_i
(\cos\bar{\theta}_i -\kappa) =0.
\end{equation}
This implies that either $1)\,\sin\bar{\theta}_i =0$ or 
$2)\,\kappa=3/4\cos\bar{\theta}_i$.

We show now that the second alternative is not feasible because it 
violates (\ref{2.63}). Using (\ref{2.82a}) and (\ref{2.83a})
and forming the function
\begin{equation}
\label{2.96}
f(\theta)=p_1(\theta)^2+p_2(\theta)^2 - 1
\end{equation}
we find that if $f^{\prime}(\theta_i)=0$ where $\bar{\theta}_i$ satisfies 
$2)$ then $f^{\prime\prime}(\theta_i) > 0$ if $\cos^2(\bar{\theta}_i) \ne 1$.
It so happens that $\cos^2(\bar{\theta}_i)= 1$ implies that the first 
alternative $1)$ is satisfied. Our argument shows that 
$|{\bf p}(\theta)| \ge 1$ in a neighborhood of $\theta_i$ and 
equality holds in that neighborhood only at $\theta_i$ if 
the second alternative $2)$ is satisfied. 

We therefore conclude that the first alternative $1)$ must be satisfied and
\begin{equation}
\label{2.97}
\bar{\theta}_i =\pi i,\,\,\ i=0,1,2,...\,\, .
\end{equation}

Using two successive values of (\ref{2.97}) in (\ref{2.93}) we obtain
\begin{equation}
\label{2.98}
2\lambda(1-\kappa)=1
\end{equation}
\begin{equation}
\label{2.99}
2\lambda(1+\kappa)=1.
\end{equation}
Clearly $\lambda=1/2$ and $\kappa=0$ so the primer vector equations 
(\ref{2.82a}) and (\ref{2.83a}) respectively become
\begin{equation}
\label{2.100}
p_1(\theta)=\frac{1}{2}\sin\bar{\theta}
\end{equation}
\begin{equation}
\label{2.101}
p_2(\theta)=\cos\bar{\theta} .
\end{equation}
Forming the function (\ref{2.96}) we 
see that $|{\bf p}(\theta)| < 1$ except at $\theta_i$ satisfying (\ref{2.97})
where equality holds, as seen in Figure $3$.

We shall pursue this problem further in our later examination of 
multi-impulse and degenerate solutions. It suffices to state at this point 
that if we substitute (\ref{2.100}) and (\ref{2.101}) into (\ref{2.60}), 
there are not unique solutions for $\alpha_1,...,\alpha_k$ for $k > 2$. 
The three-impulse stationary solutions are therefore degenerate, 
and can be replaced  by two-impulse solutions.

In seeking non-degenerate three-impulse solutions, we must consider 
solutions in which all of the velocity increments are not applied at 
stationary values.

\subsubsection{Non-Degenerate Three-Impulse Solutions}

Throughout this paper we adopt the convention that the motion of increasing 
$\theta$ is counter clockwise and that $\theta_0 < \theta_f$. If 
$-\pi \le \bar{\theta}_0 \le \pi$ there are two cases,
$-\pi < \bar{\theta}_0 < 0$ and $0 < \bar{\theta}_0 < \pi$.
It is well known that there are at most four intersections of an ellipse 
such as (\ref{2.81}) with the unit circle. For the first case the 
intersections can be represented using at most four values 
$\bar{\theta}_i,\,\, i=1,2,3,4$ where $-\pi \le \bar{\theta}_i \le 
\pi,\,\, i=1,2,3,4$. 
For the second case the four values representing intersections are 
considered on the interval $0 \le \bar{\theta}_i \le 2\pi,\,\, i=1,2,3,4$.

We present the following lemma establishing necessary conditions for
optimal non-degenerate three-impulse solutions. The primer vector loci 
for these solutions are depicted in Figure 1 or Figure 2. 
The primer vector is defined 
by (\ref{2.82a})-(\ref{2.84a}) and the function $f$ is defined by (\ref{2.96}).

{\bf Lemma}: It is necessary that an optimal 
non-degenerate three-impulse solution satisfy the following:
\begin{equation}
\label{2.102}
\theta_1= \theta_0,\,\,\, \theta_3=\theta_f
\end{equation}
\begin{equation}
\label{2.103}
\theta_2= \frac{1}{2}(\theta_0 +\theta_f)
\end{equation}
\begin{equation}
\label{2.104}
0 < \theta_f - \theta_0 < 2\pi
\end{equation}
\begin{equation}
\label{2.105}
f^{\prime}(\theta_0)= -f^{\prime}(\theta_f) < 0
\end{equation}
\begin{equation}
\label{2.106}
f(\theta) \le 0,\,\,\,\,\, \theta_0 \le \theta \le \theta_f.
\end{equation}
There is an integer $i$ such that either 
$2i\pi < \bar{\theta}_0 < (2i+1)\pi$ and
\begin{equation}
\label{2.107}
\bar{\theta}_2 =(2i+1)\pi,\,\, and\,\, \phi=\theta_2-(2i+1)\pi
\end{equation}
\begin{equation}
\label{2.108}
\kappa =-\frac{3}{8}(1-\cos\bar{\theta}_0),\,\,\,\,\,\, -\frac{3}{4} < \kappa < 0
\end{equation}
\begin{equation}
\label{2.109}
\lambda = \frac{4}{5+3\cos\bar{\theta}_0},\,\,\,\,\,\, \frac{1}{2} < \lambda  <2
\end{equation}
or $(2i-1)\pi < \bar{\theta}_0 < 2i\pi$ and
\begin{equation}
\label{2.110}
\bar{\theta}_2 =  2i\pi,\,\, and\,\, \phi=\theta_2-2i\pi
\end{equation}
\begin{equation}
\label{2.111}
\kappa =\frac{3}{8}(1+\cos\bar{\theta}_0),\,\,\,\,\,\, 0 < \kappa < \frac{3}{4}
\end{equation}
\begin{equation}
\label{2.112}
\lambda = \frac{4}{5-3\cos\bar{\theta}_0},\,\,\,\,\,\, \frac{1}{2} < \lambda <2.
\end{equation}

{\bf Proof}: If a solution is non degenerate there must be a value
$\bar{\theta}_m$ on $\bar{\theta}_0 \le \bar{\theta} \le \bar{\theta} _f$ 
where $\bar{\theta}_m \ne \pi i$ for any integer $i$ and $\bar{\theta}_m$  
satisfies (\ref{2.93}), otherwise 
all velocity increments are at stationary values and the solution is 
degenerate. For any integer $j$ the value 
$\bar{\theta}_n=2\pi j-\bar{\theta}_m$
is distinct from $\bar{\theta}_m$ and satisfies (\ref{2.93}) also. We let 
$\bar{\theta}_1$ be the smaller of the numbers $\bar{\theta}_m$ and 
$\bar{\theta}_n$ and $\bar{\theta}_3$ be the larger, and 
$\frac{1}{2}(\bar{\theta}_1+\bar{\theta}_3) = \pi j$.

There must also be a third velocity increment on a value $\bar{\theta}_2$
satisfying (\ref{2.93}) where 
$\bar{\theta}_0 < \bar{\theta}_2 < \bar{\theta}_f$.
For any integer ${\it l}$ the number $2\pi{\it l} - \bar{\theta}_2$ also  
satisfies (\ref{2.93}). Since there can be only three impulses, this 
number is distinct from $\bar{\theta}_1$ and $\bar{\theta}_3$, and we 
must have $2\pi{\it l} - \bar{\theta}_2=\bar{\theta}_2$, consequently 
$\bar{\theta}_2=\pi{\it l}$ for some integer ${\it l}$.

We consider first the case where ${\it l}$ is odd. We therefore write
${\it l}=2i+1$ for some integer $i$. Substituting $\bar{\theta}_1$ 
and $\bar{\theta}_2$ into (\ref{2.93}) and equating the left-hand sides, 
we obtain
$$
\sin^2\bar{\theta}_1+4(\cos\bar{\theta}_1-\kappa)^2=4(1+\kappa)^2.
$$
Expanding and simplifying we find that
$$
3(1-\cos^2\bar{\theta}_1) +8\kappa(1+\cos\bar{\theta}_1)=0.
$$
This says that either  $\cos\bar{\theta}_1 =-1$ or
$$
\kappa = -\frac{3}{8}(1-\cos\bar{\theta}_1). 
$$
Since $\bar{\theta}_1 \ne \pi i$ for any integer $i$, we have the latter. 
Differentiating $f(\theta)$, evaluating the derivative at $\theta_1$, 
and substituting the above formula for $\kappa$, we obtain
$$
f^{\prime}(\theta_1)= {\bf p}(\theta_1)\cdot {\bf p}^{\prime}(\theta_1)=
-\frac{3}{2}\lambda^2 \sin\bar{\theta}_1(1+\cos\bar{\theta}_1).
$$
Next we observe that since $\bar{\theta}_1+\bar{\theta}_3=2\pi j$
then $\theta_3=2\pi j-\bar{\theta}_1+\phi$ and
$$
f^{\prime}(\theta_3) = \frac{3}{2}\lambda^2 
\sin\bar{\theta}_1(1+\cos\bar{\theta}_1).
$$

If $\sin\bar{\theta}_1 < 0$ then $f^{\prime}(\theta_1) > 0$  and 
$f^{\prime}(\theta_3) < 0$. This says that the primer vector exits the 
unit disk at $\bar{\theta}_1$ and enters at $\bar{\theta}_3$ thus 
$\bar{\theta}_1$, $\bar{\theta}_2$ and $\bar{\theta}_3$ cannot support 
a three-impulse solution since $\bar{\theta}_1 < \bar{\theta}_3$ 
and (\ref{2.63}) must be satisfied.

As a consequence of this argument, it is necessary that 
$\sin\bar{\theta}_1 > 0$ 
hence $f^{\prime}(\theta_1) < 0$ and $f^{\prime}(\theta_3) > 0$. In order for
a three-impulse solution to be supported on 
$\bar{\theta}_1$, $\bar{\theta}_2$ and $\bar{\theta}_3$ and satisfy 
(\ref{2.63}) it is necessary that $\bar{\theta}_1=\bar{\theta}_0$
and $\bar{\theta}_3 =\bar{\theta}_f$. It follows that 
$\sin\bar{\theta}_0 > 0$ resulting in 
the restriction $2i\pi < \bar{\theta}_0 < (2i+1)\pi < \bar{\theta}_f < (2i+2)\pi$. 
(Recall that ${\it l}=2i+1$). The only integers 
that satisfy the restriction are ${\it l}=j=1$, establishing 
(\ref{2.102})-(\ref{2.108}) for the first case recalling the definition of
$\bar{\theta}_i$. 
Substituting (\ref{2.108}) into 
(\ref{2.93}) and solving for $\lambda$ produces (\ref{2.109}).

We consider next the case where ${\it l}$ is even. We therefore write
${\it l}=2i$ for some integer $i$. Substituting $\bar{\theta}_1$ and 
$\bar{\theta}_2$ into (\ref{2.93}) and equating the left-hand sides again,
$$
\sin^2\bar{\theta}_1+4(\cos\bar{\theta}_1-\kappa)^2=4(1-\kappa)^2.
$$
Expanding and simplifying as before we find
$$
\kappa =\frac{3}{8}(1+\cos\bar{\theta}_1).
$$
Substituting this value of $\kappa$ into $f^{\prime}(\theta_1)$ 
we get
$$
f^{\prime}(\theta_1)= {\bf p}(\theta_1)\cdot {\bf p}^{\prime}(\theta_1)=
\frac{3}{2}\lambda^2 \sin\bar{\theta}_1(1-\cos\bar{\theta}_1)
$$
and similarly
$$
f^{\prime}(\theta_3))=
-\frac{3}{2}\lambda^2 \sin\bar{\theta}_1(1-\cos\bar{\theta}_1).
$$

If $\sin\bar{\theta}_1 > 0$ then $f^{\prime}(\theta_1) > 0$,
$f^{\prime}(\theta_3) <0$
and we again find that a three-impulse solution cannot be supported on
$\bar{\theta}_1$,$\bar{\theta}_2$ and $\bar{\theta}_3$ and satisfy 
(\ref{2.63}) because $\bar{\theta}_1 < \bar{\theta}_3$.

It is therefore necessary that $\sin\bar{\theta}_1 < 0$, hence 
$f^{\prime}(\theta_1) < 0$ and $f^{\prime}(\theta_3) >0$. In order that a 
three-impulse solution be supported on $\bar{\theta}_1$,$\bar{\theta}_2$ 
and $\bar{\theta}_3$ 
and satisfy (\ref{2.63}), it is again necessary that 
$\bar{\theta}_1 =\bar{\theta}_0$
and $\bar{\theta}_3=\bar{\theta}_f$. This requires that 
$\sin\bar{\theta}_0 < 0$ resulting in 
the restriction 
$(2i-1)\pi < \bar{\theta}_0 < 2i\pi < \bar{\theta}_f < (2i+1)\pi$.
(Recall that ${\it l}=2i$).  The only integers 
satisfying this restriction are ${\it l}=j=0$ establishing 
(\ref{2.102})-(\ref{2.106}), recalling again the definition of 
$\bar{\theta}_i$; (\ref{2.110}) and (\ref{2.111}) follow as well.
Substituting (\ref{2.111}) into (\ref{2.93}) and solving for $\lambda$
reveals the formula (\ref{2.112}). $\blacksquare$

Some of the statements in this theorem can be expressed differently by 
adding and subtracting $\bar{\theta_2}$ from $\bar{\theta_0}$ in 
(\ref{2.108}), (\ref{2.109}), (\ref{2.111}) and (\ref{2.112}). 
We restate the lemma as the following:

{\bf Corollary}: It is necessary that a 
non-degenerate three-impulse solution satisfy (\ref{2.102})-(\ref{2.106}), 
and
\begin{equation}
\label{2.113}
\lambda = \frac{4}{5-3\cos(\frac{\theta_f-\theta_0}{2})}.
\end{equation}
Either 
\begin{equation}
\label{2.114}
\kappa = -\frac{3}{8}\left[1+\cos(\frac{\theta_f-\theta_0}{2})\right]
\end{equation}
and the primer vector satisfies (\ref{2.85a}) and (\ref{2.86a})
or 
\begin{equation}
\label{2.115}
\kappa = \frac{3}{8}\left[1+\cos(\frac{\theta_f-\theta_0}{2})\right]
\end{equation}
and the primer satisfies (\ref{2.82a}) and (\ref{2.83a}).

An immediate consequence of the lemma and the necessary and sufficient 
conditions (\ref{2.58})-(\ref{2.63}) is the following fundamental theorem 
of non-degenerate three-impulse solutions. 

{\bf Theorem}: The optimal impulsive rendezvous problem defined
 by (\ref{2.40}) and 
(\ref{2.47}) on the interval $\theta_0 \le \theta \le \theta_f$  
has a non-degenerate three-impulse solution 
$\{\theta_1,\theta_2,\theta_3,\Delta{\bf V}_1,\Delta{\bf V}_2,
\Delta{\bf V}_3\}$ if and only if (\ref{2.102})-(\ref{2.107}) 
and (\ref{2.113}) are satisfied, 
and either (\ref{2.85a}), (\ref{2.86a}) and (\ref{2.114}) 
are satisfied or (\ref{2.82a}), (\ref{2.83a}) (\ref{2.110}) and (\ref{2.115}) 
are satisfied, and
\begin{equation}
\label{2.116}
\Delta{\bf V}_i =-\alpha_i{\bf p}(\theta_i),\,\,\,\, i=1,2,3
\end{equation}
\begin{equation}
\label{2.117}
\displaystyle\sum_{1=1}^{3} R(\theta_i){\bf p}(\theta_i) \alpha_i =
-{\bf z}_f,
\end{equation}
\begin{equation}
\label{2.118}
\alpha_i >0,\,\,\, i=1,2,3,
\end{equation}
where $ R(\theta)$ is given from (\ref{2.52}) and ${\bf z}_f$ from 
(\ref{2.57}).

From (\ref{2.108}), (\ref{2.109}), (\ref{2.111}) and (\ref{2.112})) we see 
the bounds on $\lambda$ and $\kappa$ for which three-impulse solutions exist. 
These boundary primer vector loci are illustrated in Figure $3$.

We remark that an optimal solution in which (\ref{2.107})-(\ref{2.109}) are
satisfied has an associated antipodal solution satisfying 
(\ref{2.110})-(\ref{2.112}) instead. Similarly if a generalized boundary 
point ${\bf z}_f$ has $\kappa$ satisfying (\ref{2.114}) then its antipodal 
solution for the point$-{\bf z}_f$ has $\kappa$ satisfying (\ref{2.115}).

\subsubsection{Finding Non-Degenerate Three-Impulse Solutions}

We now apply the preceding theorem to find non-degenerate three impulse 
solutions. 

We consider the case where $\phi=\theta_2- 2i\pi$ and $\kappa$ and 
$\lambda$ are given respectively by (\ref{2.111}) and (\ref{2.112}), 
$\theta_0$ and $\theta_f$ are known and $\theta_2$ is known from
(\ref{2.103}). Setting $\hat{\theta}=\theta-\theta_2$ the primer
vector from (\ref{2.82a}) and (\ref{2.83a}) becomes
\begin{equation}
\label{2.118a}
p_1(\theta) = \lambda\sin\hat{\theta}
\end{equation}
\begin{equation}
\label{2.119}
p_2(\theta) = 2\lambda(\cos\hat{\theta} -\kappa)
\end{equation}
where $\kappa$ and $\lambda$ respectively follow from (\ref{2.111}) and
(\ref{2.112}) with $\hat{\theta}_0$ replacing $\bar{\theta}_0$.

Since (\ref{2.49}) is a fundamental matrix solution associated with $A$
defined by (\ref{2.47}) so is also the matrix
\begin{eqnarray}
\label{2.120}
\hat{\Phi}(\theta)=
\left(\begin{array}{ccc}
\cos\hat{\theta} &\sin\hat{\theta} & 1\\
-\sin\hat{\theta} &\cos\hat{\theta} &0\\
0 &0 & 1\\
\end{array}
\right).
\end{eqnarray}

Analogous to (\ref{2.52}) we define
\begin{equation}
\label{2.121}
\hat{R}(\theta)=\hat{\Phi}^{-1}(\theta)B
\end{equation}
and substitute into (\ref{2.117}) of the preceding theorem.
Setting $s=\sin\hat{\theta}_0$, $c=\cos\hat{\theta}_0$ and
$\beta_i=\lambda\alpha_i$, $i=1,2,3$ we obtain
\begin{eqnarray}
\label{2.122}
\left(\begin{array}{ccc}
\frac{1}{2}(5-3s^2-3c) &\frac{1}{2}(5-3c) & \frac{1}{2}(5-3s^2-3c) \\
\frac{3}{2}s(c-1) &0 & \frac{3}{2}s(1-c) \\
\frac{1}{2}(3-5c) &\frac{1}{2}(3c-5) & \frac{1}{2}(3-5c) \\
\end{array}
\right)
\left( \begin{array}{c} \beta_1 \\ \beta_2 \\ \beta_3 \end{array} \right)
=\left( \begin{array}{c} -z_{f_1} \\ -z_{f_2} \\ -z_{f_3} \end{array} \right).
\end{eqnarray}

The region $Z_f$ is defined by the conical set generated by the column 
vectors of the coefficient matrix. To find the region in terms of 
inequalities, we first solve (\ref{2.122}), use the identity $c^2=1-s^2$,
and factor the denominators to obtain
\begin{eqnarray}
\label{2.123}
\left( \begin{array}{c} \beta_1 \\ \beta_2 \\ \beta_3 \end{array} \right)=
\left(\begin{array}{ccc}
\frac{-1}{(5-3c)(1-c)} &\frac{1}{3(-s)(1-c)} & \frac{-1}{(5-3c)(1-c)} \\ \\
\frac{-2(5c-3)}{(5-3c)^2(1-c)} &0 & \frac{-2(3c^2-3c+2)}{(5-3c)^2(1-c)} \\ \\
\frac{-1}{(5-3c)(1-c)} &\frac{-1}{3(-s)(1-c)} & \frac{-1}{(5-3c)(1-c)} \\
\end{array}
\right)
\left( \begin{array}{c} -z_{f_1} \\ -z_{f_2} \\ -z_{f_3} \end{array} \right).
\end{eqnarray}

It follows from (\ref{2.104}) and the definition of $\hat{\theta}_0$
that all of the denominators in this matrix are positive. Multiplying
by the positive common denominators, employing (\ref{2.118}) and 
the definitions of $\beta_i$, $i=1,2,3$, we find
\begin{equation}
\label{2.124}
3(-s)z_{f_1} - (5-3c)z_{f_2} +3(-s)z_{f_3} > 0
\end{equation}
\begin{equation}
\label{2.125}
(5c-3)z_{f_1} +(3c^2-3c+2)z_{f_3} > 0
\end{equation}
\begin{equation}
\label{2.126}
3(-s)z_{f_1} + (5-3c)z_{f_2} - 3(-s)z_{f_3} > 0
\end{equation}

From these inequalities it is seen that the subregion of $Z_f$ that 
admits three-impulse solutions described by (\ref{2.110})-(\ref{2.112})
satisfies
\begin{equation}
\label{2.127}
z_{f_1} +z_{f_3} > \frac{5-3c}{-3s}|z_{f_2}|
\end{equation}
\begin{equation}
\label{2.128}
(5c-3)z_{f_1} +(3c^2-3c+2)z_{f_3} > 0.
\end{equation}

The antipodal region that admits three-impulse solutions by 
(\ref{2.107})-(\ref{2.109}) is found from the fact that $\phi$ is decreased 
by $\pi$ from the preceding solutions, resulting in primer vectors having 
opposite signs. For this reason the left hand side of (\ref{2.117}) 
changes sign. The consequence is that all inequalities reverse in 
(\ref{2.124})-(\ref{2.128}). The two separate $Z_f$ regions admitting
three-impulse solutions therefore have symmetry with respect to the origin.

In the calculation of an optimal trajectory, the actual velocity increments
$\Delta {\bf v}_i$ are calculated from the transformed velocity increments
$\Delta {\bf V}_i$. Solving for these from (\ref{2.37}) and (\ref{2.38})
we get
\begin{equation}
\label{2.134}
\Delta v_{\theta_i} = \frac{h_i^3}{\mu r_i} \Delta V_{2i}
\end{equation}
\begin{equation}
\label{2.135}
\Delta v_{r_i} = h_i\Delta V_{1i}+\frac{v_{ri}h_i^2}{\mu} \Delta V_{2i}
\end{equation}
for $i=1,2,3$.

{\bf Example:} Select $\theta_0=\pi/2$ and $\theta_f=3\pi/2$. From 
(\ref{2.102}) and (\ref{2.103}) we see that $\theta_1=\pi/2$,
$\theta_2=\pi$ and $\theta_3=3\pi/2$. We shall use 
(\ref{2.110})-(\ref{2.112}).
If we use (\ref{2.107})-(\ref{2.109}) instead we get a solution antipodal to
this one. Using $\phi= \pi- 2i\pi$ we obtain $\kappa =3/8$, $\lambda=4/5$ 
and $\hat{\theta}_0=-\pi/2$. Eq. (\ref{2.123}) becomes
\begin{equation}
\label{2.131}
\beta_1=\frac{1}{5}z_{f_1} -\frac{1}{3}z_{f_2} +\frac{1}{5}z_{f_3} > 0
\end{equation}
\begin{equation}
\label{2.132}
\beta_2=\frac{-6}{25}z_{f_1} +\frac{4}{25}z_{f_3} > 0
\end{equation}
\begin{equation}
\label{2.133}
\beta_3=\frac{1}{5}z_{f_1} +\frac{1}{3}z_{f_2} +\frac{1}{5}z_{f_3} > 0
\end{equation}
where (\ref{2.127}) and (\ref{2.128}) become
\begin{equation}
\label{2.140}
z_{f_1} +z_{f_3} > \frac{5}{3}|z_{f_2}|,\,\,\, -3z_{f_1}+2z_{f_3} > 0.
\end{equation}
For the antipodal subregion the inequalities in (\ref{2.140}) are reversed.

We observe that
\begin{equation}
\label{2.141}
p_1(\theta)=-\frac{4}{5}\sin \theta,\,\,\, p_2(\theta)=-\frac{8}{5}
(\cos\theta+\frac{3}{8}).
\end{equation}
Substituting into (\ref{2.116}) we find 
\begin{eqnarray}
\label{2.142}
\Delta {\bf V}_1= \alpha_1
\left(\begin{array}{c}
4/5 \\
3/5
\end{array}
\right)
\end{eqnarray}
\begin{eqnarray}
\label{2.143}
\Delta {\bf V}_2= \alpha_2
\left(\begin{array}{c}
0 \\
-1
\end{array}
\right)
\end{eqnarray}
\begin{eqnarray}
\label{2.144}
\Delta {\bf V}_3= \alpha_3
\left(\begin{array}{c}
-4/5 \\
3/5
\end{array}
\right)
\end{eqnarray}
The antipodal solution reverses the directions of these velocity increments.
The actual velocity increments follow from (\ref{2.134}) and (\ref{2.135}).
The vector ${\bf z}_f$ is calculated from (\ref{2.50}) and (\ref{2.57}).

These calculations were performed for the specific boundary conditions
$r_0= 8000km$, $v_{r_0}=-0.831929 km/sec$, $v_{\theta_0}=7.487362 km/sec$,
and $r_f=6545.455 km$, $v_{r_f}= -0.679267 km/sec$, 
$v_{\theta_f}= 7.471940 km/sec$. The result is 
$$
{\bf z}_f=(-2.777777\times 10^{-5}\;\;  0\;\; 5.555555\times 10^{-5})^T \,km^{-1}
$$
which clearly satisfies (\ref{2.140}) and shows that the optimal rendezvous
for these boundary conditions requires three impulses. The three velocity
increments are found to be: $\Delta v_{\theta_1}= -0.23538 km/sec$,
and $\Delta v_{r_1}= -0.287688 km/sec$ at $\theta_1=\pi/2$,
$\Delta v_{\theta_2}= -1.313191 km/sec$ and 
$\Delta v_{r_2}= -7.691605\times 10^{-2} km/sec$ at $\theta_2=\pi$
where $r(\theta_2)$ is approximately $7384 km$, and
$\Delta v_{\theta_3}= -0.193261 km/sec$ and $\Delta v_{r_3}=0.310108 km/sec$
at $\theta_3=3\pi/2$. The fictitious nominal circular orbit has a  
radius of approximately
$7272km$. Simulation of the optimal three-impulse rendezvous trajectory
connecting the initial and terminal conditions and showing the
application of the three velocity impulses is depicted in $Fig.\,9$.
The maximum deviation of the rendezvous trajectory from the nominal
circular orbit is approximately $10\%$ of the nominal radius.

\setcounter{equation}{0}
\section{Two-Impulse Solutions} 

A two-impulse solution is a solution for which $k=2$  
in (\ref{2.58})-(\ref{2.63}). Recalling that $f$ is defined by (\ref{2.96})
it follows that a non-degenerate
two-impulse solution has exactly two zeros of $f$ on the interval 
$\theta_0 \le \theta \le \theta_f$.
 
The two-impulse solutions can be classified as either 
non-stationary solutions or stationary solutions. 

The non-stationary solutions are non-degenerate and fall naturally into 
several categories. We recall that $\bar{\theta}=
\theta+\phi$ where $\phi$ is an arbitrary constant. The following definitions
apply in general to optimal k-impulse solutions but are especially relevant 
for $k=2$. These definitions are illuminated by Figures $10-17$.

An optimal solution is called a {\bf non-stationary two-intersection solution} 
if 
$\theta_1 < \theta_2$, $f(\theta_1)=f(\theta_2)=0$, 
$f^{\prime}(\theta_1)\ne 0$, $f^{\prime}(\theta_2) \ne 0$, 
$0 < \bar{\theta}_2-\bar{\theta}_1 < 2\pi$,
there are no other zeros of $f$ on
the interval $\bar{\theta}_0 \le \bar{\theta} \le  \bar{\theta}_f$,
and there is an integer $i$ such that $\bar{\theta}_1+\bar{\theta}_2=2\pi i$.
If $0 < \bar{\theta}_0 < \pi$ then $i=1$, if $-\pi < \bar{\theta}_0 < 0$
then $i=0$. Primer vector loci for two-intersection solutions are depicted
in Figures $10$ and $11$. 

An optimal solution is  called a {\bf non-stationary four-intersection
solution} if $\theta_1 <  \theta_2$, $f(\theta_1)=f(\theta_2)=0$, 
$0 < \bar{\theta}_2-\bar{\theta}_1 < \pi$, and there is an integer $i$ 
such that $i\pi< \bar{\theta}_1 <\bar{\theta}_2 < (i+1)\pi$.
If $-\pi < \bar{\theta}_0 < \pi$ then $i=0$ or $i=-1$. Primer vector loci 
for four-intersection solutions can be seen in Figures $12$ and $13$.

An optimal solution is called a {\bf non-stationary three-intersection 
solution} if
$\theta_1 <  \theta_2$, $f(\theta_1)=f(\theta_2)=0$, and either
$f^{\prime}(\theta_1) = 0$ or $f^{\prime}(\theta_2) = 0$ but not both.
Examples of primer vector loci for three intersection solutions are 
presented in Figures $14$ and $15$.

An optimal solution is called a {\bf stationary two-intersection} solution if 
$\theta_1 <  \theta_2$, $f(\theta_1)=f(\theta_2)$,
$f^{\prime}(\theta_1)= f^{\prime}(\theta_2) = 0$,  
$\bar{\theta}_2-\bar{\theta}_1 =\pi$ and there are no other zeros of $f$
on the interval $\bar{\theta}_0 \le \bar{\theta} \le \bar{\theta}_f$.
Primer vector loci of stationary two-intersection solutions are presented in 
Figure $16$. If $\bar{\theta}_2 - \bar{\theta}_1 > \pi$ it is called a {\bf stationary 
multi-impulse solution} and is degenerate. Primer vector loci of stationary
multi-impulse solutions are depicted in Figure $17$. An optimal 
multi-impulse solution may have arbitrarily many impulses if 
$\theta_f - \theta_0$ is sufficiently large but always has an equivalent 
one-impulse or two-impulse solution having the same cost as will be shown in
the subsequent investigation of degenerate solutions.

The following theorem shows that there is a partition of the various 
categories of two-impulse solutions.

{\bf Theorem}: An optimal non-degenerate two-impulse solution is one and
only one of the following:
\begin{enumerate}
\item  a stationary two-intersection solution
\item  a non-stationary two-intersection solution
\item  a non-stationary three-intersection solution
\item a non-stationary four-intersection solution.
\end{enumerate}

{\bf Proof}: Consider an optimal non-degenerate two-impulse solution
having the primer vector given by (\ref{2.82}) and (\ref{2.83}).
Forming the function (\ref{2.96}) we obtain
$$
f(\theta)=\lambda^2[\sin^2\bar{\theta} +4(\cos\bar{\theta} -\kappa)^2 -1.
$$
Differentiating this function,
$$
f^{\prime}(\theta)= -2\lambda^2\sin\bar{\theta}(3\cos\bar{\theta}-4\kappa).
$$
If the velocity impulses are at $\bar{\theta}_1$ and $\bar{\theta}_2$,
then $f(\theta_1)=f(\theta_2)=0$ where $\bar{\theta}_0 \le \bar{\theta}_1
< \bar{\theta}_2 \le \bar{\theta}_f$. Since all solutions considered are 
two-impulse solutions then $k=2$ in (\ref{2.58})-(\ref{2.63}) and 
there are no other  zeroes of $f$ on the interval
$\bar{\theta}_0 \le \bar{\theta} \le \bar{\theta}_f$.

1) If $\bar{\theta}_1$ and $\bar{\theta}_2$ are integer multiples of $\pi$ 
then $f^{\prime}(\theta_1)=f^{\prime}(\theta_2)=0$ and $\bar{\theta}_2-
\bar{\theta}_1$ is an integer multiple of $\pi$.
Since there are no other zeros on the interval $\bar{\theta}_0 \le
\bar{\theta} \le \bar{\theta}_f$, then $\bar{\theta}_2-\bar{\theta}_1=\pi$.
 This solution 
satisfies the definition of a stationary two-intersection solution.

2) If neither $\bar{\theta}_1$ nor $\bar{\theta}_2$ are integer multiples 
of $\pi$ and there is an integer $i$ such that $\bar{\theta}_1+
\bar{\theta}_2=2\pi i$ then $0 < \bar{\theta}_2-\bar{\theta}_1 < 2\pi$.
If not, then $\bar{\theta}_2-\bar{\theta}_1 \ge  2\pi$ and there is
an integer $j$ such that either $\bar{\theta}^{\prime}=2\pi j-
\bar{\theta}_2$ or $\bar{\theta}^{\prime}=2\pi j-\bar{\theta}_1$.
In either case
$\bar{\theta}_1 <\bar{\theta}^{\prime} < \bar{\theta}_2$ and 
$f(\bar{\theta}^{\prime})=0$, establishing more than two zeros of $f$
on the interval $\theta_0 < \theta < \theta_ f$.  

We show now that $f^{\prime}(\theta_1) \ne 0$ and $f^{\prime}(\theta_2) \ne 0$.
We argue by contradiction. Suppose $f^{\prime}(\theta_1)= 0$. Since 
$\bar{\theta}_1$ is not an integer multiple of $\pi$ it is necessary that 
$\kappa=\frac{3}{4}\cos \bar{\theta}_1$. Differentiating 
$f^{\prime}(\theta)$ we find
$$
f^{\prime\prime}(\theta)=-2\lambda^2(6\cos^2\bar{\theta}-4\kappa
\cos\bar{\theta}-3).
$$
Evaluating at $\bar{\theta}_1$ and substituting for $\kappa$ into this
expression we obtain
$$
f^{\prime\prime}(\theta_1)=6\lambda^2\sin^2\bar{\theta}_1.
$$
This is clearly positive since $\bar{\theta}_1$ is not an integer
multiple of $\pi$. Since $f(\theta_1)=0$, then $f(\theta) > 0$
on an interval where $\bar{\theta} >\bar{\theta}_1$. This violates 
(\ref{2.63}) establishing the contradiction. 
Similarly if we suppose
$f^{\prime}(\theta_2)= 0$ we find that $f(\theta) > 0$ on an interval where
$\bar{\theta} < \bar{\theta}_2$, establishing the contradiction. 
As a consequence $f^{\prime}(\theta_1) \ne 0$,
$f^{\prime}(\theta_2) \ne 0$, and the definition of
 a non-stationary two-intersection solution is satisfied.

3) If $\bar{\theta}_1$ is not an integer multiple of $\pi$ but 
$\bar{\theta}_2$ is an integer multiple of $\pi$, clearly 
$f^{\prime}(\theta_2)=0$ and $f^{\prime}(\theta_1)\ne 0$.
If $f^{\prime}(\theta_1)= 0$ the argument above establishes a contradiction.
It follows that the definition of a 
non-stationary three-intersection solution is satisfied.

If $\bar{\theta}_1$ is an integer multiple of $\pi$ but $\bar{\theta}_2$
is not an integer multiple of $\pi$, clearly $f^{\prime}(\theta_1)=0$
and the above argument shows that $f^{\prime}(\theta_2) \ne 0$,
so that in this case also, the definition of a non-stationary
three-intersection solution is satisfied.

4) Finally, if neither $\bar{\theta}_1$ nor $\bar{\theta}_2$ are integer
multiples of $\pi$ and if $\bar{\theta}_1+ \bar{\theta}_2 \ne 2\pi i$ for 
any integer $i$ then there is an integer $i$ such that
$i\pi < \bar{\theta}_1 < \bar{\theta}_2 < (i+1)\pi$. The reason is 
as follows. If this statement is not true and if $i$ is the largest 
integer such that $i\pi < \bar{\theta}_1$ then $\bar{\theta}_2 \ge (i+1)\pi$.
It follows then that there is an integer $j$ such that if
$\bar{\theta}^{\prime}=2\pi j- \bar{\theta}_1 \ne  \bar{\theta}_2 $ or
$\bar{\theta}^{\prime}=2\pi j- \bar{\theta}_2 \ne  \bar{\theta}_2 $
then $\bar{\theta}_1 < \bar{\theta}^{\prime} <\bar{\theta}_2$ and
$f(\bar{\theta}^{\prime})=0$. This establishes at least three zeros on 
$\bar{\theta}_0 < \bar{\theta} < \bar{\theta}_f$ and a contradiction.

Having established that there is an integer $i$ such that 
$i\pi <\bar{\theta}_1 <\bar{\theta}_2 < (i+1)\pi$ we subtract
$\bar{\theta}_1$ in this inequality and obtain 
$0 < \bar{\theta}_2 - \bar{\theta}_1 < \pi$. 
The definition of a non-stationary four-intersection solution is 
therefore satisfied.  $\blacksquare$

\setcounter{equation}{0}
\subsection{Non-Stationary Two-Intersection Solutions}

First we consider the category of solutions in which the primer vector loci
demonstrate the geometry represented in Figures $10$ and $11$, 
the non-tangential 
intersection with the unit circle. In the following we specify that
$\theta_0 < \theta_f$. We define $\bar{\theta_r}=
\theta_r-\phi$ for any subscript $r$.

\subsubsection{ Fundamental Theorem of Non-Stationary Two-Intersection Solutions}

{\bf Theorem}: The optimal impulsive rendezvous defined by (\ref{2.40}) 
and (\ref{2.47}) on the interval $\theta_0 \le \theta \le \theta_f$
satisfying the convention $-\pi \le \bar{\theta_0} \le \pi$ has a 
non-stationary two-intersection solution $\{\theta_1,\theta_2,\Delta {\bf V}_1,
\Delta {\bf V}_2\}$ if and only if
\begin{equation}
\label{3.01}
\theta_1= \theta_0,\,\,\, \theta_2=\theta_f,
\end{equation}
\begin{equation}
\label{3.02}
0 < \theta_f - \theta_0 < 2\pi,
\end{equation}
\begin{equation}
\label{3.03}
f(\theta)  < 0,\,\,\,\,\, \theta_0 < \theta < \theta_f,
\end{equation}
\begin{equation}
\label{3.04}
f^{\prime}(\theta_0)= -f^{\prime}(\theta_f) < 0,
\end{equation}
and either
\begin{equation}
\label{3.05}
0 < \bar{\theta}_0 < \pi,\,\,\, \bar{\theta}_f=2\pi-\bar{\theta}_0,\,\,\,
and\; \phi=\frac{\theta_0+\theta_f}{2}-\pi,
\end{equation}
or
\begin{equation}
\label{3.06}
-\pi < \bar{\theta}_0 < 0,\,\,\, \bar{\theta}_f=-\bar{\theta}_0,\,\,\,
and\; \phi=\frac{\theta_0+\theta_f}{2},
\end{equation}
and, in addition,
\begin{equation}
\label{3.07}
\Delta {\bf V}_1=-{\bf p}(\theta_1)\alpha_1,\,\,\,
\Delta {\bf V}_2=-{\bf p}(\theta_2)\alpha_2,
\end{equation}
\begin{equation}
\label{3.08}
R(\theta_1){\bf p}(\theta_1)\alpha_1+
R(\theta_2){\bf p}(\theta_2)\alpha_2=-{\bf z}_f,
\end{equation}
\begin{equation}
\label{3.09}
\alpha_1 > 0,\,\,\, \alpha_2 >0.
\end{equation}

{\bf Proof}: First we show that (\ref{3.01})-(\ref{3.09}) are necessary.

If the optimization problem has a non-stationary two-intersection solution,
then, $0 <  \bar{\theta}_2-\bar{\theta}_1 < 2\pi$ where 
$f(\theta_1)=f(\theta_2)=0$ and there is an integer
$i$ such that $\bar{\theta}_1+\bar{\theta}_2 =2\pi i$  and
$\bar{\theta}_0 \le \bar{\theta}_1 \le \bar{\theta}_2 \le \bar{\theta}_f$.
Since $f^{\prime}(\theta_1) \ne 0$ and
 $f^{\prime}(\theta_2) \ne 0$, it follows from (\ref{2.62}) 
that $\bar{\theta}_1=\bar{\theta}_0$ and $\bar{\theta}_2=\bar{\theta}_f$
and neither of these is an integer multiple of $\pi$. From (\ref{2.63}) 
we have 
$f(\theta) \le 0$, $\bar{\theta}_0 \le \bar{\theta} \le \bar{\theta}_f$.
Since there are not more than two zeros of $f$ on this interval we must have
$f(\theta) < 0$, $\bar{\theta}_0 < \bar{\theta} < \bar{\theta}_f$. 
We have established (\ref{3.01}) and (\ref{3.03}); (\ref{3.04}) follows 
from the continuity of $f$, the fact that there are only two zeros of $f$ 
on the interval $\bar{\theta}_0 \le \bar{\theta} \le \bar{\theta}_f$,
$f^{\prime}(\theta_0) \ne 0$, $f^{\prime}(\theta_f) \ne 0$, and 
$\bar{\theta}_1$ and $\bar{\theta}_2=2\pi i- \bar{\theta}_1$ 
can be substituted into the expression for $f^{\prime}(\theta)$.

Since $\bar{\theta}_0$ is non-stationary we have two cases, either
$0 < \bar{\theta}_0 < \pi$ or $-\pi < \bar{\theta}_0 < 0$.

In the former case we observe that $ \bar{\theta}_m=2\pi-\bar{\theta}_0$
is a zero of $f$ satisfying $0 < \theta_m-\theta_0 < 2\pi$. This requires
that $\bar{\theta}_f \le \bar{\theta}_m$ and (\ref{3.02}) is satisfied. 
Solving $\bar{\theta}_f=2\pi-\bar{\theta}_0$ for $\phi$ we find that
$\phi=\frac{\theta_0+\theta_f}{2}-\pi$ establishing (\ref{3.05}).

In the latter case we note that $\bar{\theta}_n=-\bar{\theta}_0$ is a zero 
of $f$ satisfying $0 < \bar{\theta}_n-\bar{\theta}_0 < 2\pi$. This requires 
that $\bar{\theta}_f \le \bar{\theta}_n$ and (\ref{3.02}) is also satisfied 
in this case. Solving $\bar{\theta}_f=-\bar{\theta}_0$ for $\phi$ we find that
$\phi=\frac{\theta_0+\theta_f}{2}$ establishing (\ref{3.06}) in this case.

The expressions (\ref{3.07})-(\ref{3.09}) follow from the fact that an
optimal solution satisfies (\ref{2.58})-(\ref{2.60}) then (\ref{3.03}) 
shows the solution is non-degenerate and the inequalities in (\ref{3.09})
are strict.

To show that (\ref{3.01})-(\ref{3.09}) are also sufficient we note that
they imply (58)-(63) of Reference $1$ with $k=2$ 
showing optimality, where (\ref{3.03}) shows that $f$ has exactly two roots on 
$\bar{\theta_0} \le \bar{\theta} \le \bar{\theta}_f$. The expressions 
(\ref{3.01}),(\ref{3.02}) and (\ref{3.04})
show that $f^{\prime}(\theta_1) \ne 0$ and $f^{\prime}(\theta_2) \ne 0$;
(\ref{3.01}) and (\ref{3.05}) or (\ref{3.06}) show that
$\bar{\theta}_1+\bar{\theta}_2=2\pi i$ where $i=1$ or $i=0$.

These show that the optimal rendezvous is a non-stationary two-intersection
solution  completing the proof.  $\blacksquare$

{\bf Remark}: The convention $-\pi \le \bar{\theta}_0 \le \pi$ is
convenient as an assumption in this theorem but unnecessary. See
the previous theorem for three-impulse solutions. Without this 
assumption (\ref{3.05}) and (\ref{3.06}) would assert the existence of 
an integer $i$ such that $2i\pi <\bar{\theta}_0 <(2i+1)\pi$ and
$(2i-1)\pi <\bar{\theta}_0 <2i\pi$ respectively.
 
\subsubsection{Finding Non-Stationary Two-Intersection Solutions}

Since $\lambda > 0$ we set $\beta_1=\lambda\alpha_1$, $\beta_2=\lambda\alpha_2$
and (\ref{3.09}) becomes
\begin{equation}
\label{3.10}
\beta_1 >0,\,\,\, \beta_2 >0.
\end{equation}

We observe that $\Phi(\bar{\theta})$ is a fundamental matrix solution
associated with $A$ so $ R({\bar\theta}_1)$ and $R({\bar\theta}_2)$ may be
inserted in (\ref{3.08}). We select (\ref{3.06}) so that 
$\bar{\theta}_f=-\bar{\theta}_0$ for the following development.
There is no need to repeat this development for (\ref{3.05}) because 
it leads to the antipodal solution.

Setting $s_0=sin(\bar{\theta}_0)$ and $c_0=cos(\bar{\theta}_0)$ (\ref{3.08})
becomes
\begin{eqnarray}
\label{3.11}
\left(\begin{array}{cc}
s_0^2+4c_0(c_0-\kappa) & s_0^2+4c_0(c_0-\kappa)  \\ \\
-s_0c_0+4s_0(c_0-\kappa) & s_0c_0-4s_0(c_0-\kappa) \\ \\
-4c_0(c_0-\kappa) &-4c_0(c_0-\kappa) \\
\end{array}
\right)
\left( \begin{array}{c} \beta_1 \\ \beta_2 \\ \end{array} \right)=
\left( \begin{array}{c} -z_{f_1} \\ -z_{f_2} \\-z_{f_3} \end{array} \right).
\end{eqnarray}

For brevity we let 
\begin{equation}
\label{3.12}
q_0=4(c_0-\kappa)
\end{equation}
and obtain the following three equations in the unknowns $\beta_1$,
$\beta_2$ and $q_0$
\begin{equation}
\label{3.13}
(s_0^2+c_0q_0)(\beta_1+\beta_2)=-z_{f_1}
\end{equation}
\begin{equation}
\label{3.14}
s_0(c_0-q_0)(\beta_1-\beta_2)=z_{f_2}
\end{equation}
\begin{equation}
\label{3.15}
q_0(\beta_1+\beta_2)=z_{f_3}.
\end{equation}
Solving these equations we find that
\begin{equation}
\label{3.16}
\beta_1=-\frac{1}{2}\left[\frac{z_{f_1}}{s_0^2+c_0q_0}-\frac{z_{f_2}}{s_0(c_0-q_0)}\right]
\end{equation}
\begin{equation}
\label{3.17}
\beta_2=-\frac{1}{2}\left[\frac{z_{f_1}}{s_0^2+c_0q_0}+\frac{z_{f_2}}{s_0(c_0-q_0)}\right]
\end{equation}
\begin{equation}
\label{3.18}
q_0=-\frac{s_0^2z_{f_3}}{z_{f_1}+c_0z_{f_3}}.
\end{equation}
These equations are subject to (\ref{3.10}). The expression for $\kappa$ 
is found from (\ref{3.12}) and (\ref{3.18}). Setting $f(\bar{\theta}_0)=0$
we find 
\begin{equation}
\label{3.19}
\lambda=2(4s_0^2+q_0^2)^{1/2}.
\end{equation}
Substituting (\ref{3.18}) into (\ref{3.16}) and (\ref{3.17}) we obtain
\begin{equation}
\label{3.20}
\beta_1=-\frac{z_{f_1}+c_0z_{f_3}}{2s_0^2}\left(1-\frac{s_0z_{f_2}} {c_0z_{f_1}+z_{f_3}}\right)
\end{equation}
\begin{equation}
\label{3.21}
\beta_2=-\frac{z_{f_1}+c_0z_{f_3}}{2s_0^2}\left(1+\frac{s_0z_{f_2}} {c_0z_{f_1}+z_{f_3}}\right).
\end{equation}
In view of the inequality (\ref{3.10}) we find it necessary and sufficient 
for boundary conditions to satisfy
\begin{equation}
\label{3.22}
z_{f_1}+c_0z_{f_3} < 0,\,\,\, |s_0z_{f_2}| < |c_0z_{f_1}+z_{f_3}|
\end{equation}
to have non-stationary two-intersection solutions satisfying (\ref{3.06}).
For antipodal solutions satisfying (\ref{3.05}) the necessary and sufficient
conditions become
\begin{equation}
\label{3.23}
z_{f_1}+c_0z_{f_3} > 0,\,\,\, |s_0z_{f_2}| < |c_0z_{f_1}+z_{f_3}|.
\end{equation}

{\bf Example}: Select $\bar{\theta}_0=-\pi/2$, $\bar{\theta}_f=\pi/2$,
$c_0 =0$ and $s_0=-1$. From (\ref{3.18}) we find that $q_0=z_{f_3}/z_{f_1}$; it
follows from (\ref{3.12}) that $\kappa=-\frac{z_{f_3}}{4z_{f_1}}$.
We also find from (\ref{3.19}) that 
$\lambda=2(4+\frac{z_{f_3}^2}{z_{f_1}^2})^{1/2}$ although this formula will not be needed.

For this example the region (\ref{3.22}) of admissible solutions becomes
\begin{equation}
\label{3.24}
z_{f_1} < 0,\,\,\, |z_{f_2}| < |z_{f_3}|.
\end{equation}
The expressions (\ref{3.20}) and (\ref{3.21}) respectively become
\begin{equation}
\label{3.25}
\beta_1=-\frac{z_{f_1}}{2}\left(1+\frac{z_{f_2}}{z_{f_3}}\right)
\end{equation}
\begin{equation}
\label{3.26}
\beta_2=-\frac{z_{f_1}}{2}\left(1-\frac{z_{f_2}}{z_{f_3}}\right).
\end{equation}

The optimal velocity increments therefore follow from (\ref{3.07})
\begin{eqnarray}
\label{3.27}
\Delta {\bf V}_1=\left(\begin{array}{c}
-1 \\ -2\kappa
\end{array}
\right)\beta_1,\,\,\,
\Delta {\bf V}_2=\left(\begin{array}{c}
1 \\ -2\kappa
\end{array}
\right)\beta_2\,\,\,.
\end{eqnarray}
The actual velocity increments are obtained from (\ref{2.134}) and 
(\ref{2.135}). For antipodal solutions the first inequality of 
(\ref{3.24}) and the directions of (\ref{3.27}) are reversed.

This example was simulated for
$r_0=6545km $, $v_{r_0}=-0.784km/sec$, $\dot{{\theta}_0}=1.3\times 10^{-3} 
rad/sec$, $r_f=8000km$, $v_{r_f}=-0.710km/sec$, 
$\dot{{\theta}_f}=0.798\times 10^{-3} rad/sec$, boundary values which 
satisfy (23). The resulting optimal rendezvous orbit is depicted in Figure 
$18$. The arrows in the figure indicate the application 
of the optimal velocity increments 
$\Delta V_1=(\Delta V_x,\Delta V_y)= (0.4795,-0.8282)^Tkm/sec$, 
$\Delta V_2=(\Delta V_x,\Delta V_y)= (-0.0142,0.2986)^Tkm/sec$.

\subsection{Non-Stationary Four-Intersection Solutions}

Next we consider the category of solutions in which the primer vector loci 
demonstrate the geometry represented in Figures $12$ and $13$, the four 
intersections. Again we specify that $\theta_0 < \theta_f$.

\subsubsection{Fundamental Theorem of Non-Stationary Four-Intersection
Solutions}

The following theorem is fundamental for this type of two-impulse solutions.

{\bf Theorem}: The optimal impulsive rendezvous problem defined by 
(\ref{2.40}) and (\ref{2.47}) on the interval 
$\theta_0 \le \theta \le \theta_f$ 
satisfying the convention $-\pi \le \bar{\theta}_0 \le \pi$ has a 
non-stationary four-intersection solution $\{\theta_1,\theta_2,\Delta{\bf V}_1,
\Delta{\bf V}_2\}$ if and only if 
\begin{equation}
\label{3.28}
\theta_1 = \theta_0,\,\,\,\theta_2=\theta_f,
\end{equation}
\begin{equation}
\label{3.29}
either\,\,\, 0 < \bar{\theta}_0 < \bar{\theta}_f < \pi\,\,\,\
or\,\,\, -\pi < \bar{\theta}_0 < \bar{\theta}_f < 0,
\end{equation}
and
\begin{equation}
\label{3.30}
f(\theta) < 0,\,\,\,\, \theta_0 \le \theta \le \theta_f,
\end{equation}
\begin{equation}
\label{3.31}
f^{\prime}(\theta_0) < 0,\,\,\,f^{\prime}(\theta_f) > 0, 
\end{equation}
\begin{equation}
\label{3.32}
\kappa =\frac{3}{8}(\cos\bar{\theta}_0 +\cos\bar{\theta}_f),\,\,\,
-\frac{3}{4} < \kappa < \frac{3}{4},
\end{equation}
\begin{equation}
\label{3.33}
\lambda=[\sin^2\bar{\theta}_0+4(\cos\bar{\theta}_0 -\kappa)^2]^{-1/2},\,\,\,
\frac{1}{2} < \frac{1}{2(1+|\kappa|)} \le \lambda  \le 
(\frac{3}{3-4\kappa^2})^{1/2} < 2
\end{equation}
and (\ref{3.07})-(\ref{3.09}) are satisfied.

{\bf Proof}: First we show that (\ref{3.28})-(\ref{3.33}) and 
(\ref{3.07})-(\ref{3.09}) are necessary.

We suppose that the optimization problem has a non-stationary 
four-intersection solution. It follows that $f(\theta_1)=f(\theta_2)=0$
where $f$ is defined by (\ref{2.96}),
$\bar{\theta}_0 \le \bar{\theta}_1 < \bar{\theta}_2 \le \bar{\theta}_f$,
there is an integer $i$ such that
$i\pi < \bar{\theta}_1 <\bar{\theta}_2 <(i+1)\pi$, $0<\bar{\theta}_2-
\bar{\theta}_1 <\pi$, and $k=2$ in (\ref{2.58})-(\ref{2.63}) from the 
definition of non-stationary four-intersection solution. 
By (\ref{2.62}) there
are no other zeros of $f$ on $\bar{\theta}_0 \le \theta \le \bar{\theta}_f$,
hence $\bar{\theta}_1=\bar{\theta}_0$ and $\bar{\theta}_2=\bar{\theta}_f$.
From the convention $-\pi \le \bar{\theta}_0 \le \pi$ it follows that either 
$i=0$ and $0 < \bar{\theta}_0 <\bar{\theta}_f <\pi$ or $i=-1$ and 
$-\pi < \bar{\theta}_0 <\bar{\theta}_f < \pi$. Since $\bar{\theta}_0$
and $\bar{\theta}_f$ are the only zeros of $f$ on 
$\bar{\theta}_0 \le \bar{\theta} \le \bar{\theta}_f$, (\ref{2.63})
implies that $f(\theta) < 0$ on the interval $\bar{\theta}_0  < \bar{\theta} <
\bar{\theta}_f$. We have established (\ref{2.28})-(\ref{2.30}).

Using (\ref{2.82}),(\ref{2.83}) and (\ref{2.96}) we have
$$
f(\theta)= \lambda^2[\sin^2\bar{\theta}+4(\cos\bar{\theta}-\kappa)^2]-1 
$$
and
$$
f^{\prime}(\theta)=-2\lambda^2\sin\bar{\theta}\,(3\cos\bar{\theta}-4\kappa).
$$
Setting $f(\theta_0)=f(\theta_f)$ we get
$$
(\cos\bar{\theta}_f-\cos\bar{\theta}_0)[3(\cos\bar{\theta}_0+
\cos\bar{\theta}_f)-8\kappa]=0.
$$ 

Since $0 <\bar{\theta}_f-\bar{\theta}_0 < \pi$ then $\cos\bar{\theta}_f \ne
\cos\bar{\theta}_0$ establishing (\ref{3.32}). If $f^{\prime}(\theta_0)=0$ then
$\kappa=3/4\cos\bar{\theta}_0$. Since 
$$
f^{\prime\prime}(\theta)=-2\lambda^2(6\cos^2\theta -4\kappa\cos\bar{\theta}-3)
$$
then $f^{\prime\prime}(\theta_0) =6\lambda^2\sin^2\bar{\theta}_0 > 0$
showing that $f(\theta) > 0$ on an interval where $\bar{\theta} >
\bar{\theta}_0$ contradicting (\ref{2.63}) and establishing that
$f^{\prime}(\theta_0) \ne 0$. A similar argument shows that
$f^{\prime}(\theta_f) \ne 0$. Either $f^{\prime}(\theta_0) > 0$ or
$f^{\prime}(\theta_f) < 0$ implies that (\ref{3.30}) is false establishing
(\ref{2.31}).

The above arguments have shown (\ref{3.28})-(\ref{3.32}).
Solving the equation $f(\theta_0)=0$ for $\lambda$ and minimizing and 
maximizing this expression with respect to $\theta_0$ establishes (\ref{3.33}).
Finally (\ref{3.07})-(\ref{3.09}) follow from (\ref{2.38}) and 
(\ref{2.58})-(\ref{2.60}) by setting $k=2$.

We now show that (\ref{3.28})-(\ref{3.33}) and (\ref{3.07})-(\ref{3.09})
are sufficient. These conditions show that (\ref{2.58})-(\ref{2.63})  
are satisfied for $k=2$ resulting in an optimal two-impulse solution 
of the problem. This solution is non-degenerate because (\ref{3.31}) 
implies that $\theta_0$ and $\theta_f$ are non-stationary. It is a 
four-intersection solution because (\ref{3.28}) and (\ref{3.29}) reveal 
that $i=-1$ or $i=0$ and $0 < \bar{\theta}_2 -\bar{\theta}_1 < \pi$ 
follows from (\ref{3.29}) after setting $\bar{\theta}_1=\bar{\theta}_0$
and $\bar{\theta}_2=\bar{\theta}_f$. $\blacksquare$

\subsubsection{Finding Non-Stationary Four-Intersection Solutions}

Since $\lambda > 0$ we again set $\beta_1= \lambda \alpha_1$,
$\beta_2= \lambda\alpha_2$ and (\ref{3.09}) must be satisfied. For brevity 
we set $c_0=\cos\bar{\theta}_0$, $s_0=\sin\bar{\theta}_0$, 
$c_f=\cos\bar{\theta}_f$ and $s_f=\sin\bar{\theta}_f$. Since 
$\Phi(\bar{\theta})$ is a fundamental matrix solution associated with $A$ 
we again 
use $R(\bar{\theta}_1)$ and $R(\bar{\theta}_2)$  in (\ref{3.08}).
Setting $\bar{\theta}_1=\bar{\theta}_0$, $\bar{\theta}_2=\bar{\theta}_f$ ,
substituting ${\bf p}({\theta}_0)$ and ${\bf p}({\theta}_f)$ into
(\ref{3.08}) and utilizing (\ref{3.32}), after rearranging we obtain:
\begin{eqnarray}
\label{3.34}
\left(\begin{array}{cc}
1+\frac{3}{2}c_0(c_0-c_f) & 1-\frac{3}{2}c_f(c_0-c_f)  \\ \\
\frac{3}{2}s_0(c_0-c_f)  &-\frac{3}{2}s_f(c_0-c_f) \\ \\
-c_0-\frac{3}{2}(c_0-c_f) & -c_f+\frac{3}{2}(c_0-c_f)\\
\end{array}
\right)
\left( \begin{array}{c} \beta_1 \\ \beta_2 \\ \end{array} \right)=
\left( \begin{array}{c} -z_{f_1} \\ -z_{f_2} \\-z_{f_3} \end{array} \right).
\end{eqnarray}
Solving these equations, we find that
\begin{equation}
\label{3.35}
\beta_1 =-\frac{s_f}{D}z_{f_1}- \frac{\frac{2}{3(c_0-c_f)}-c_f}{D}z_{f_2}
\end{equation}
\begin{equation}
\label{3.36}
\beta_2 =-\frac{s_0}{D}z_{f_1}+ \frac{\frac{2}{3(c_0-c_f)}+c_0}{D}z_{f_2}
\end{equation}
if
\begin{equation}
\label{3.37}
z_{f_3}= \frac{[9(s_0c_0+s_fc_f)-15(c_0s_f+c_fs_0)]}{6D}z_{f_1}-
\frac{[9c_0^2-30c_0c_f+9c_f^2+16]}{6D}z_{f_2}
\end{equation}
where
\begin{equation}
\label{3.38}
D=s_0+s_f+\frac{3}{2}(c_0-c_f)(c_0s_f-c_fs_0).
\end{equation}
The velocity increments from (\ref{3.07}) become
\begin{eqnarray}
\label{3.39}
\Delta {\bf V}_1= \left(\begin{array}{c}
-s_0 \\
-\frac{5}{4}c_0+\frac{3}{4}c_f \\
\end{array}
\right)\beta_1,\,\,\,\,
\Delta {\bf V}_2=\left( \begin{array}{c} 
-s_f \\ 
\frac{3}{4}c_0-\frac{5}{4}c_f \\ 
\end{array} \right)\beta_2.
\end{eqnarray}
In the case where $0 < \bar{\theta}_0 < \pi$, then $D> 0$ and a 
non-degenerate four-intersection solution is found only in a ${\bf z}_f$ 
region where
\begin{equation}
\label{3.40}
s_fz_{f_1}+ \left[\frac{2}{3(c_0-c_f)}-c_f\right]z_{f_2} <0,\,\,\,\,
s_0z_{f_1}- \left[\frac{2}{3(c_0-c_f)}+c_0\right]z_{f_2} <0,
\end{equation}
and $z_{f_3}$ satisfies (\ref{3.37}). Geometrically this says that there 
are solutions restricted to a sector described by (\ref{3.40}) of the 
plane described by (\ref{3.37}).

In the case where $-\pi < \bar{\theta}_0 < 0$ then $D <0$ and the 
inequalities in (\ref{3.40}) are reversed but (\ref{3.37}) remains valid.
Geometrically this reverses the sector in the plane with every vector 
${\bf z}_f$ in the sector being replaced by $-{\bf z}_f$. This is the 
antipodal case.

Of some interest is the special case where $\kappa=0$ so that $c_0=-c_f$,
$s_0=s_f$ resulting in a primer locus having symmetry about a vertical line
through the origin. The equations (\ref{3.35})-(\ref{3.37}) simplify:
\begin{equation}
\label{3.41}
\beta_1= -\frac{z_{f_1}}{2(1+3c_0^2)}-\frac{z_{f_2}}{6s_0c_0}
\end{equation}
\begin{equation}
\label{3.42}
\beta_2= -\frac{z_{f_1}}{2(1+3c_0^2)}+\frac{z_{f_2}}{6s_0c_0}
\end{equation}
\begin{equation}
\label{3.43}
z_{f_3}=-\frac{4}{3s_0}z_{f_2}.
\end{equation}
The velocity increments (\ref{3.39}) also simplify somewhat in this case.
This type of solution exists only for the ${\bf z}_f$ region where
\begin{equation}
\label{3.44}
z_{f_2}< -\frac{3s_0c_0}{1+3c_0^2}z_{f_1},\,\,\, 
z_{f_2} > \frac{3s_0c_0}{1+3c_0^2}z_{f_1}
\end{equation}
and $z_{f_3}$ satisfies (\ref{3.43}), consequently
\begin{equation}
\label{3.45}
z_{f_1}< 0,\,\,\, \frac{3s_0c_0}{1+3c_0^2}z_{f_1} < z_{f_2} < 
-\frac{3s_0c_0}{1+3c_0^2}z_{f_1}
\end{equation}
and
\begin{equation}
\label{3.46}
\frac{4s_0c_0}{1+3c_0^2}z_{f_1} < s_0z_{f_3} < 
-\frac{4s_0c_0}{1+3c_0^2}z_{f_1}.
\end{equation}

{\bf Example}: Select $\bar{\theta}_0=\pi/6$ and $\bar{\theta}_f=5\pi/6$,
then $c_0=-c_f=\sqrt{3}/2$, $s_0=s_f=1/2$, 
$\beta_1=-2/13z_{f_1}-2\sqrt{3}/9z_{f_2}$,
$\beta_2=-2/13z_{f_1}+2\sqrt{3}/9z_{f_2}$ and 
\begin{eqnarray}
\Delta {\bf V}_1= \left(\begin{array}{c}
-\frac{1}{2} \\
-\sqrt{3} \\
\end{array}
\right)\beta_1,\,\,\,\,
\Delta {\bf V}_2\left( \begin{array}{c} 
-\frac{1}{2} \\
\sqrt{3} \\ 
\end{array} \right)\beta_2.
\end{eqnarray} 
We select values of $y_0$ and $y_f$ so that $z_{f_1} < 0$ and 
$z_{f_2}=z_{f_3}=0$. These conditions are satisfied by the boundary
values $r_0=8000km$, $v_{r_0}=0 km/sec$, $\dot{{\theta}_0}=8.37\times 
10^{-4} rad/sec$, $r_f=6545km$, $v_{r_f}=-0.866km/sec$, 
$\dot{{\theta}_f}=1.251\times 10^{-3} rad/sec$. Simulation of this example
with these boundary values was performed, and the optimal rendezvous
orbit is presented in Figure $19$. The imposition of the velocity increments
$\Delta V_1=(\Delta V_x,\Delta V_y)= (0.0382,-0.1216)^Tkm/sec$, 
$\Delta V_2=(\Delta V_x,\Delta V_y)= (-0.0362,-0.1007)^Tkm/sec$
is indicated by the arrows.

\subsection{ Two-Impulse Non-Stationary Three-intersection Solutions}

Three-impulse non-stationary three-intersection solutions 
are discussed
in Sec. $3.3.2$. If $\theta_f$ is somewhat decreased or $\theta_0$ 
increased from that of a non-degenerate three-impulse solution as shown
in Figures $14$ and $15$ the result is a two-impulse non-stationary 
three-intersection solution.

\subsubsection{Fundamental Theorem of Two-impulse Non-Stationary Three-intersection Solutions}

{\bf Theorem}: The optimal impulsive rendezvous problem defined by 
(\ref{2.40}) and (\ref{2.47}) on the interval 
$\theta_0 \le \theta \le \theta_f$ 
satisfying the convention $-\pi \le \bar{\theta}_0 \le \pi$ has a 
 two-impulse non-stationary three-intersection solution
$\{\theta_1,\theta_2,\Delta{\bf V}_1,\Delta{\bf V}_2\}$ if and only if
$0 < \bar{\theta}_0  < \pi$ and
\begin{equation}
\label{3.47}
\bar{\theta}_1=\bar{\theta}_0,\,\bar{\theta}_2=\pi \le \bar{\theta}_f <
2\pi-\bar{\theta}_0,f^{\prime}(\theta_0) < 0,\, \kappa=\frac{3}{8}
(\cos\bar{\theta}_0-1),\,\lambda=\frac{4}{5+3\cos\bar{\theta}_0},
\end{equation}
or
\begin{equation}
\label{3.48}
2\pi-\bar{\theta}_f <\bar{\theta}_0 < \bar{\theta}_1=\pi,\,
\bar{\theta}_2=\bar{\theta}_f ,\,f^{\prime}(\theta_f) > 0,\, 
\kappa=\frac{3}{8} (\cos\bar{\theta}_f-1),\,
\lambda=\frac{4}{5+3\cos\bar{\theta}_f},
\end{equation}

or else $-\pi < \bar{\theta}_0 < 0$ and
\begin{equation}
\label{3.49}
\bar{\theta}_1=\bar{\theta}_0,\,\bar{\theta}_2=0 \le \bar{\theta}_f <
-\bar{\theta}_0,\,f^{\prime}(\theta_0) < 0,\, \kappa=\frac{3}{8}
(\cos\bar{\theta}_0+1),\,\lambda=\frac{4}{5-3\cos\bar{\theta}_0},
\end{equation}
or
\begin{equation}
\label{3.50}
-\pi < -\bar{\theta}_f <\bar{\theta}_0 < \bar{\theta}_1=0,\,
\bar{\theta}_2=\bar{\theta}_f ,\,f^{\prime}(\theta_f) > 0,\, 
\kappa=\frac{3}{8} (\cos\bar{\theta}_f+1),\,
\lambda=\frac{4}{5-3\cos\bar{\theta}_f},
\end{equation}
and in either case
\begin{equation}
\label{3.51}
f(\theta)\le 0, \bar{\theta}_0 < \bar{\theta} <\bar{\theta}_f,
\end{equation}
and (\ref{3.07})-(\ref{3.09}) are satisfied.

{\bf Proof}: First we show that (\ref{3.47})-(\ref{3.51}) and 
(\ref{3.07})-(\ref{3.09}) are necessary.

If the optimization problem has a two-impulse non-stationary three-intersection
solution then $f(\theta_1)=f(\theta_2)=0$ where $\bar{\theta}_0 \le
\bar{\theta}_1 < \bar{\theta}_2 \le \bar{\theta}_f$ and either
$f^{\prime}(\theta_1)=0$ or $f^{\prime}(\theta_2)=0$a but not both,
and $-\pi \le \bar{\theta}_0 \le \pi$. We observe that
$$
f(\theta) =\lambda^2[\sin^2\bar{\theta}+4(\cos\bar{\theta}-\kappa)^2] -1
$$
$$
f^{\prime}(\theta)=-2\lambda^2\sin\bar{\theta}(3\cos\bar{\theta}-4\kappa)
$$
$$
f^{\prime\prime}(\theta)=-2\lambda^2(6\cos^2\bar{\theta}-4\kappa
\cos\bar{\theta} -3)
$$
Equating $f(\theta_1)$ and $f(\theta_2)$ and performing some manipulations 
we obtain
$$
(\cos\bar{\theta}_2-\cos\bar{\theta}_1)[3(\cos\bar{\theta}_1+
\cos\bar{\theta}_2)-8\kappa]=0.
$$
We shall show that $\cos\bar{\theta}_1 \ne \cos\bar{\theta}_2$.

Suppose that $f^{\prime}(\theta_1) \ne 0$ and $f^{\prime}(\theta_2)=0$.
If these are reversed the following argument is also valid with cyclic 
interchange of $\theta_1$ and $\theta_2$.

Setting $f^{\prime}(\theta_2)=0$ implies either $\sin\bar{\theta}_2=0$
or $\kappa=\frac{3}{4}\cos\bar{\theta}_2$ but the latter implies
$f^{\prime\prime}(\theta_2)=6\lambda^2\sin^2\bar{\theta}_2 > 0$, 
which implies that
$f(\theta) > 0$ on an interval where $\bar{\theta} < \bar{\theta}_2$
because $f(\theta_2)=0$, contradicting (\ref{2.63}). It follows
that $\sin\bar{\theta}_2=0$. If $\cos\bar{\theta}_1=\cos\bar{\theta}_2$
then $\sin\bar{\theta}_1=0$ also setting $f^{\prime}(\theta_1)=0$
contrary to the original supposition consequently $\kappa=
\frac{3}{8}(\cos\bar{\theta}_1+\cos\bar{\theta}_2)$.

Continuing $f^{\prime}(\theta_2)=0$ implies $\sin\bar{\theta}_2=0$
which implies that $\bar{\theta}_2=0$ or $\bar{\theta}_2=\pi$.

If $\bar{\theta}_2=\pi$ then $0 <\bar{\theta}_1< \bar{\theta}_2$ since
$f^{\prime}(\theta_1)\ne  0$. Setting $k=2$ in (\ref{2.62}), we 
see that $\bar{\theta}_1=\bar{\theta}_0$, so $0 < \bar{\theta}_0 < \pi$,
$\bar{\theta}_2 \le \theta_f$, and $\bar{\theta}_f < 2\pi -\bar{\theta}_0$
because $f(2\pi-\bar{\theta}_0)=0$. Clearly $f^{\prime}(\theta_0) < 0$
because $f^{\prime}(\theta_0) > 0$ with $f(\theta_0)=0$ establishes an
interval where $\bar{\theta} >\bar{\theta}_0$ and (\ref{2.63}) 
is violated. Substituting $\cos\bar{\theta}_2=-1$ in the above expression 
for $\kappa$ reveals $\kappa=\frac{3}{8}(\cos\bar{\theta}_0-1)$. The 
expression for $\lambda$ comes from setting $f(\theta_2)=0$ substituting 
for $\kappa$ and solving for $\lambda$. We have established (\ref{3.47}).

If $\bar{\theta}_2=0$ the proof is analogous; 
$-\pi <\bar{\theta}_1< \bar{\theta}_2$ since $f(\theta_1) \ne 0$.
Again setting $k=2$ in (\ref{2.62}) we get $\bar{\theta}_1=
\bar{\theta}_0$ so $-\pi <\bar{\theta}_0 < 0$, $\bar{\theta}_2 \le 
\bar{\theta}_f$ and $\bar{\theta}_f < -\bar{\theta}_0$. Again (\ref{2.63}) 
implies that $f^{\prime}(\theta_0) < 0$. Substituting
$\cos\bar{\theta}_2=1$ into the above expression for $\kappa$ reveals that 
$\kappa=\frac{3}{8}(\cos\bar{\theta}_0+1)$. Again setting $f(\theta_2)=0$
substituting for $\kappa$ and evaluating $\lambda$ completes (\ref{3.49}).

The expressions (\ref{3.48}) and (\ref{3.50}) follow from 
$f^{\prime}(\theta_1) = 0$ and $f(\theta_2) \ne 0$ by repeating the 
preceding argument with $\theta_1$ and $\theta_2$ interchanged. These arguments lead to an interchange of $\bar{\theta}_0$ and $\bar{\theta}_f$ and
a reversal of all inequalities.

The expression (\ref{3.51}) is a restatement of (\ref{2.63})
and (\ref{3.07})-(\ref{3.09}) follow from (\ref{2.58})-(\ref{2.60}) 
by setting $k=2$. The inequalities in (\ref{3.09}) are strict
because the solution is a non-degenerate two-impulse solution.

To show that (\ref{3.47})-(\ref{3.51}) and (\ref{3.07})-(\ref{3.09}) are 
sufficient we note that they imply (\ref{2.58})-(\ref{2.63}) with $k=2$
resulting in an optimal two-impulse solution. In each of 
(\ref{3.47})-(\ref{3.50}) $\bar{\theta}_0 \le \bar{\theta}_1 <
\bar{\theta}_2 \le \bar{\theta}_f$, and $f(\theta_1)=f(\theta_2)=0$ and
either $f^{\prime}(\theta_1) \ne 0$ or $f^{\prime}(\theta_2) \ne 0$
but not both, hence the solution is a non-stationary three-intersection 
solution and not degenerate. $\blacksquare$

\subsubsection{Finding Two-Impulse Non-Stationary Three-Intersection Solutions}

First we consider the case where $0 < \bar{\theta}_0 < \pi$,
$\bar{\theta}_0=\bar{\theta}_1 < \bar{\theta}_2 =\pi < \bar{\theta}_f
< 2\pi -\bar{\theta}_0$. Again we set $c_0=\cos\bar{\theta}_0$,
$s_0=\sin\bar{\theta}_0$ and observe that $\Phi(\bar{\theta})$ is a 
fundamental matrix solution associated with A. Substituting 
$\kappa=\frac{3}{8}(c_0-1)$ into ${\bf p}(\theta_0)$ and ${\bf p}(\theta_2)$
the expression (\ref{3.08}) becomes
\begin{eqnarray}
\label{3.52}
\left(\begin{array}{cc}
\frac{1}{2}(3c_0^2+3c_0+2) & \frac{1}{2}(3c_0 +5)  \\ \\
\frac{3}{2}s_0(c_0+1)  & 0 \\ \\
-\frac{1}{2}(5c_0+3) & \frac{1}{2}(3c_0+5) \\
\end{array}
\right)
\left( \begin{array}{c} \beta_1 \\ \beta_2 \\ \end{array} \right)=
\left( \begin{array}{c} z_{f_1} \\ z_{f_2} \\z_{f_3} \end{array} \right).
\end{eqnarray}

A solution exists if
\begin{equation}
\label{3.53}
\beta_1=\frac{2z_{f_2}}{3s_0(c_0+1)},
\end{equation}
\begin{equation}
\label{3.54}
\beta_2=\frac{6s_0(c_0+1)z_{f_1}-2(3c_0^2+3c_0+2)z_{f_2}}
{3s_0(c_0+1)(3c_0+5)},
\end{equation}
and
\begin{equation}
\label{3.55}
z_{f_3}=z_{f_1}-\frac{(3c_0+5)z_{f_2}}{3s_0}.
\end{equation}

It is necessary from (\ref{3.09}) that $\beta_1 > 0$ and $\beta_2 > 0$,
and since the denominators in (\ref{3.53}) and (\ref{3.54}) must
be positive, it is necessary that the ${\bf z}_f$ region satisfy (\ref{3.55})
and
\begin{equation}
\label{3.56}
z_{f_1} > \frac{(3c_0^2+3c_0+2)z_{f_2}}{3s_0(c_0+1)}, \,\,\, z_{f_2} > 0.
\end{equation}
Geometrically, this region is a sector of a plane through the origin
(56). If ${\bf z}_f$ is contained in this sector then (\ref{3.07}) 
determines optimal velocity increments
\begin{eqnarray}
\label{3.57}
\Delta {\bf V}_1= \left(\begin{array}{c}
s_0 \\
\frac{5c_0+3}{4} \\
\end{array}
\right)\beta_1,\,\,\,\,
\Delta {\bf V}_2=-\left( \begin{array}{c} 
0 \\
\frac{3c_0+5}{4} \\ 
\end{array} \right)\beta_2.
\end{eqnarray} 
We note also that if $-\pi < \bar{\theta}_0 < 0$,
$\bar{\theta}_0 =\bar{\theta}_1 < \bar{\theta}_2=0 < \bar{\theta}_f
< -\bar{\theta}_0$ then we have an antipodal solution where ${\bf z}_f$
is replaced by $-{\bf z}_f$ in (\ref{3.52})-(\ref{3.56}).

Next we consider the case where $2\pi-\bar{\theta}_f < \bar{\theta}_0
< \bar{\theta}_1=\pi$, $\bar{\theta}_2=\bar{\theta}_f$. We set 
$c_f=\cos\bar{\theta}_f$, $s_f=\sin\bar{\theta}_f$ and substitute
$\kappa=\frac{3}{8}(c_f-1)$ in ${\bf p}(\theta_1)$ and
${\bf p}(\theta_f)$. The expression (\ref{3.08}) becomes
\begin{eqnarray}
\label{3.58}
\left(\begin{array}{cc}
\frac{1}{2}(3c_f+5) & \frac{1}{2}(3c_f^2+3c_f+2)  \\ \\
0 & \frac{3}{2}s_f(c_f+1)  \\ \\
\frac{1}{2}(3c_f+5) & -\frac{1}{2}(5c_f+3) \\
\end{array}
\right)
\left( \begin{array}{c} \beta_1 \\ \beta_2 \\ \end{array} \right)=
\left( \begin{array}{c} -z_{f_1} \\ -z_{f_2} \\-z_{f_3} \end{array} \right).
\end{eqnarray}

A solution exists if
\begin{equation}
\label{3.59}
\beta_1=\frac{-6s_f(c_f+1)z_{f_1}+2(3c_f^2+3c_f+2)z_{f_2}}
{3s_f(c_f+1)(3c_f+5)},
\end{equation}
\begin{equation}
\label{3.60}
\beta_2=\frac{-2z_{f_2}}{3s_f(c_f+1)},
\end{equation}
and
\begin{equation}
\label{3.61}
z_{f_3} = z_{f_1}-\frac{3c_f+5}{3s_f}z_{f_2}.
\end{equation}
Since $\beta_1 > 0$, $\beta_2 > 0$, and the denominators in (\ref{3.59}) 
and (\ref{3.60}) are negative, The ${\bf z}_f$ region must satisfy 
(\ref{2.61}) and
\begin{equation}
\label{3.62}
z_{f_1} < \frac{3c_f^2+3c_f+2}{3s_f(c_f+1)}z_{f_2},\,\,\, z_{f_2} > 0.
\end{equation}
Geometrically this is a sector of the plane (\ref{3.61}). If ${\bf z}_f$ 
is contained in this sector, then optimal velocity increments are determined
by (\ref{3.07})
\begin{eqnarray}
\label{3.63}
\Delta {\bf V}_1= \left(\begin{array}{c}
0 \\
\frac{3c_f+5}{4} \\
\end{array}
\right)\beta_1,\,\,\,\,
\Delta {\bf V}_2=-\left( \begin{array}{c} 
s_f \\
\frac{5c_f+3}{4} \\ 
\end{array} \right)\beta_2.
\end{eqnarray}

We note that if $-\pi < \bar{\theta}_0 < 0$,
$-\bar{\theta}_f < \bar{\theta}_0 < \bar{\theta}_1=0$,  $\bar{\theta}_2
= \bar{\theta}_f$ then we have an antipodal solution with ${\bf z}_f$
replaced by $-{\bf z}_f$ in (\ref{3.58})-(\ref{3.62}).

{\bf Example}: Select $\bar{\theta}_0=\pi/2$, $\bar{\theta}_2=\pi$
and $\bar{\theta}_f=\frac{4\pi}{3}$. It follows that $c_0=0$, $s_0=1$,
$c_f=-\frac{1}{2}$ $s_f=-\frac{\sqrt{3}}{2}$, (\ref{3.56}) becomes
\begin{equation}
\label{3.64}
z_{f_1} > \frac{2}{3}z_{f_2},\,\,\, z_{f_2} > 0
\end{equation}
and (\ref{3.55}) becomes
\begin{equation}
\label{3.65}
z_{f_3} = z_{f_1} - \frac{5}{3}z_{f_2}.
\end{equation}
Substituting into (57) of Reference $1$ we obtain
\begin{equation}
\label{3.66}
z_{f_1} =- \frac{1}{2}y_{f_1}+ \frac{\sqrt{3}}{2}y_{f_2}
+\frac{1}{2}y_{f_3}+y_{0_2},
\end{equation}
\begin{equation}
\label{3.67}
z_{f_2} =- \frac{\sqrt{3}}{2}y_{f_1}- \frac{1}{2}y_{f_2}
+\frac{\sqrt{3}}{2}y_{f_3} -y_{0_1}+ y_{0_3},
\end{equation}
\begin{equation}
\label{3.68}
z_{f_3} = y_{f_3} - y_{0_3}.
\end{equation}
Combining (\ref{2.64})-(\ref{2.68}), we find
\begin{equation}
\label{3.69}
(2\sqrt{3}-6)y_{f_1}+(3\sqrt{3}+2)y_{f_2}+(-2\sqrt{3}+3)y_{f_3}+
4y_{0_1}+6y_{0_2}-4y_{0_3} > 0,
\end{equation}
\begin{equation}
\label{3.70}
-\sqrt{3}y_{f_1}-y_{f_2}+\sqrt{3}y_{f_3}- 2y_{0_1}+2y_{0_3} > 0,
\end{equation}
\begin{equation}
\label{3.71}
(3-5\sqrt{3})y_{f_1}-(3\sqrt{3}+5)y_{f_2}+(5\sqrt{3}+5)y_{f_3}-
10y_{0_1}-6y_{0_2}+4y_{0_3} = 0.
\end{equation}
These inequalities are satisfied by the boundary values
$r_0=7200km$, $v_{r_0}=0 km/sec$, $\dot{{\theta}_0}=1.033\times 
10^{-3} rad/sec$, $r_f=10375km$, $v_{r_f}=-3.547km/sec$, 
$\dot{{\theta}_f}=4.746\times 10^{-4} rad/sec$.
This example was simulated for these boundary values, and the resulting 
optimal rendezvous orbit is presented in Figure $20$. The optimal velocity 
increments, represented by the arrows on the figure are
$\Delta V_1= (\Delta V_x,\Delta V_y)= (-0.3759,0.5269)^Tkm/sec$, 
$\Delta V_2= (\Delta V_x,\Delta V_y)= (-1.691,0.9413)^Tkm/sec$.

\subsection{Non-Degenerate Stationary Two-Impulse Solutions}

A k-impulse solution is called {\bf stationary} if $f(\theta_i)=0$,
$i=1,\ldots,k$ for distinct $\theta_i$ on the interval $\theta_0 \le \theta 
\le \theta_f$ implies $f^{\prime}(\theta_i)=0$, $i=1,\ldots,k$. It is called 
{\bf non-degenerate} if there is no {\it l}-impulse solution for 
${\it l} < k$. It can be shown from the periodicity of (2.60) 
that stationary solutions for $k >2$ are degenerate.

Primer vector loci of stationary solutions are presented in Figures $16$ 
and $17$. Figure $16$ depicts a primer vector locus for a stationary 
non-degenerate two-impulse solution, whereas Figure $17$ represents 
a primer vector locus of a stationary degenerate multi-impulse solution.

Degenerate solutions are presented in a paper that follows.

\subsubsection{ Fundamental Theorem of Non-Degenerate Stationary 
Two-Impulse Solutions}

We shall accept the convention $-\pi < \bar{\theta}_0 \le \pi$. If 
$\bar{\theta}_0$ is outside this interval, a simple change of variable
can be employed.

{\bf Theorem}: The optimal impulsive rendezvous defined by (\ref{2.40}) 
and (\ref{2.47}) on the interval $\bar{\theta}_0 \le \bar{\theta} \le
\bar{\theta}_f$ satisfying the convention $-\pi < \bar{\theta}_0 \le \pi$
has a non-degenerate stationary two-impulse solution
$\{\theta_1, \theta_2, \Delta {\bf V}_1, \Delta{\bf V}_2 \}$ if and only if
\begin{equation}
\label{3.73}
\kappa =0, \lambda=\frac{1}{2}
\end{equation}
\begin{equation}
\label{3.74}
\pi \le \theta_f-\theta_0 < 3\pi,
\end{equation}
and if $-\pi <\bar{\theta}_0 \le 0$ then
\begin{equation}
\label{3.75}
\bar{\theta}_1=0,\,\,\,\bar{\theta}_2=\pi,\,\,\, \phi=\theta_1=\theta_2-\pi,
\,\,\, \pi \le \bar{\theta}_f < 2\pi,
\end{equation}
but if $0 < \bar{\theta}_0 \le \pi$ then
\begin{equation}
\label{3.76}
\bar{\theta}_1=\pi,\,\,\,\bar{\theta}_2=2\pi,\,\,\, \phi=\theta_1-\pi
=\theta_2-2\pi,\,\,\, 2\pi \le \bar{\theta}_f,
\end{equation}
and in either case
\begin{equation}
\label{3.77}
f(\theta)=-\frac{3}{4}\sin^2\bar{\theta} \le 0,\,\,\,
\bar{\theta}_0  \le \bar{\theta} \le \bar{\theta}_f
\end{equation}
and (\ref{3.07})-(\ref{3.09}) are satisfied.

{\bf Proof}: First we show that (\ref{3.73})-(\ref{3.77}) and
(\ref{3.07})-(\ref{3.09}) are necessary.
It is shown in (\ref{2.96}) that a stationary solution must be supported on 
integer multiples of $\pi$; non-degeneracy requires this support  at the
two smallest multiples of $\pi$ greater than or equal $\bar{\theta}_0$,
consequently $\bar{\theta}_f$ must be placed after the second smallest
multiple of $\pi$ and previous to the third. The results (\ref{3.74})-
(\ref{3.76}) follow and (\ref{3.73}) is found from (\ref{2.97}) and 
(\ref{2.98}). Substituting
(\ref{3.73}) into the expression for the primer vector, we have
$$
f(\theta)=\frac{1}{4}\sin^2\bar{\theta}+\cos^2\bar{\theta}-1
$$
which results in (\ref{3.77}). We obtain (\ref{3.07})-(\ref{3.09}) by 
setting $k=2$ in (\ref{2.58})-(\ref{2.60}); since the solution is
non-degenerate the inequalities in (\ref{3.09}) must be strict.

It is seen that (\ref{3.73})-(\ref{3.77}) and (\ref{3.07})-(\ref{3.09})
are sufficient because they
show that this solution is stationary and (\ref{2.58})-(\ref{2.63})
are satisfied for $k=2$. This solution is non-degenerate
because the inequalities in (\ref{3.09}) are strict. $\blacksquare$

We now apply this theorem in order to find non-degenerate stationary 
solutions. We investigate two cases, $-\pi <\bar{\theta}_0 \le 0$ and 
$0 < \bar{\theta}_0 \le \pi$ .

\subsubsection{Finding Non-Degenerate Stationary Solutions}
 
Case 1: Suppose $-\pi <\bar{\theta}_0 \le 0$ then $\bar{\theta}_1=0$,
$\bar{\theta}_2=\pi$ and $\pi \le \bar{\theta}_f < 2\pi$. Observing
that ${\bf p}(\theta_1)^T=(0,1)$ and ${\bf p}(\theta_2)^T=(0,-1)$,
using $R(\bar{\theta})$ from (\ref{2.52}), we see that (\ref{3.08})
becomes
\begin{eqnarray}
\label{3.78}
\left(\begin{array}{cc}
2 & 2  \\ \\
0 & 0  \\ \\
-2 & 2 \\
\end{array}
\right)
\left( \begin{array}{c} \alpha_1 \\ \alpha_2 \\ \end{array} \right)=
\left( \begin{array}{c} -z_{f_1} \\ -z_{f_2} \\-z_{f_3} \end{array} \right).
\end{eqnarray}
The solution of this system of equations is
\begin{equation}
\label{3.79}
\alpha_1=-\frac{1}{4}(z_{f_1}-z_{f_3})
\end{equation}
\begin{equation}
\label{3.80}
\alpha_2=-\frac{1}{4}(z_{f_1}+z_{f_3})
\end{equation}
where
\begin{equation}
\label{3.81}
z_{f_2}=0.
\end{equation}
From (\ref{3.09}), it is necessary that
\begin{equation}
\label{3.82}
z_{f_1} < -|z_{f_3}|.
\end{equation}
Optimal two-impulse solutions for this case are found for boundary 
conditions satisfying (\ref{3.81}) and (\ref{3.82}). Geometrically 
the ${\bf z}_f$ region is restricted to a sector (\ref{3.82}) of the plane
(\ref{3.81}). If these boundary conditions are satisfied, the optimal 
velocity increments are applied at $ \bar{\theta}_1=0$, 
$ \bar{\theta}_2=\pi$ and are calculated from (\ref{3.07}):
\begin{eqnarray}
\label{3.83}
\Delta {\bf V}_1= \left(\begin{array}{c}
0 \\
-\alpha_1 \\
\end{array}
\right),\,\,\,\,
\Delta {\bf V}_2=\left( \begin{array}{c} 
0 \\
\alpha_2 \\ 
\end{array} \right).
\end{eqnarray}

Case 2: Suppose $0 <\bar{\theta}_0 \le \pi$ then $\bar{\theta}_1=\pi$,
$\bar{\theta}_2=2\pi$ and $2\pi \le \bar{\theta}_f < 3\pi$. We note
that ${\bf p}(\theta_1)^T=(0,-1)$ and ${\bf p}(\theta_2)^T=(0,1)$.
In this case (\ref{3.08}) becomes
\begin{eqnarray}
\label{3.84}
\left(\begin{array}{cc}
2 & 2  \\ \\
0 & 0  \\ \\
2 & -2 \\
\end{array}
\right)
\left( \begin{array}{c} \alpha_1 \\ \alpha_2 \\ \end{array} \right)=
\left( \begin{array}{c} -z_{f_1} \\ -z_{f_2} \\-z_{f_3} \end{array} \right).
\end{eqnarray}
The solution of this system of equations is
\begin{equation}
\label{3.85}
\alpha_1=-\frac{1}{4}(z_{f_1}+z_{f_3})
\end{equation}
\begin{equation}
\label{3.86}
\alpha_2=-\frac{1}{4}(z_{f_1}-z_{f_3})
\end{equation}
where again (\ref{3.81}) is satisfied. From (\ref{3.09}), it is necessary 
also for this case that (\ref{3.82}) is satisfied.

Optimal two-impulse solutions for this case also follow from the boundary 
conditions satisfying (\ref{3.81}) and (\ref{3.82}). For these boundary
conditions the optimal velocity increments are applied at
$ \bar{\theta}_1=\pi$, and $ \bar{\theta}_2=2\pi$ and are calculated 
from (\ref{3.07}):
\begin{eqnarray}
\label{3.87}
\Delta {\bf V}_1= \left(\begin{array}{c}
0 \\
\alpha_1 \\
\end{array}
\right),\,\,\,\,
\Delta {\bf V}_2=\left( \begin{array}{c} 
0 \\
-\alpha_2 \\ 
\end{array} \right).
\end{eqnarray}

{\bf Example}: Select $ \bar{\theta}_0=0$, $ \bar{\theta}_f=\pi$ so that
$ \bar{\theta}_1= \bar{\theta}_0$, $ \bar{\theta}_2= \bar{\theta}_f$
and $\phi=\theta_0=\theta_f-\pi$. Using (\ref{2.50}) and (\ref{2.56})
with $\bar{\theta}$ in (\ref{3.49}) we have
$$
z_{f_1}=-y_{f_1}+y_{f_3}-y_{0_1}+ y_{0_3}
$$
$$
z_{f_2}=-y_{f_2}-y_{0_2}=0
$$
$$
z_{f_3}=y_{f_3}- y_{0_3}.
$$
The condition (\ref{3.81}) becomes
$$
y_{0_2}+y_{f_2}=0
$$
and (\ref{3.82}) becomes
$$
-y_{f_1}+y_{f_3}-y_{0_1}+ y_{0_3} < -|y_{f_3}-y_{0_3}|.
$$
If $y_{f_3} >y_{0_3}$ this becomes
$$
2y_{f_3} < y_{0_1}+y_{f_1},
$$ 
but if $y_{f_3} < y_{0_3}$ it becomes
$$
2y_{0_3} < y_{0_1}+y_{f_1},
$$
and if $y_{f_3} =y_{0_3}$ then
$$
z_{f_1} = 2y_{0_3}-(y_{0_1}+y_{f_1}) < 0.
$$ 
The boundary values
$r_0=7200km$, $v_{r_0}=0 km/sec$, $\dot{{\theta}_0}=1.1558\times 
10^{-3} rad/sec$, $r_f=7200km$, $v_{r_f}=0 km/sec$, 
$\dot{{\theta}_f}=1.1554\times 10^{-3} rad/sec$. 
satisfy the first of the last three inequalities, and are used for
simulation of this example. The resulting optimal rendezvous is shown
in Figure $21$. The optimal velocity increments 
$\Delta V_1=(\Delta V_x,\Delta V_y)=(1.7756\times 10^{-3},-0.8779)^Tkm/sec$,
$\Delta V_2= (\Delta V_x,\Delta V_y)=(-4.3162\times 10^{-4},-0.8038)^Tkm/sec$ 
are indicated by arrows in the figure. 

\setcounter{equation}{0}
\section{One-Impulse Solutions}

For certain boundary conditions a minimizing solution may contain only 
one impulse. This situation is investigated here.

\subsection{Fundamental Theorem of Non-Degenerate One-Impulse Solutions}

We apply (\ref{2.58})-(\ref{2.63}) where $k=1$. If a one-impulse minimizing
solution $\{\theta_1,\Delta{\bf V}_1\}$ is non-degenerate then
$\Delta{\bf V}_1 \ne 0$. For this case applications are facilitated without
use of the vector ${\bf z}_f$.

{\bf Theorem}: The optimal impulsive rendezvous problem defined by
(\ref{2.40}) and (\ref{2.47}) has a non-degenerate one-impulse solution
$\{\theta_1,\Delta{\bf V}_1\}$ if and only if
\begin{equation}
\label{3.88}
\theta_1=\theta_0\,\, or\,\, f^{\prime}(\theta_1)=0\,\, or\,\,
 \theta_1=\theta_f,
\end{equation}
\begin{equation}
\label{3.89}
f(\theta_1)=0\,\, and \,\, f(\theta) \le 0\,\, on\,\, 
\theta_0 \le \theta \le \theta_f,
\end{equation}
\begin{equation}
\label{3.90}
\alpha_1=|\Delta{\bf V}_1|,
\end{equation}
\begin{equation}
\label{3.91}
\frac{\Delta{\bf V}_1}{|\Delta{\bf V}_1|}=-p(\theta_1),
\end{equation}
\begin{equation}
\label{3.92}
B\Delta{\bf V}_1= \Phi(\theta_1-\theta_f){\bf y}_f-
\Phi(\theta_1-\theta_0){\bf y}_0.
\end{equation}
 
{\bf Proof}: Setting $k=1$ and $\Delta{\bf V}_1 \ne 0$ in (\ref{2.58}),
(\ref{2.59}) and (\ref{2.61})-(\ref{2.63}), these expressions become 
equivalent to 
(\ref{3.88})-(\ref{3.91}). In (\ref{2.60})  we replace
$\theta_1$ by $\bar{\theta}_1$ obtaining $R(\bar{\theta}_1)\Delta{\bf V}_1=
{\bf z}_f$. Multiplying on the left by $\Phi(\bar{\theta}_1)$ in view of 
(\ref{2.52}) the expression (\ref{3.92}) emerges. $\blacksquare$ 

\subsection{Finding Non-Degenerate One-Impulse Solutions}

The expression (\ref{3.88}) is resolved into two situations, the 
non-stationary solutions where $\theta_1=\theta_0$ or $\theta_1=\theta_f$,
or the stationary solutions where $f^{\prime}(\theta_1)=0$.

\subsubsection{Non-Stationary Solutions}

In the following we set $\theta_d=\theta_f-\theta_0$, $c_d=\cos\theta_d$,
and $s_d=\sin\theta_d$.

Case 1: Suppose $\theta_1=\theta_0$ then (\ref{3.92}) becomes
\begin{equation}
\label{3.93}
B\Delta{\bf V}_1= \Phi(-\theta_d){\bf y}_f-{\bf y}_0.
\end{equation}
Utilizing (\ref{2.40}) and (\ref{2.41}) and solving (\ref{3.93}) we obtain
\begin{equation}
\label{3.94}
\Delta V_{11}= -s_dy_{f_1}-c_dy_{f_2}+y_{0_2},
\end{equation}
\begin{equation}
\label{3.95}
\Delta V_{12}= -\frac{1}{2}(y_{f_3}-y_{0_3}),
\end{equation}
where
\begin{equation}
\label{3.96}
y_{0_1} =c_dy_{f_1}-s_dy_{f_2}+y_{f_3}.
\end{equation}
Expressions (\ref{3.94})-(\ref{3.95}) are necessary and sufficient to 
have one-impulse solutions supported at $\theta_0$.

Case 2: Suppose $\theta_1=\theta_f$ then (\ref{3.92}) becomes
\begin{equation}
\label{3.97}
B\Delta{\bf V}_1= {\bf y}_f - \Phi(\theta_d){\bf y}_0.
\end{equation}
Solving,
\begin{equation}
\label{3.98}
\Delta V_{11}= -y_{f_2}-s_dy_{0_1}+c_dy_{0_2},
\end{equation}
\begin{equation}
\label{3.99}
\Delta V_{12}= -\frac{1}{2}(y_{f_3}-y_{0_3}),
\end{equation}

where
\begin{equation}
\label{3.100}
y_{f_1} =c_dy_{0_1}+s_dy_{0_2}+y_{0_3}.
\end{equation}
Expressions (\ref{3.98})-(\ref{3.100}) are necessary and sufficient to 
have one-impulse solutions supported at $\theta_f$.

\subsubsection{Stationary Solutions}
In the following we set $c_0=\cos\bar{\theta}_0$, $s_0=\sin\bar{\theta}_0$,
$c_f=\cos\bar{\theta}_f$, and $s_f=\sin\bar{\theta}_f$.

Case 1: Suppose $\bar{\theta}_1=0$ and $-\pi < \bar{\theta}_0 <0 < 
\bar{\theta}_f <\pi$, then (\ref{3.92}) becomes
\begin{equation}
\label{3.101}
B\Delta{\bf V}_1= \Phi(-\bar{\theta}_f){\bf y}_f -
\Phi(-\bar{\theta}_0){\bf y}_0.
\end{equation}
Solving, noting (\ref{3.91}) and the fact that $p_1(\theta_1)=0$ if
$\bar{\theta}_1$ is a stationary point, we obtain
\begin{equation}
\label{3.102}
\Delta V_{11}= 0,
\end{equation}
\begin{equation}
\label{3.103}
\Delta V_{12}= -\frac{1}{2}(y_{f_3}-y_{0_3}),
\end{equation}
where
\begin{equation}
\label{3.104}
c_0y_{0_1}-s_0y_{0_2}+y_{0_3}= c_fy_{f_1}-s_fy_{f_2}+y_{f_3},
\end{equation}
\begin{equation}
\label{3.105}
s_0y_{0_1}+c_0y_{0_2}= s_fy_{f_1}+c_fy_{f_2}.
\end{equation}
Expressions (\ref{3.102})-(\ref{3.105}) are necessary and sufficient to 
have stationary one-impulse solutions at $\bar{\theta}_1=0$.

Case 2: Suppose $\bar{\theta}_1=\pi$ and $0 < \bar{\theta}_0 <\pi < 
\bar{\theta}_f < 2\pi$, then (\ref{3.92}) becomes
\begin{equation}
\label{3.106}
B\Delta{\bf V}_1= \Phi(\pi-\bar{\theta}_f){\bf y}_f -
\Phi(\pi-\bar{\theta}_0){\bf y}_0.
\end{equation}
Solving, as before  we find
\begin{equation}
\label{3.107}
\Delta V_{11}= 0,
\end{equation}
\begin{equation}
\label{3.108}
\Delta V_{12}= -\frac{1}{2}(y_{f_3}-y_{0_3}),
\end{equation}
where
\begin{equation}
\label{3.109}
c_0y_{0_1}-s_0y_{0_2}-y_{0_3}= c_fy_{f_1}-s_fy_{f_2}-y_{f_3},
\end{equation}
\begin{equation}
\label{3.110}
s_0y_{0_1}-c_0y_{0_2}= s_fy_{f_1}-c_fy_{f_2}.
\end{equation}
Expressions (\ref{3.107})-(\ref{3.110}) are necessary and sufficient to
have stationary one-impulse solutions at $\bar{\theta}_1=\pi$.

\section*{Conclusions}

        A planar impulsive rendezvous problem can be solved for initial
and terminal positions and velocities near those associated with a
nominal circular orbit. The rendezvous trajectories produced are exact
for this restricted two-body problem and approach optimality as the
boundary conditions approach boundary conditions of a nominal circular
orbit. 

By replacing the cost function (\ref{2.11}) by the related cost function
(\ref{2.48}), it was found that velocity impulses that minimize (\ref{2.48}) 
produce a close approximation of a minimum of (\ref{2.11}) if the boundary 
conditions are near those of a nominal circular orbit.

The boundary conditions are incorporated into a vector ${\bf z}_f$
defined by (\ref{2.57}) which determines the number of velocity impulses 
required and from which these velocity increments can be calculated. It 
was found that a non-degenerate rendezvous trajectory requires three 
impulses if and only if the boundary conditions are such that ${\bf z}_f$ or 
$-{\bf z}_f$ satisfies (\ref{2.127}) and (\ref{2.128}). It was found that the 
proper placement of the velocity impulses is at the initial, terminal, 
and mid-point values of the true anomaly. The calculations are 
simple enough that the three velocity increments could be found from 
a hand calculator that contains trigonometric functions, if necessary. 
An example of a three-impulse rendezvous problem with boundary conditions 
that lead to quick solution was presented. Simulation of the rendezvous 
trajectory that resulted showed that throughout the flight the 
deviation of the radial distance 
remained within ten percent of the radius of a nominal circular 
orbit.

It was found that non-degenerate two-impulse solutions of an 
optimal impulsive rendezvous near a circular orbit can be classified 
according to the ways the locus of a primer vector can intersect the 
unit circle. These optimal two-impulse solutions fall into four categories.

For each of these categories necessary and sufficient conditions
for an optimal solution were presented. Distinct regions of the boundary
conditions that admit optimal two-impulse solutions for each category 
were displayed. A closed-form solution of the optimal velocity 
increments for each
category can be found. Some examples and simulations were presented.

Necessary and sufficient conditions for optimal non-degenerate one-impulse
solutions were also found. These optimal one-impulse solutions consist
of four types. Distinct regions of boundary conditions admitting the optimal
one-impulse solutions were also displayed. For each type a closed-form
optimal velocity increment can be found.

Although this work emphasizes non-degenerate 
rendezvous in the vicinity of a circular orbit, it also presents 
a framework for studies of stationary, degenerate, singular
rendezvous and optimal rendezvous beyond the vicinity of a circular orbit.
Presentation of results of these studies is to follow.
In these results we will show that degenerate and singular solutions are
more than interesting curiosities. Not only do they provide important
understanding to the subject of impulsive rendezvous, but they also
comprise important solutions to some impulsive rendezvous problems.

\newpage
  
\section*{References}

\begin{itemize}

\item[1] Hohmann, W., {\it Die Erreichbarkeit der Himmelskorper}, 
Oldenbourg: Munich, Germany, 1925, {\it The Attainability of Heavenly bodies}, 
NASA Tech. Translation F-44, 1960.

\item[2] Oberth, H., {\it Wege Zur Raumschiffahrt}, R. Oldenbourg, Munich, 
Germany, 1929.

\item[3] Contensou, P., Note sur la Cinematique Generale du Mobile 
Dirige a la Theorie du Vol Plane, {\it Communication a'l' Association 
Technique, Maritime et Aeronautique}, Vol. 45, 1946.

\item[4] Contensou, P., Application des Methodes de la Mecanique du Mobile 
Dirige a la Theorie du Vol Plane, {\it Communication a'l' Association 
Technique, Maritime et Aeronautique}, Vol. 45, 1950.

\item[5] Lawden, D. F., {\it Optimal Trajectories for Space Navigation}, Butterworths, London, England, 1963.

\item[6] Edelbaum, T. N., How many impulses? {\it Astronautics and 
Aeronautics}, Vol. 5, no. 11, 1967, pp. 64-69.

\item[7] Bell, D. J., Optimal Space Trajectories, A Review of  Published 
Work, {\it The Aeronautical Journal of the Royal Aeronautical Society}; Vol. 72, No. 686, 1968, pp. 141-146.

\item[8] Robinson, A.C., A Survey of Methods and Results in the 
Determination for Fuel-Optimal Space Maneuvers, A.A.S. Paper 68-091, 
AAS/AIAA Specialist Conference, Sept., 1968.

\item[9] Gobetz, F.W., and Doll, J.R., A Survey of Impulse Trajectories, 
{\it AIAA Journal}, Vol. 7, No. 5, 1969, pp. 801-834.

\item[10] Marec, J.P. {\it Optimal Space Trajectories}, Elsevier, 
New York, 1979.

\item[11] Shternfeld, A., {\it Soviet Space Sciences}, Basic Books, Inc., 
New York, New York, 1959, pp. 109-111.

\item[12] Hoelker, R. F., and Silber, R., The Bi-elliptical Transfer 
Between Co-Planar Circular Orbits, {\it Proceedings of the 4th Symposium 
on Ballistic Missile and Space Technology}, Los Angeles, CA, 1959.

\item[13] Edelbaum, T. N., Some Extensions of the Hohmann Transfer Maneuver, 
{\it Journal of the American Rocket Society}, Vol. 29, No. 11, 1959, 
pp. 864-865.

\item[14] Marchal, C., Transferts Optimaux Entre Orbites Elliptiques 
(Duree Indifferente), {\it Astronautica Acta}, Vol. 11, No. 6, 1965, 
pp. 432-445.

\item[15] Pontani, M., Simple Method to Determine Globally Optimal Orbital 
Transfers, {\it Journal of Guidance, Control, and Dynamics}, 
Vol. 32, No. 3, 2009.

\item[16] Ting, L., Optimum Orbital Transfer by Several Impulses, 
{\it Astronautica Acta}, Vol. 6, No., 1960, pp. 256-265.

\item[17]  Neustadt, L. W., Optimization, a Moment Problem, and 
Nonlinear Programming, {\it SIAM Journal on Control}, Vol. 2, No. 1 , 
1964, pp. 33-53.

\item[18] Stern, R. G., and Potter, J. E., Optimization of Midcourse 
Velocity Corrections, {\it Report RE-17}, Experimental Astronomy 
Laboratory, MIT, Cambridge, MA, 1965.

\item[19] Prussing, J.E., Optimal Impulsive Linear systems: Sufficient 
Conditions and Maximum Number of Impulses, 
{\it Journal of the Astronautical Sciences}, Vol. 43, No. 2, 1995, pp. 195-206.

\item[20] Carter, T. E., and Brient, J., Linearized Impulsive 
Rendezvous Problem, {\it Journal of Optimization Theory and 
Applications}, Vol. 86, No. 3, 1995, pp. 553-584.

\item[21] Prussing, J. E., Optimal Four-Impulse Fixed-Time Rendezvous in 
the Vicinity of a Circular Orbit, {\it AIAA Journal}, Vol. 7, No. 5, 
1969, pp. 928-935.

\item[22] Carter, T., and Alvarez, S., Quadratic-Based Computation of 
Four-Impulse Optimal Rendezvous Near Circular Orbit, 
{\it Journal of Guidance, Control, and Dynamics}, Vol. 23, No. 1, 
2000, pp. 109-117.

\item[23] Lion, P.M., and Handelsman, M., Primer Vector on Fixed-Time 
Impulsive Trajectories, {\it AIAA Journal}, Vol. 6, No. 1, 1968, pp. 127-132.

\item[24] Humi, M., and Carter, T., Models of Motion in a Central Force 
Field with Quadratic Drag, {\it Journal of Celestial Mechanics and 
Dynamical Astronomy}, Vol. 84, No. 3, 2002, pp. 245-262.

\item[25] Carter, T., and Humi, M., Two-Body Problem with Drag and 
High Tangential Speeds, {\it Journal of Guidance, Control and  Dynamics}
, Vol. 31, No.3, 2008, pp. 641-646.

\item[26] Carter,T.E., Optimal Impulsive Space Trajectories Based on Linear
Equations, {\it Journal of Optimization Theory and
Applications}, Vol. 70, No. 2, 1991, pp. 277-297.

\item[27] Carter, T. E., Necessary and Sufficient Conditions for Optimal 
Impulsive Rendezvous with Linear Equations  of Motions, 
{\it Dynamics and Control}, Vol. 10, No.3, 2000, pp. 219-227.

\item[28] Clohessy, W.H., and Wiltshire, R. S., Terminal Guidance System 
for Satellite Rendezvous, {\it Journal of the Aerospace Sciences}, 
Vol. 27, No. 9, 1960, pp. 653-658, 674.

\item[29] Breakwell, John V., Minimum Impulse Transfer, Preprint 63-416,
AIAA Astrodynamics Conference, New-Haven, Conn, Aug 19-23, 1963.

\end{itemize}

\newpage
  
\begin{center}
\section*{List of Captions}
\end{center}

Fig. 1 - Primer locus and unit circle for three-impulse solution.

Fig. 2 - Primer locus and unit circle for antipodal three-impulse
solution.

Fig. 3 - Primer locus and unit circle for a type of two-impulse solution

Fig. 4 - Primer locus and unit circle for another type of two-impulse
solution

Fig. 5 - Primer locus and unit circle for antipodal two-impulse solution \\
\indent\indent \indent \indent associated with Fig 4.

Fig. 6 - Primer locus and unit circle for a type of one-impulse solution.

Fig. 7 - Primer locus and unit circle for another type of one-impulse
solution.

Fig. 8 - Primer locus and unit circle for yet another type of one-impulse
solution.

Fig. 9  - A three-impulse rendezvous trajectory.

Fig. 10 - Primer vector loci for two-intersection solutions where 
$0 < \bar{\theta}_0 < \pi$, $\bar{\theta}_0+\bar{\theta}_f=2\pi$.

Fig. 11 - Primer vector loci for two-intersection solutions where 
$-\pi < \bar{\theta}_0 < 0$, $\bar{\theta}_0+\bar{\theta}_f=0$.

Fig. 12 -Primer vector loci for four-intersection solutions where
$0 < \bar{\theta}_0  < \bar{\theta}_f < \pi$.

Fig. 13 -Primer vector loci for four-intersection solutions where
$-\pi < \bar{\theta}_0  < \bar{\theta}_f < 0$.

Fig. 14 - Primer vector locus of a two-impulse three-intersection
solution where \\ 
\indent $0 <\bar{\theta}_0=\bar{\theta}_1 < \pi =\bar{\theta}_2 <
\bar{\theta}_f <2\pi -\bar{\theta}_0$.

Fig. 15 - Primer vector locus of a two-impulse three-intersection
solution where \\
\indent $0 < 2\pi - \bar{\theta}_f < \bar{\theta}_0 < \pi= 
\bar{\theta}_1 < \bar{\theta}_2=\bar{\theta}_f$.

Fig. 16 - Primer vector locus for a stationary non-degenerate two-impulse 
solution where\\
\indent $-\pi < \bar{\theta}_0\le 0$, $\pi \le \bar{\theta_f} < 2\pi$, 
$\pi \le \bar{\theta}_f -\bar{\theta}_0 < 3\pi$.

Fig. 17 - Primer vector locus for a stationary degenerate multi-impulse
solution having \\
\indent an equivalent two-impulse representation where 
$0 < \bar{\theta}_0 <\bar{\theta_f}$ or 
$-\pi < \bar{\theta}_0 <\bar{\theta_f}$, \\
\indent $3\pi < \theta_f - \theta_0$.

Fig. 18 - Rendezvous orbit (red) of a non-stationary two-intersection solution \\
$\bar{\theta}_0=-\pi/2$, $\bar{\theta}_f=\pi/2$, $r_0=6545km $, 
$v_{r_0}=-0.784km/sec$, $\dot{{\theta}_0}=1.3\times 10^{-3} rad/sec$ \\
$r_f=8000km$, $v_{r_f}=-0.710km/sec$, $\dot{{\theta}_f}=0.798\times 10^{-3} 
rad/sec$ \\
Orbits satisfying initial conditions (green) and
terminal conditions(blue) are displayed.

Fig. 19 - Rendezvous orbit (red) of a four-intersection solution \\
$\bar{\theta}_0=\pi/6$, $\bar{\theta}_f=5\pi/6$, 
$r_0=8000km$, $v_{r_0}=0 km/sec$, $\dot{{\theta}_0}=8.37\times 
10^{-4} rad/sec$ \\
$r_f=6545km$, $v_{r_f}=-0.866km/sec$, $\dot{{\theta}_f}=1.25\times 10^{-3} 
rad/sec$. \\
Orbits satisfying initial conditions (green) and 
terminal conditions(blue) are displayed.

Fig. 20 - Rendezvous orbit (red) of a two impulse nonstationary 
three-intersection \\
solution $\bar{\theta}_0=\pi/2$, $\bar{\theta}_2=\pi$, $\bar{\theta}_f=4\pi/3$, 
$r_0=7200km$, $v_{r_0}=0 km/sec$, $\dot{{\theta}_0}=1.033\times 
10^{-3} rad/sec$ \\
$r_f=10375km$, $v_{r_f}=-3.547km/sec$, $\dot{{\theta}_f}=4.746\times 10^{-3} 
rad/sec$. \\
Orbits satisfying initial conditions (green) and terminal conditions(blue) 
are displayed.

Fig. 21 - Rendezvous orbit (red) of a nondegenerate stationary solution \\
$\bar{\theta}_0=0$, $\bar{\theta}_f=\pi$, 
$r_0=7200km$, $v_{r_0}=0 km/sec$, $\dot{{\theta}_0}=1.1558\times 
10^{-3} rad/sec$ \\
$r_f=7200km$, $v_{r_f}=0 km/sec$, $\dot{{\theta}_f}=1.1556\times 10^{-3} 
rad/sec$. \\
Orbits satisfying initial conditions (green) and terminal conditions(blue) 
are displayed.

\newpage

\begin{figure}[htb]
\centering
\centerline{\psfig{file=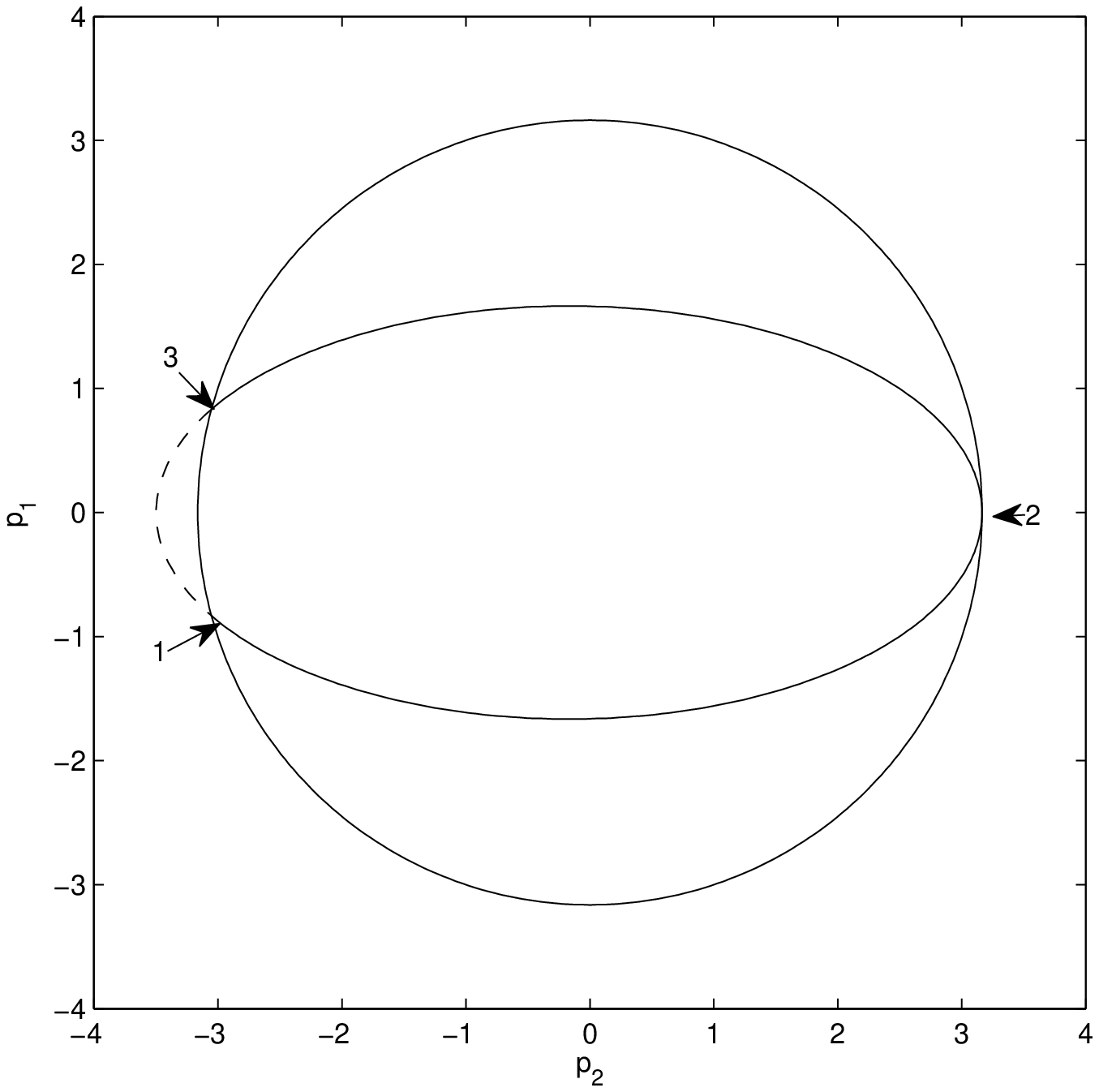,height=4in,width=5.5in}}
\label{Figure 1a}
\caption{}
\end{figure}

\newpage

\begin{figure}[htb]
\centering
\centerline{\psfig{file=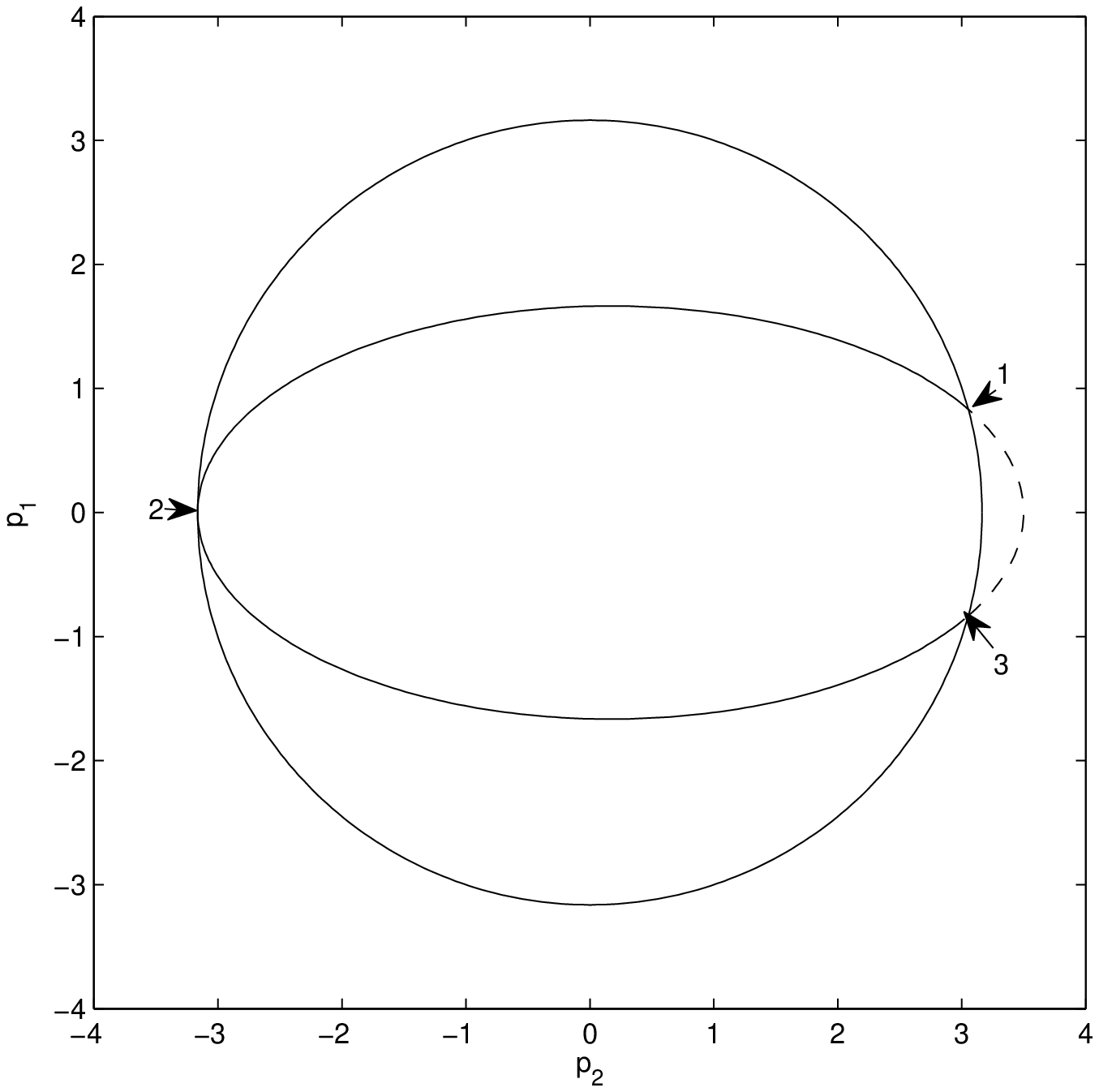,height=4in,width=5.5in}}
\caption{}
\end{figure}

\newpage

\begin{figure}[htb]
\centering
\centerline{\psfig{file=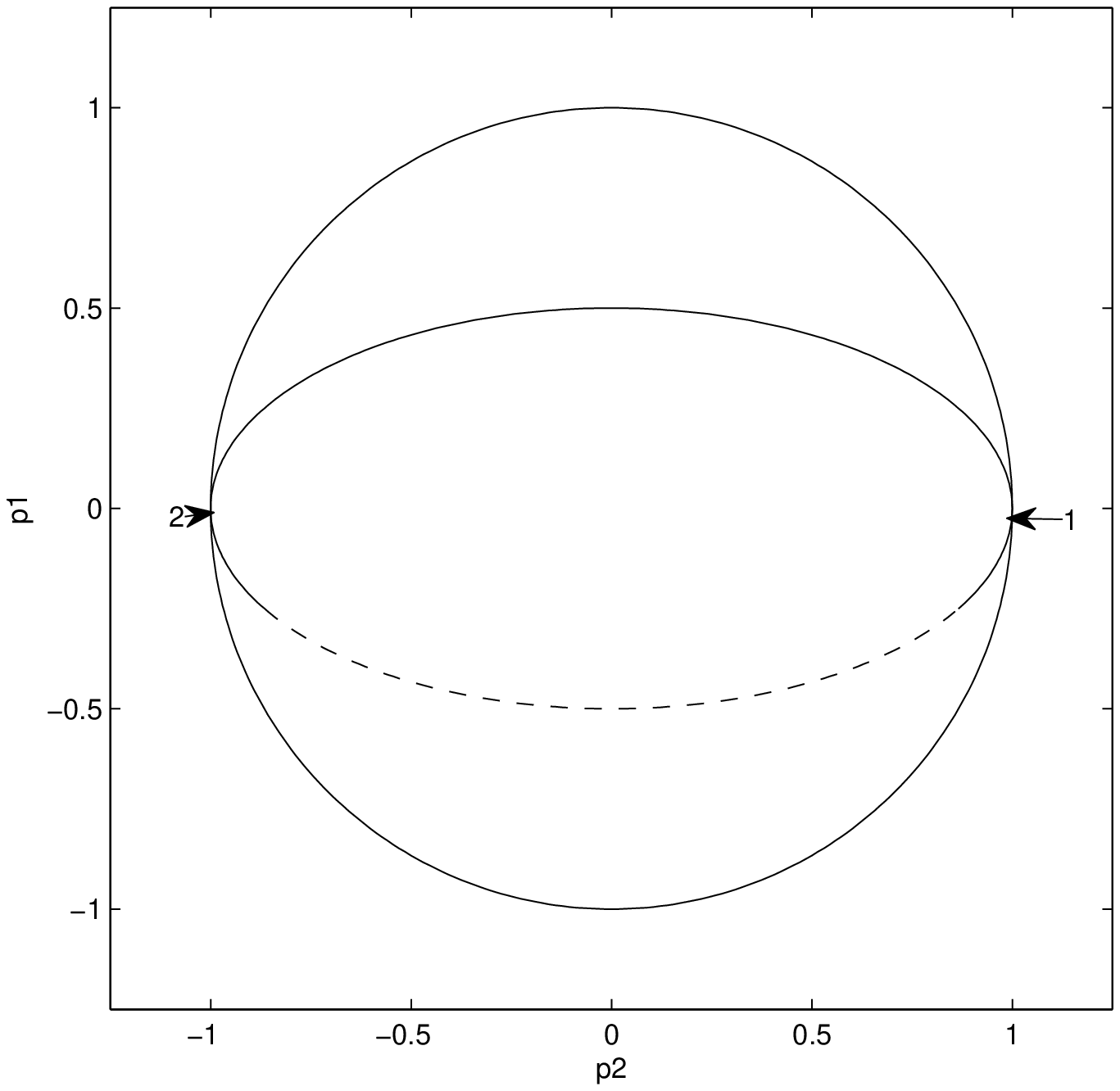,height=4in,width=5.5in}}
\caption{}
\end{figure}

\newpage

\begin{figure}[htb]
\centering
\centerline{\psfig{file=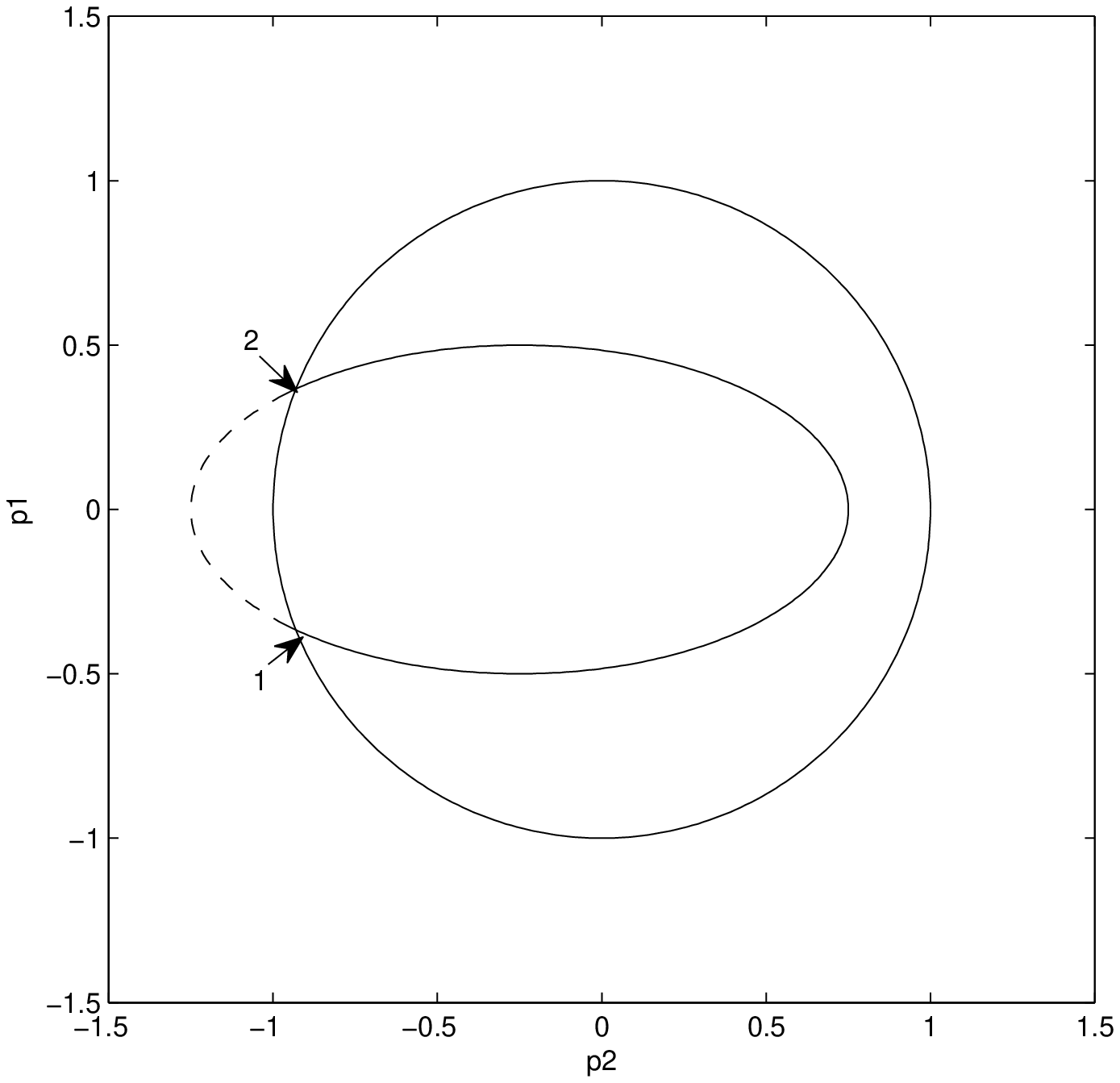,height=4in,width=5.5in}}
\caption{}
\end{figure}

\newpage

\begin{figure}[htb]
\centering
\centerline{\psfig{file=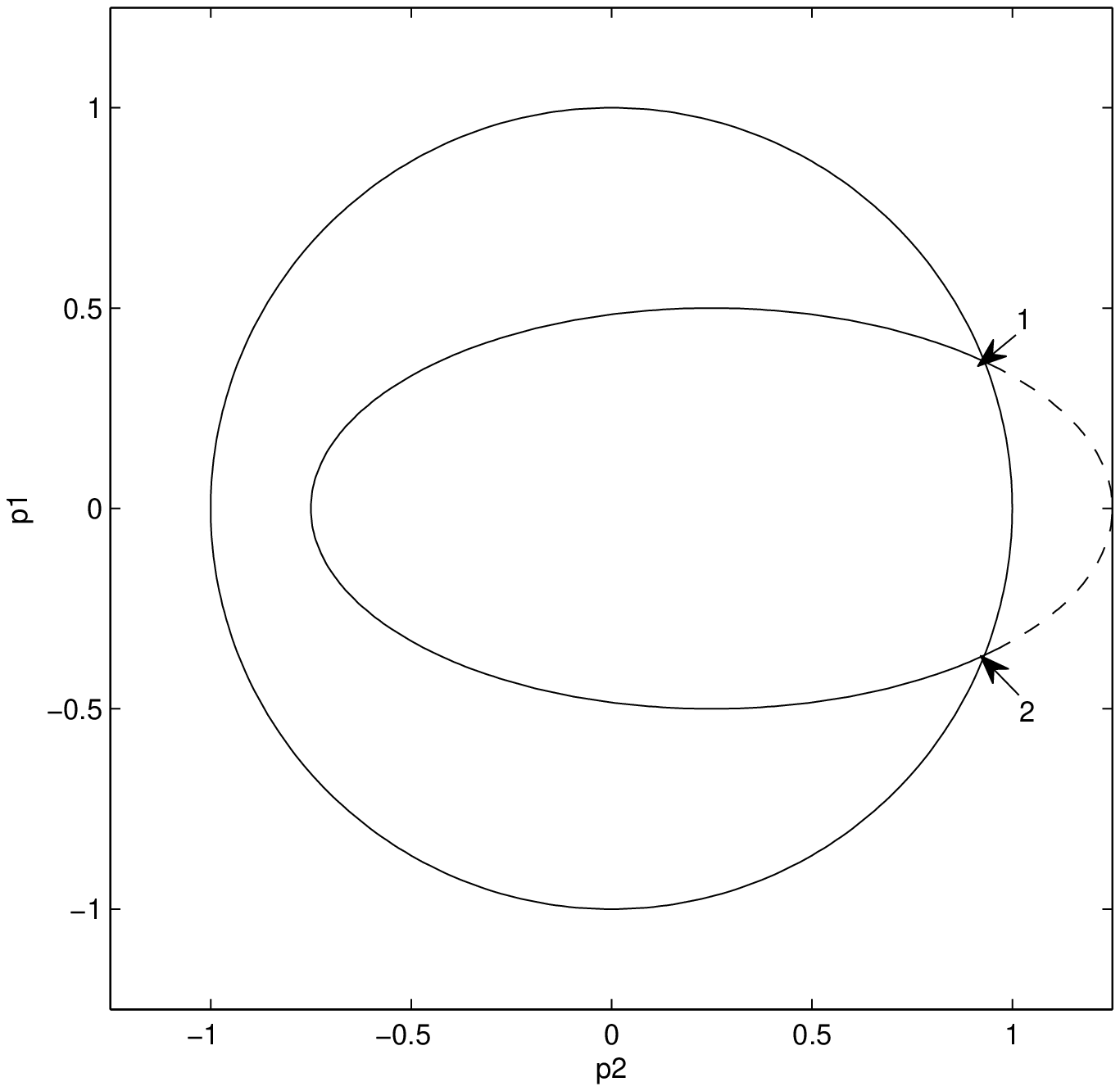,height=4in,width=5.5in}}
\caption{}
\end{figure}

\newpage

\begin{figure}[htb]
\centering
\centerline{\psfig{file=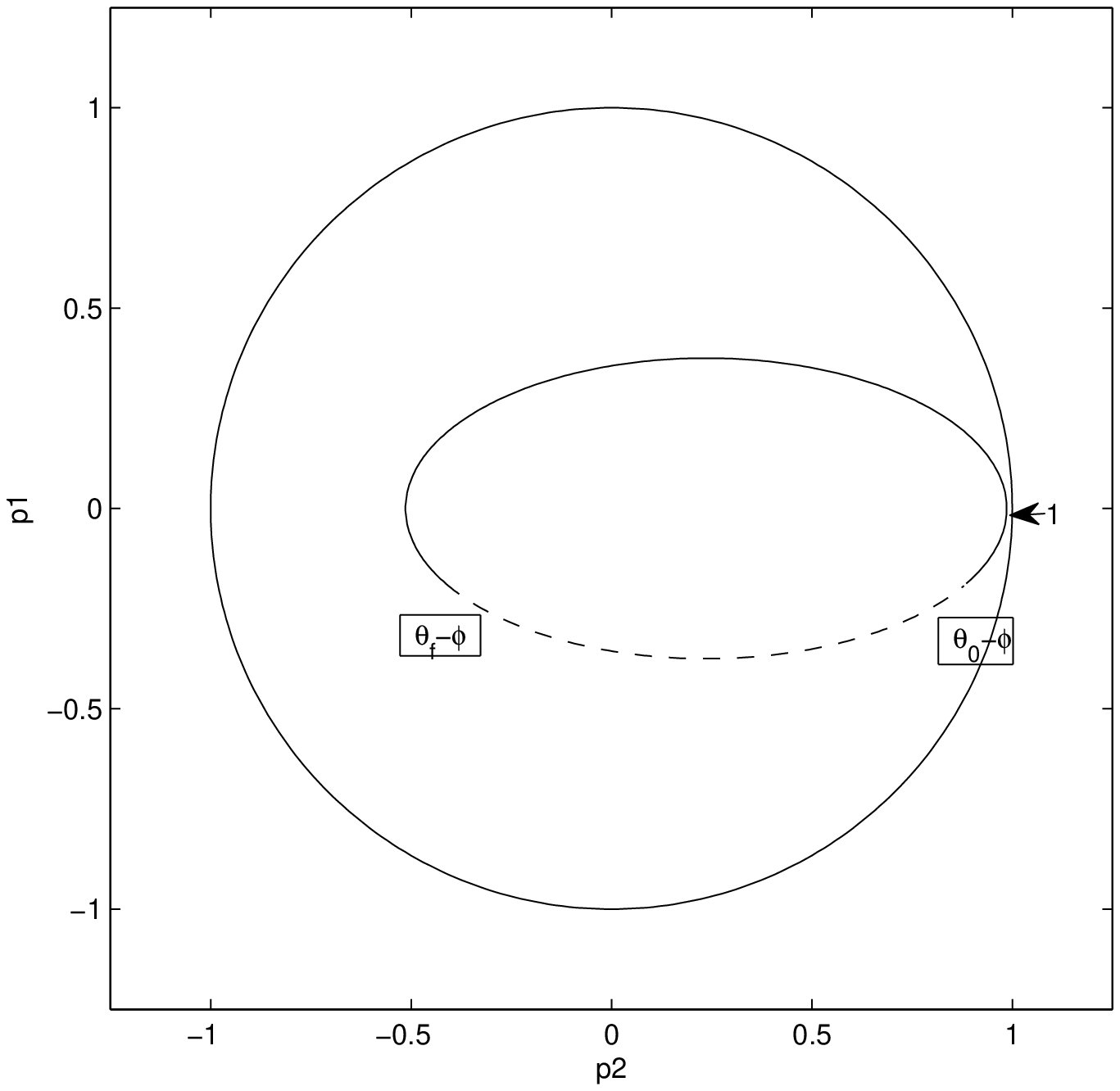,height=4in,width=5.5in}}
\caption{}
\end{figure}

\newpage

\begin{figure}[htb]
\centering
\centerline{\psfig{file=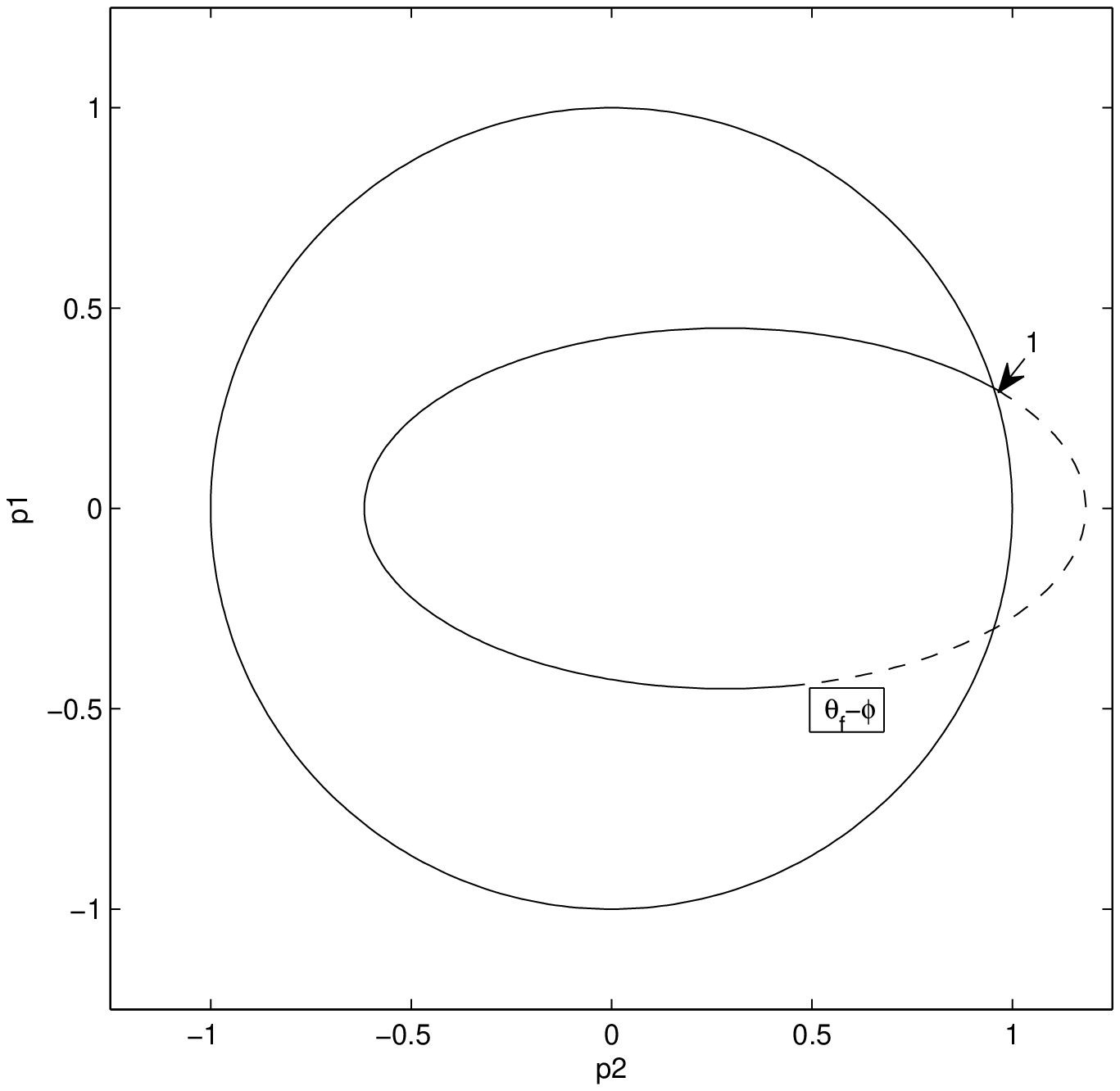,height=4in,width=5.5in}}
\caption{}
\end{figure}

\newpage

\begin{figure}[htb]
\centering
\centerline{\psfig{file=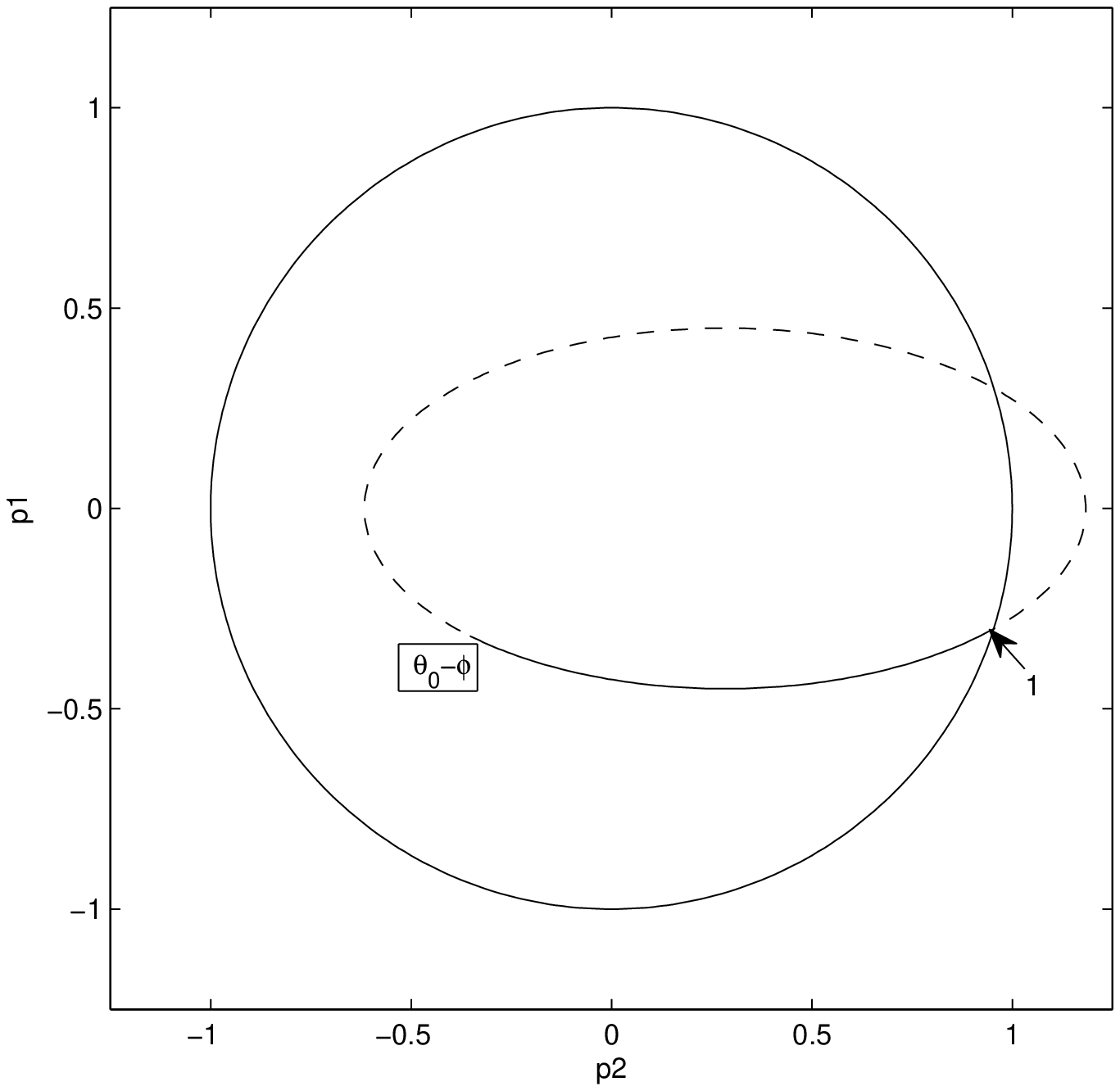,height=4in,width=5.5in}}
\caption{}
\end{figure}

\newpage

\begin{figure}[htb]
\centering
\centerline{\psfig{file=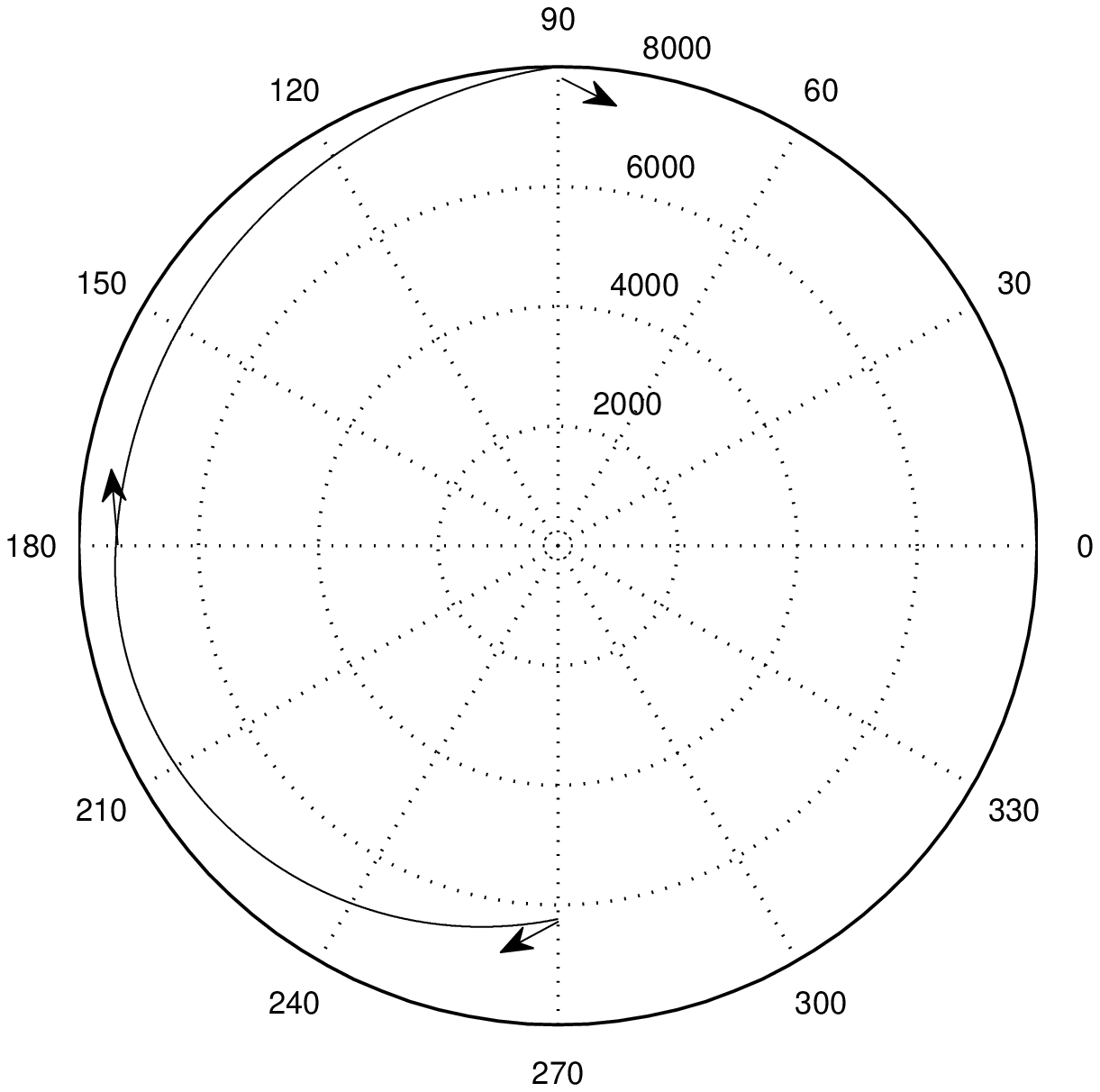,height=4in,width=5.5in}}
\caption{}
\end{figure}

\begin{figure}[htb]
\centering
\centerline{\psfig{file=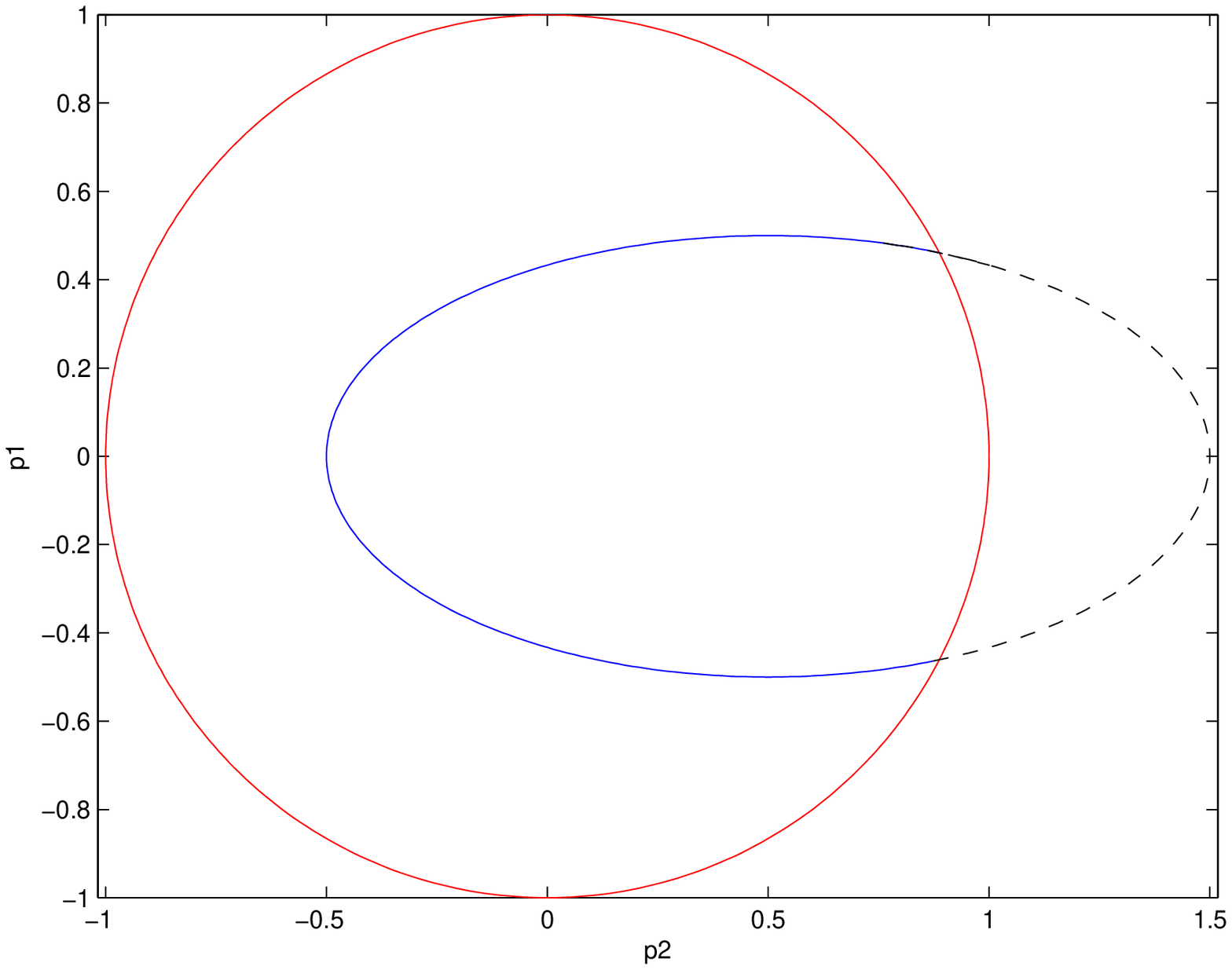,height=4in,width=5.5in}}
\caption{}
\end{figure}

\newpage

\begin{figure}[htb]
\centering
\centerline{\psfig{file=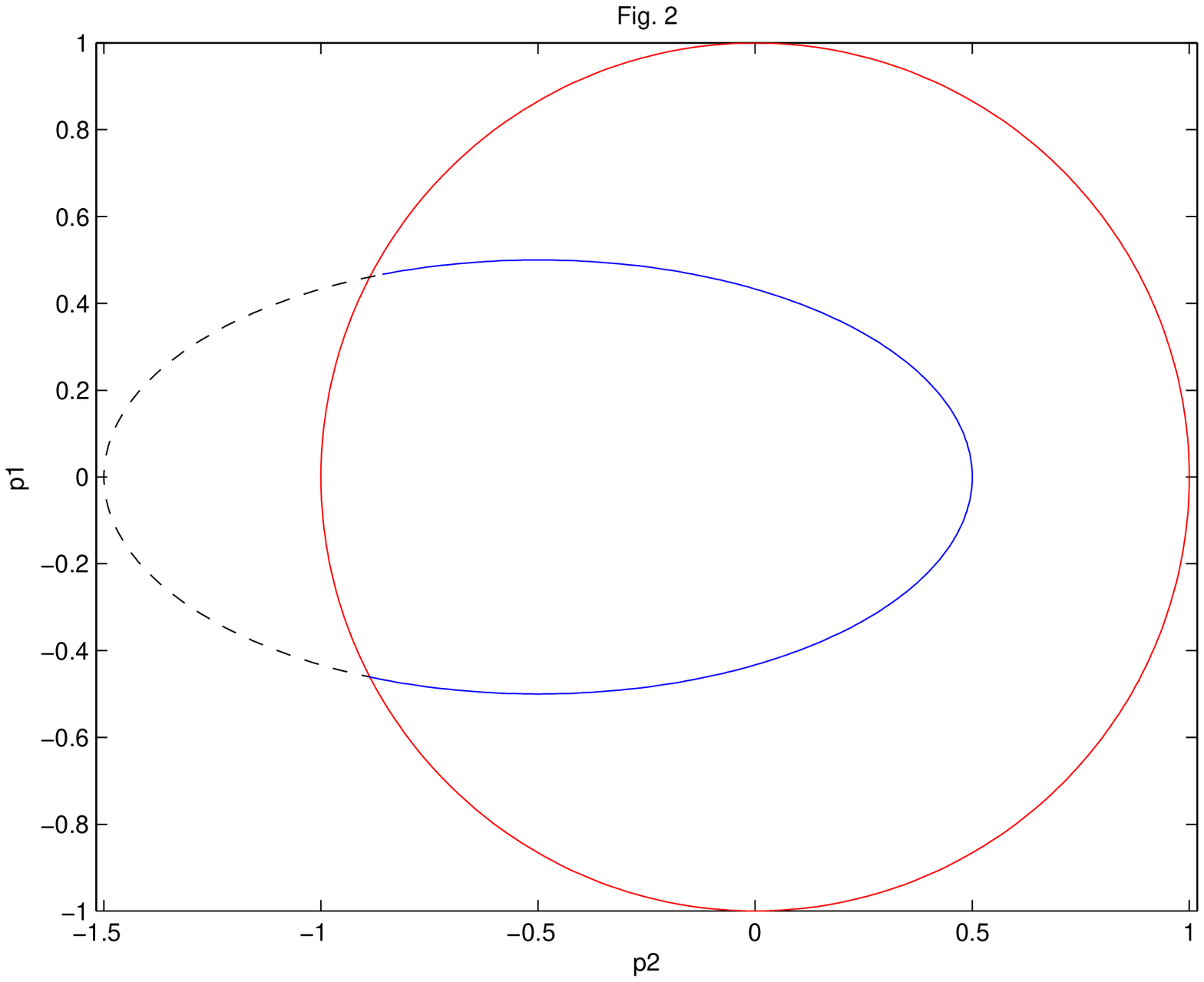,height=4in,width=5.5in}}
\caption{}
\end{figure}

\newpage

\begin{figure}[htb]
\centering
\centerline{\psfig{file=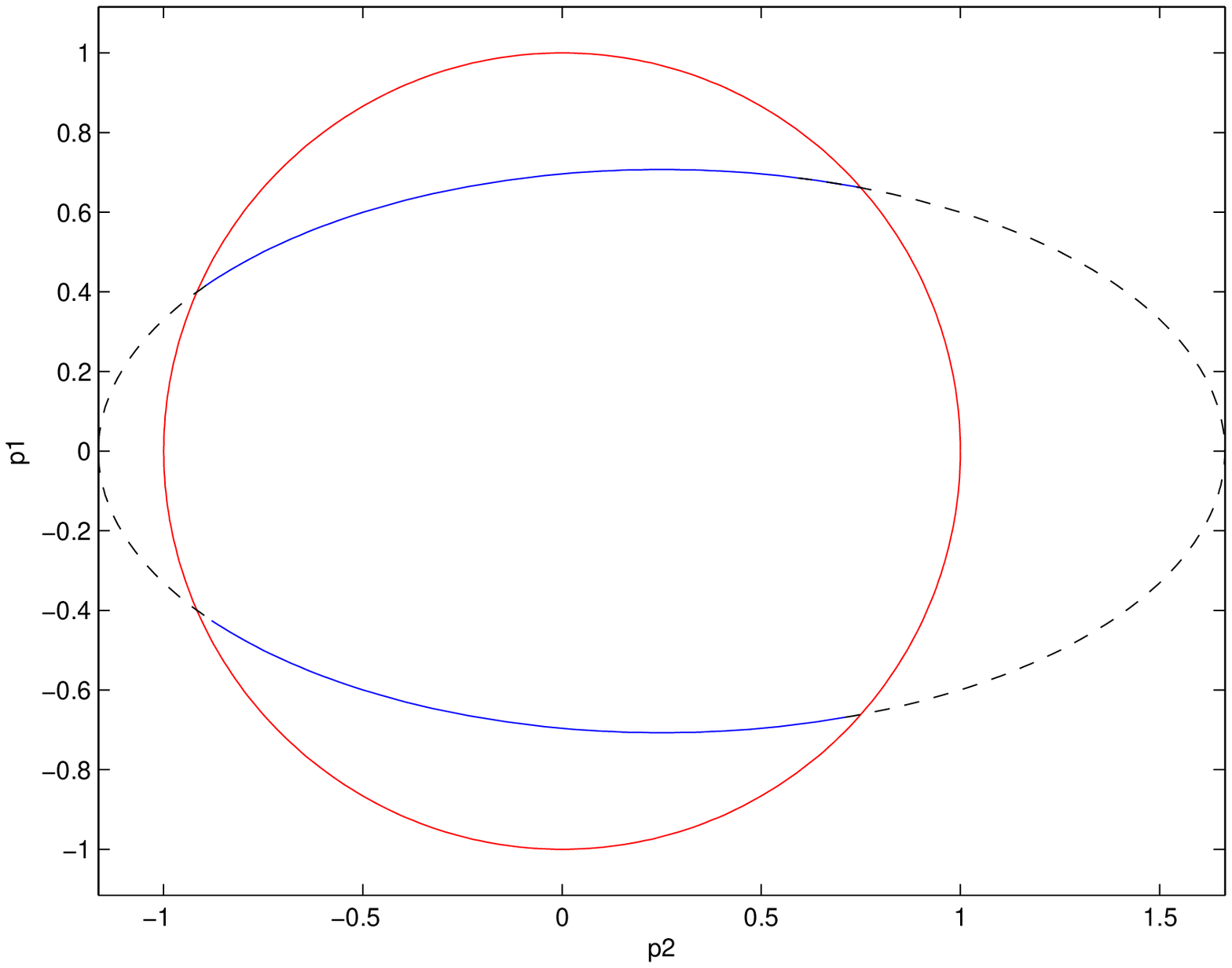,height=4in,width=5.5in}}
\caption{}
\end{figure}

\newpage

\begin{figure}[htb]
\centering
\centerline{\psfig{file=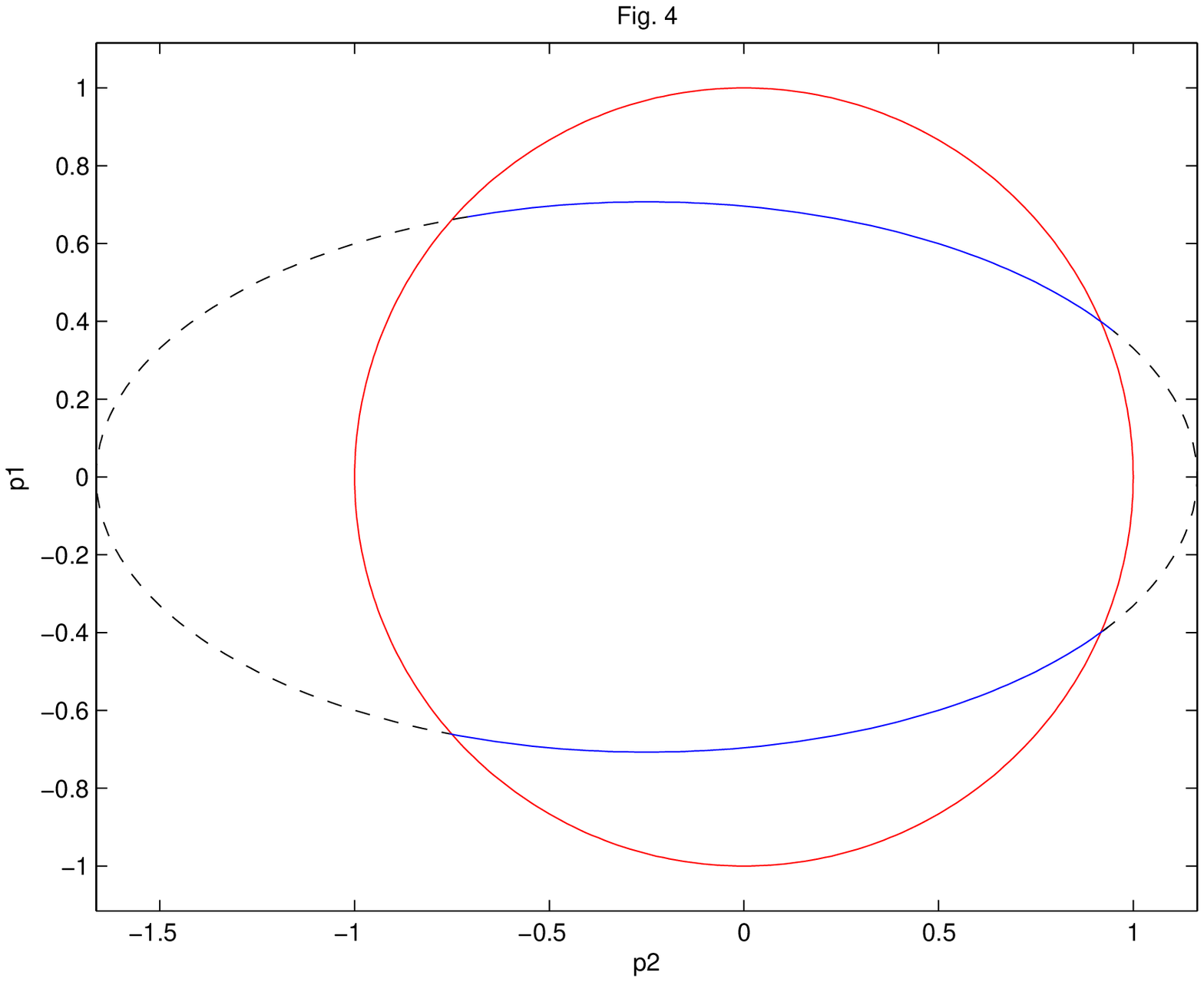,height=4in,width=5.5in}}
\caption{}
\end{figure}

\newpage

\begin{figure}[htb]
\centering
\centerline{\psfig{file=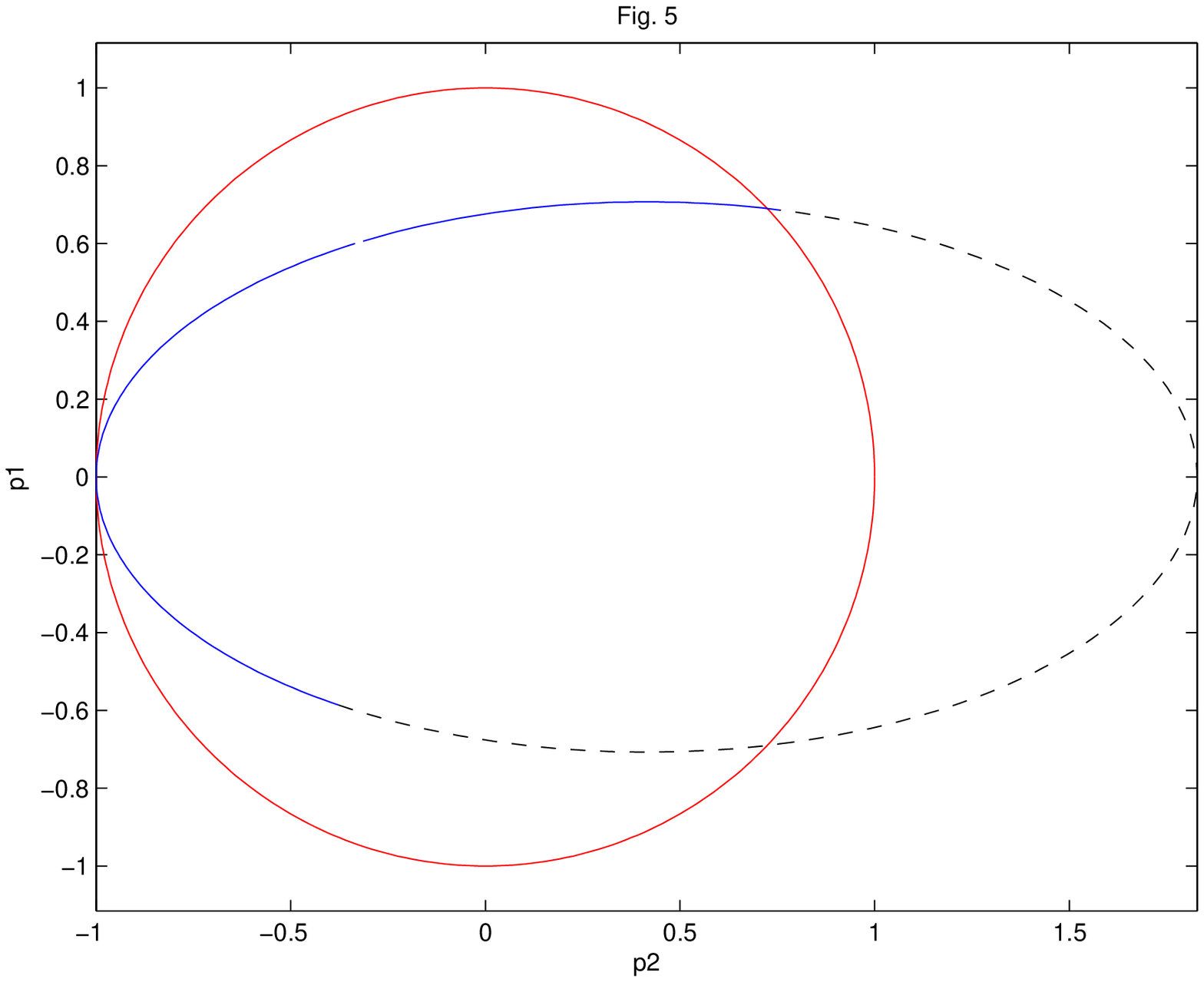,height=4in,width=5.5in}}
\caption{}
\end{figure}

\newpage

\begin{figure}[htb]
\centering
\centerline{\psfig{file=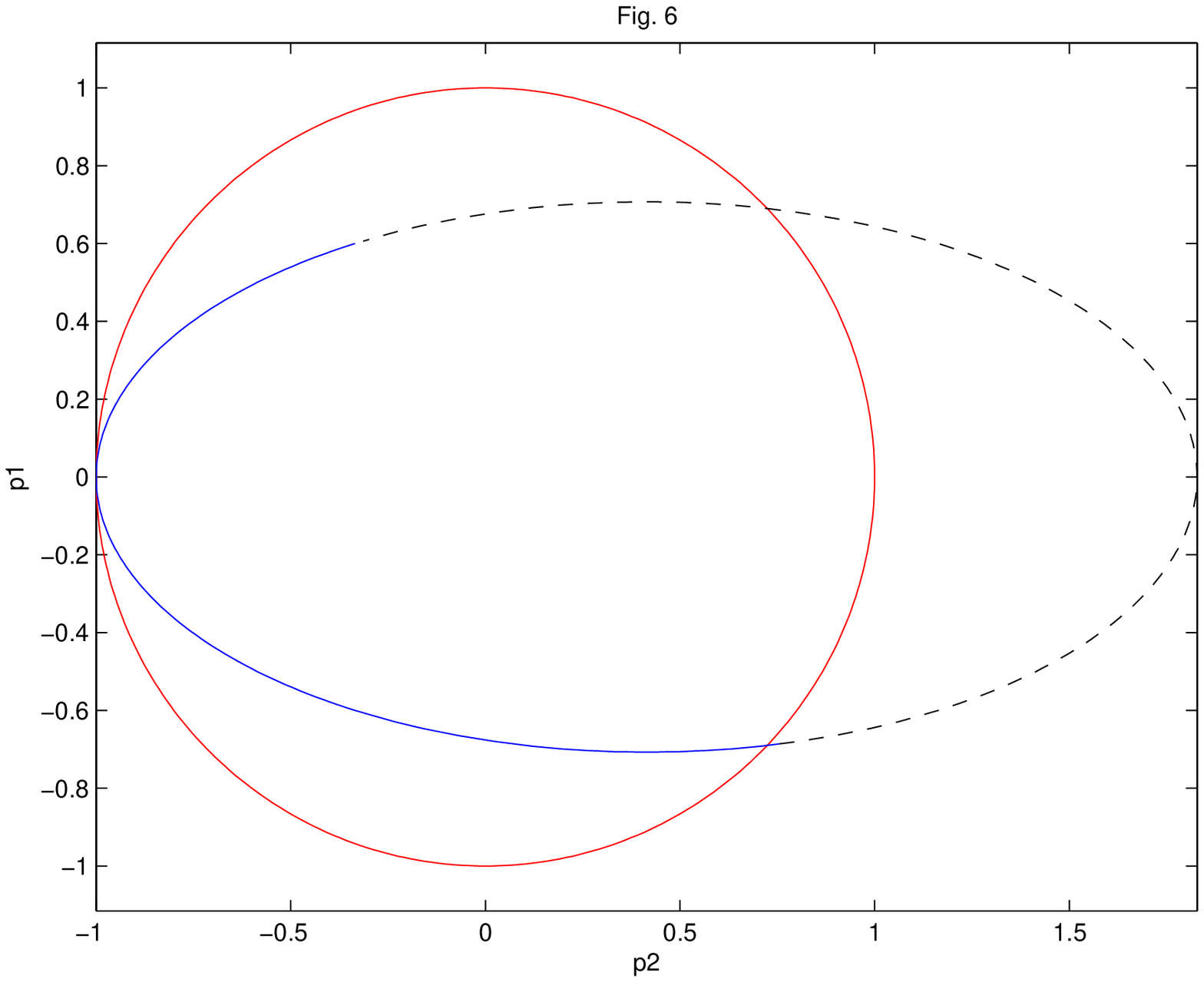,height=4in,width=5.5in}}
\caption{}
\end{figure}

\newpage

\begin{figure}[htb]
\centering
\centerline{\psfig{file=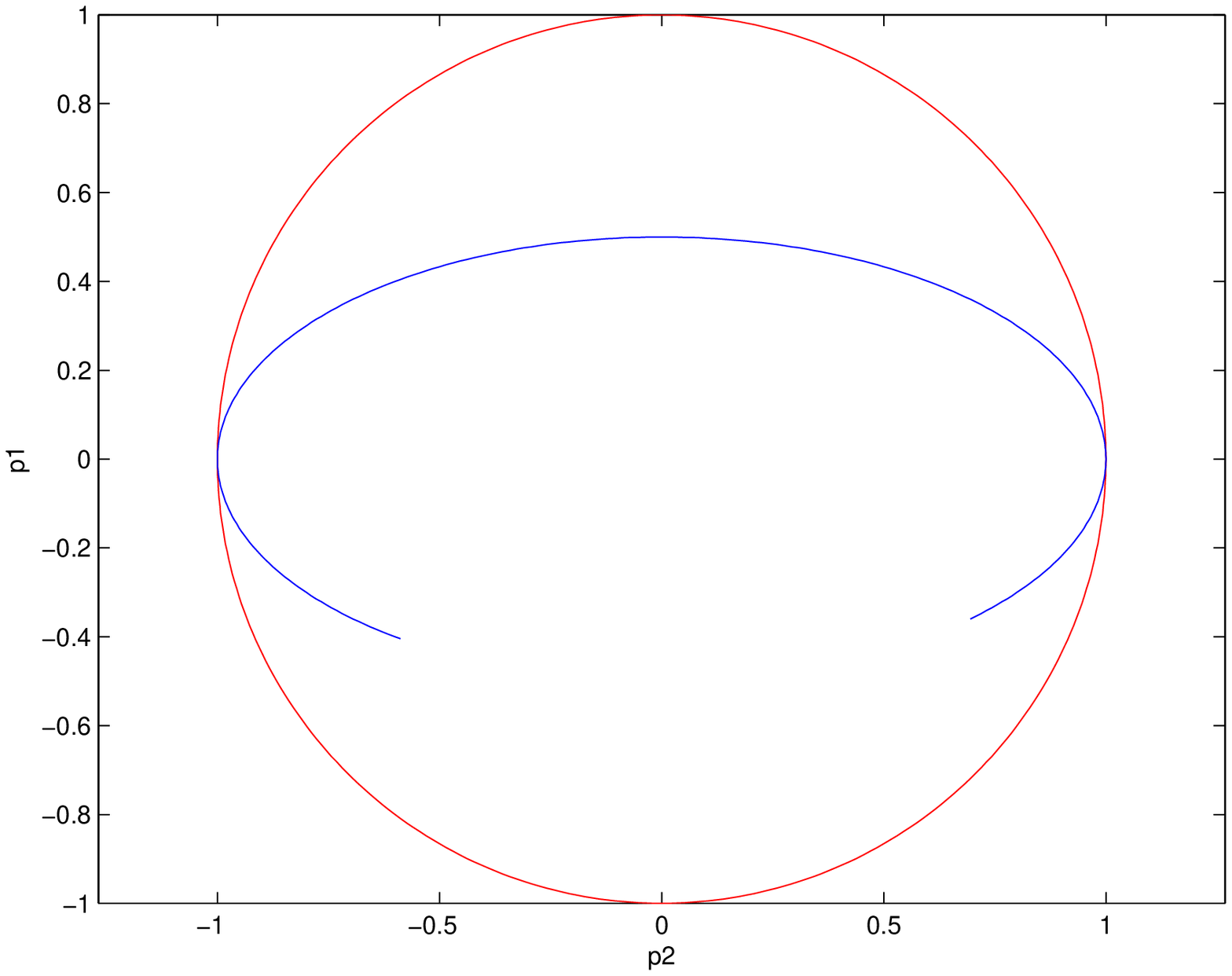,height=4in,width=5.5in}}
\caption{}
\end{figure}

\newpage

\begin{figure}[htb]
\centering
\centerline{\psfig{file=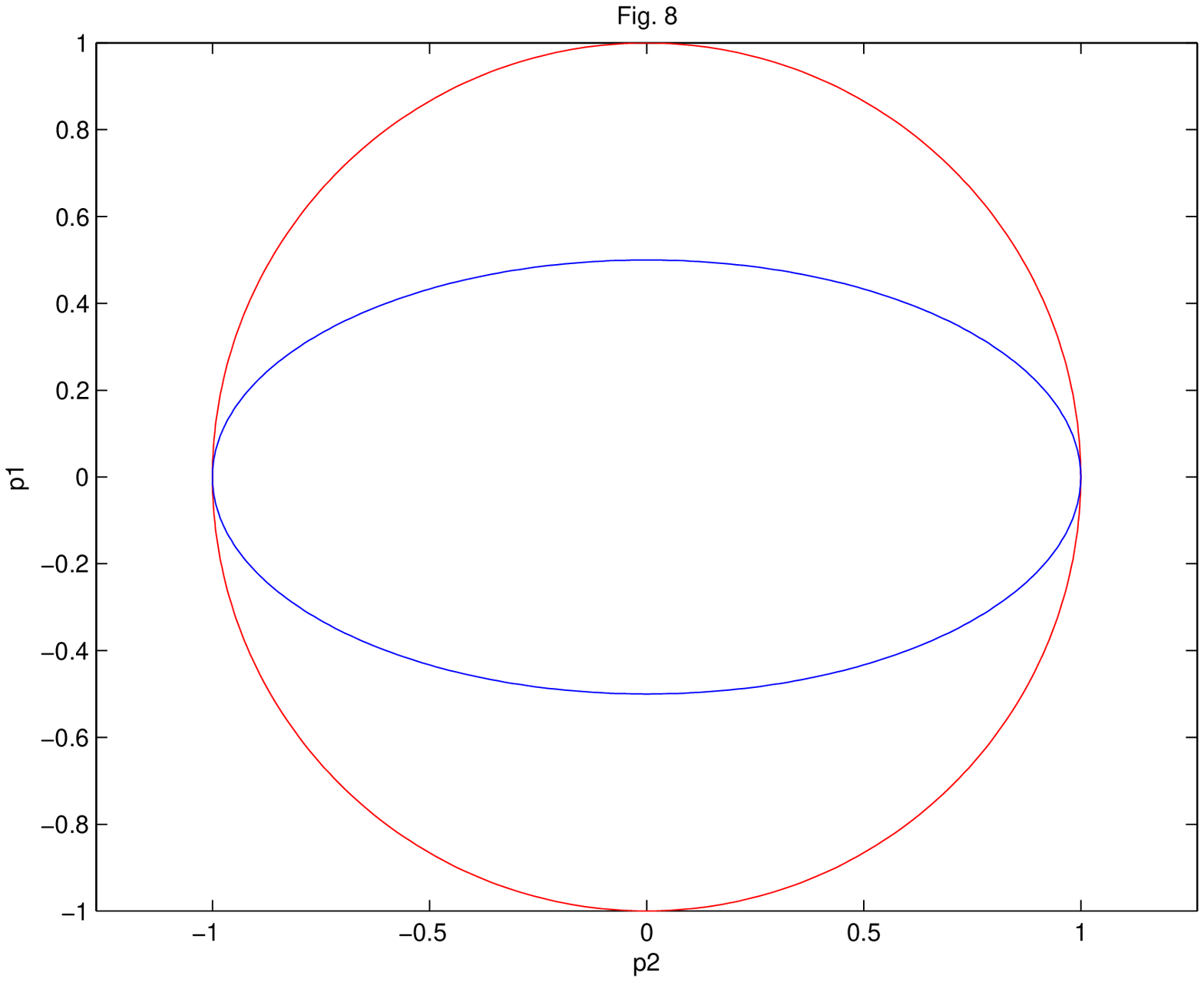,height=4in,width=5.5in}}
\caption{}
\end{figure}

\newpage

\begin{figure}[htb]
\centering
\centerline{\psfig{file=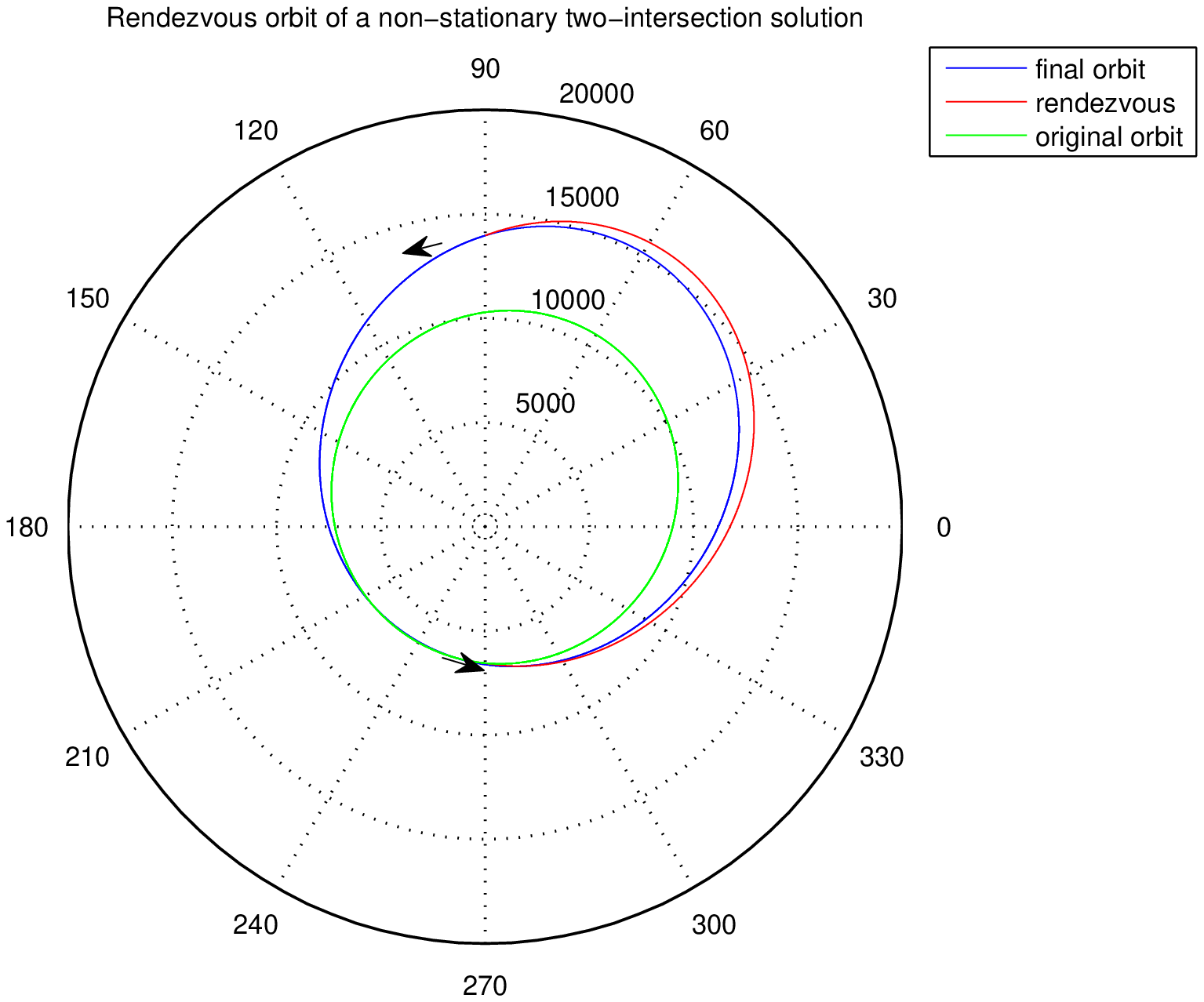,height=4in,width=5.5in}}
\caption{}
\end{figure}

\newpage

\begin{figure}[htb]
\centering
\centerline{\psfig{file=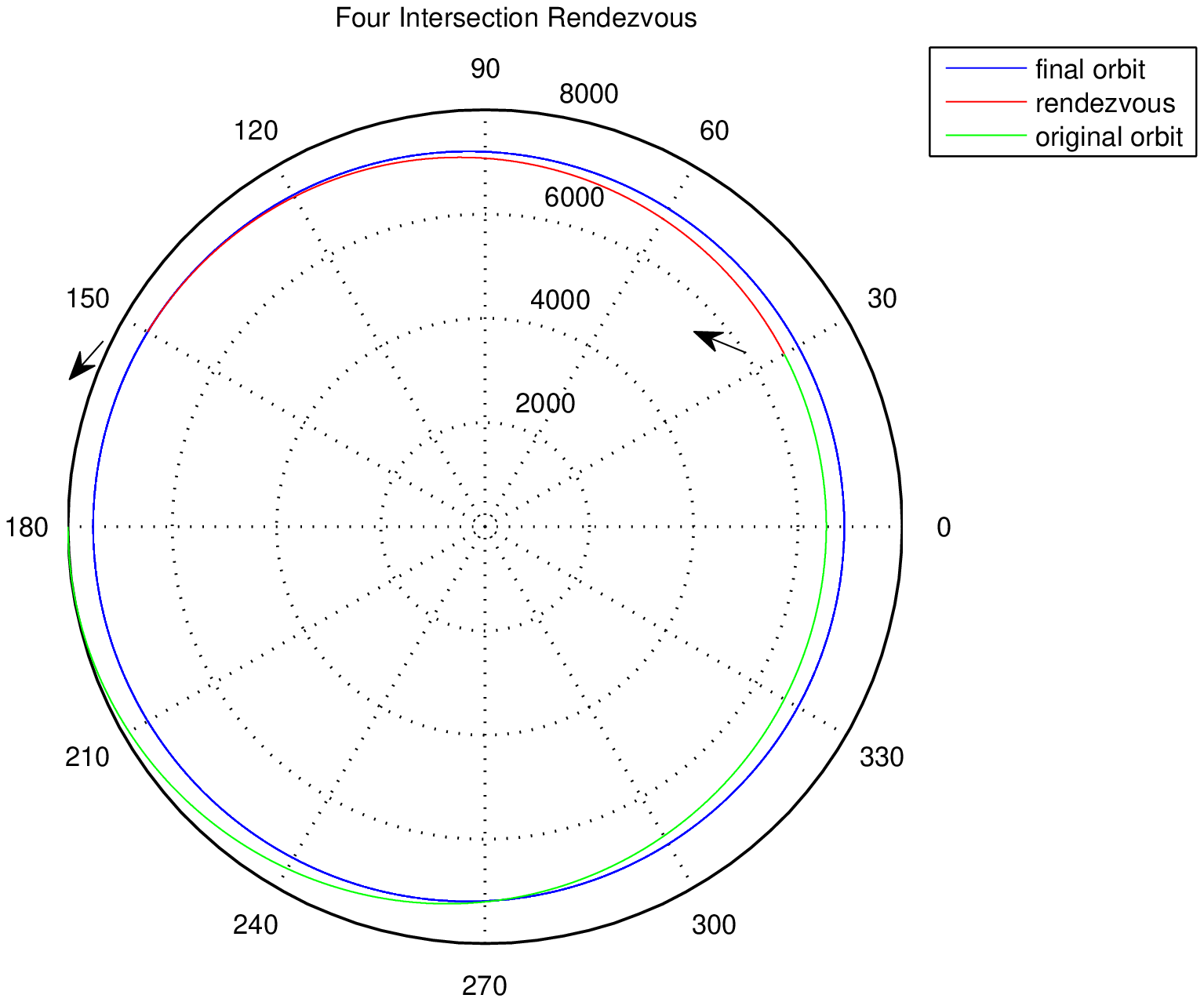,height=4in,width=5.5in}}
\caption{}
\end{figure}
\newpage

\begin{figure}[htb]
\centering
\centerline{\psfig{file=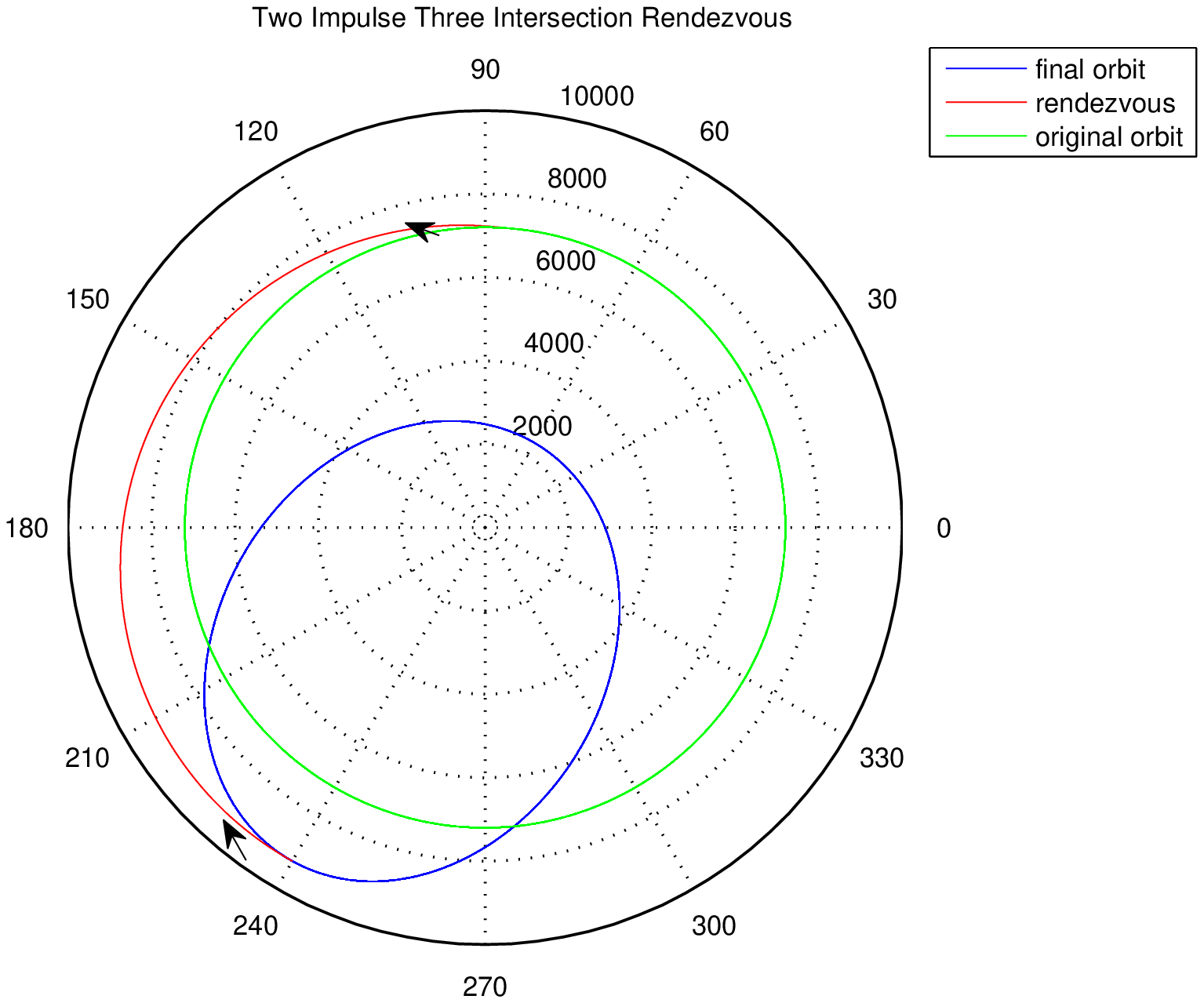,height=4in,width=5.5in}}
\caption{}
\end{figure}
\newpage

\begin{figure}[htb]
\centering
\centerline{\psfig{file=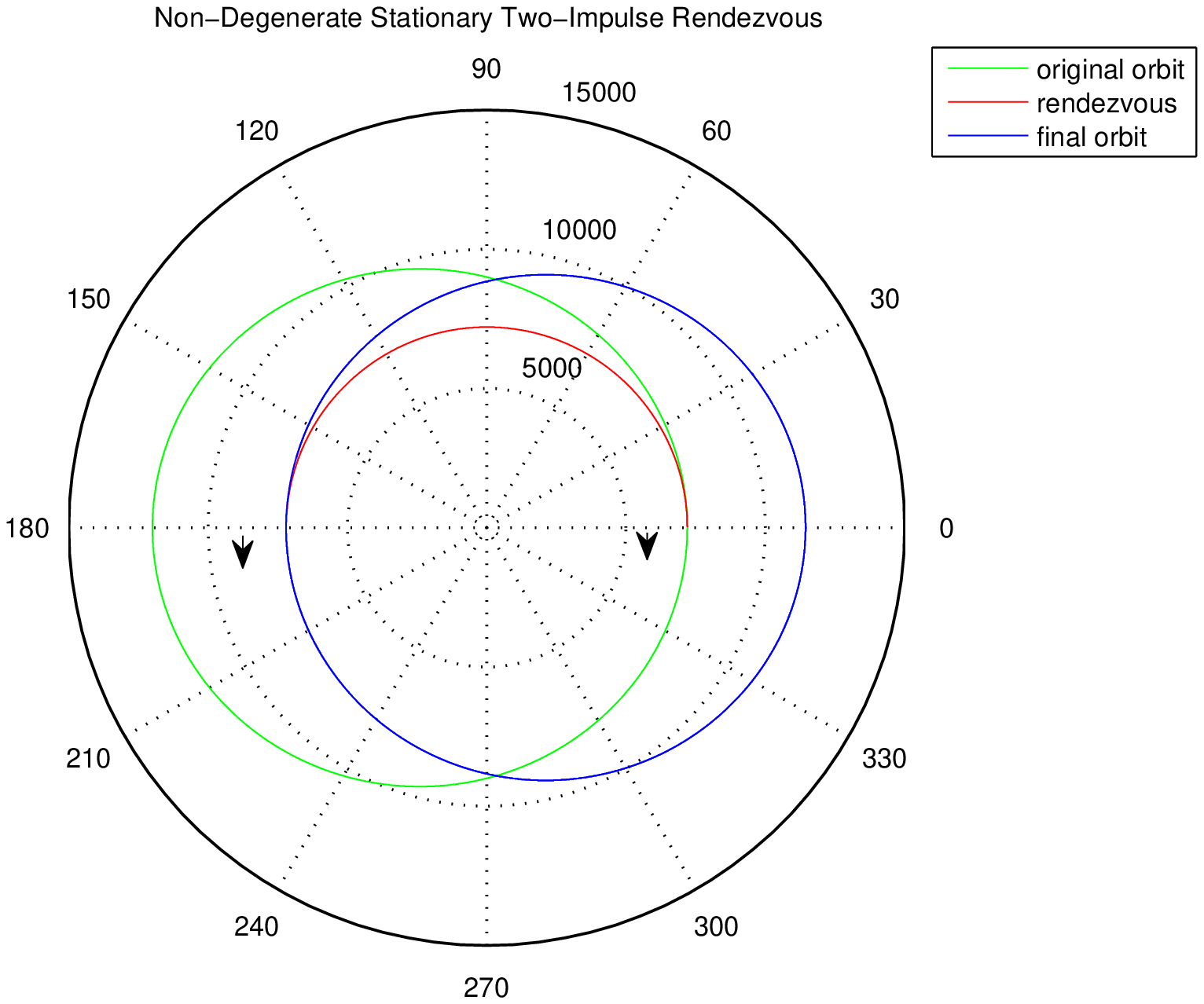,height=4in,width=5.5in}}
\caption{}
\end{figure}

\end{document}